\documentclass[12pt, draftclsnofoot, onecolumn]{IEEEtran}
\usepackage[T1]{fontenc}
\usepackage[linesnumbered,,vlined,ruled,commentsnumbered]{algorithm2e}
\usepackage{hyperref}
\usepackage{bm}
\usepackage{latexsym, graphicx, textcomp}
\usepackage{epsfig,upgreek}
\usepackage{amsfonts, amssymb, amsthm,makecell,enumerate}
\usepackage{float,color,dblfloatfix}
\usepackage{fancyhdr}
\usepackage{amsmath}
\usepackage{amsfonts} 
\usepackage{mathrsfs}
\usepackage{pifont}
\usepackage{amssymb}
\usepackage{graphicx}
\usepackage{multicol}
\usepackage{cases}
\usepackage{cite}
\usepackage{booktabs}
\usepackage{setspace}
\usepackage{subfigure}
\usepackage{pifont}
\usepackage{setspace}
\usepackage{float}
\usepackage[marginal]{footmisc}

\def\footnoterule{
\kern-5pt
\hbox to \columnwidth{\vrule width 0.618\columnwidth height 0.4pt\hfill}
\kern4.6pt}

\DeclareMathOperator*{\argmin}{arg\,min}
\DeclareMathOperator*{\argmax}{arg\,max}

\def\ge{\geqslant}

\def\le{\leqslant}
\def\leq{\leqslant}

\graphicspath{{figures/}}

\newtheorem{pf*}{Proof}

\ifCLASSINFOpdf
\else

\fi
\allowdisplaybreaks
\begin{document}

\title{Unifying Futures and Spot Market: Overbooking-Enabled Resource Trading in Mobile Edge Networks}

\author{Minghui Liwang,~\IEEEmembership{Member, IEEE,} Ruitao Chen,
        Xianbin Wang,~\IEEEmembership{Fellow, IEEE}, Xuemin (Sherman) Shen, \IEEEmembership{Fellow, IEEE}
% <-this % stops a space
\noindent
\thanks{Minghui Liwang, Ruitao Chen, and Xianbin Wang are with the Department of Electrical and Computer Engineering, Western University, London, Ontario N6A 5B9, Canada. Xuemin (Sherman) Shen is with the Department of Electrical and Computer Engineering, University of Waterloo, Ontario N2L 3G1, Canada. 

\noindent
E-mail: \{mliwang, rchen328, xianbin.wang\}@uwo.ca, sshen@uwaterloo.ca

\noindent
Corresponding author: Xianbin Wang

}

% <-this % stops a space
}

%% The paper headers
%{Shell \MakeLowercase{\textit{et al.}}: Integrating Futures and Spot Market: Overbooking-enabled Resource Trading in Edge Computing-assisted Mobile Networks}

\maketitle

\begin{abstract}
\noindent
Securing necessary resources for edge computing processes via effective resource trading becomes a critical technique in supporting computation-intensive mobile applications. Conventional onsite spot trading could facilitate this paradigm with proper incentives, which, however, incurs excessive decision-making latency/energy consumption, and further leads to underutilization of dynamic resources. Motivated by this, a hybrid market unifying futures and spot is proposed to facilitate resource trading among an edge server (seller) and multiple smart devices (buyers) by encouraging some buyers to sign a forward contract with seller in advance, while leaving the remaining buyers to compete for available resources with spot trading. Specifically, overbooking is adopted to achieve substantial utilization and profit advantages owing to dynamic resource demands. By integrating overbooking into futures market, mutually beneficial and risk-tolerable forward contracts with appropriate overbooking rate can be achieved relying on analyzing historical statistics associated with future resource demand and communication quality, which are determined by an alternative optimization-based negotiation scheme. Besides, spot trading problem is studied via considering uniform/differential pricing rules, for which two bilateral negotiation schemes are proposed by addressing both non-convex optimization and knapsack problems. Experiential results demonstrate that the proposed mechanism achieves mutually beneficial player’s utilities, while outperforming baseline methods on critical indicators, e.g., decision-making latency, resource usage, etc.

\end{abstract}

\begin{IEEEkeywords}
  Futures trading, spot trading, overbooking, forward contract, mobile edge networks
\end{IEEEkeywords}
\vfill
\IEEEpeerreviewmaketitle

\section{Introduction}

\noindent
\IEEEPARstart{T}{he} evolution of wireless technologies and explosive proliferation of smart devices have enabled a wide range of mobile applications~\cite{1,2}, such as online gaming, augmented/virtual reality and healthcare monitoring, etc., which have attracted significant number of users. However, many of the aforementioned applications are computation-intensive and require complicated onboard processing, posing great challenges to smart devices with limited computing resources and capability. Besides, limited battery power supply presents another major difficulty for intensive data processing, exchange, and decision-making on a single mobile device, that may hinder the application completion in real-time~\cite{3,4,5}. One feasible solution to overcome these challenges is cloud computing, which, however, may potentially incur delays, and burdens on cloud servers as well as backhaul links~\cite{1}. To further address these drawbacks, edge computing~\cite{3,4,6} has become a popular paradigm by exploring distributed computing/storage/communication capability at the edge of mobile networks, and thus offers flexible and cost-effective computing services for resource-constrained smart devices.
\vspace{-0.4cm}

\subsection{Motivation}

\noindent
Ensuring the needed resources for edge-assisted computing processes often relies on a certain form of resource trading, where a smart device can offload a certain amount of task data to the edge server, via wireless link to the nearby access point (e.g., base station, etc.), while paying for the obtained resources and computing services. However, conventional trading mechanisms (e.g., onsite spot trading) may face challenges caused by the dynamic nature of the resource trading market under mobile edge network architecture.

\noindent
\textbf{Motivation of futures-based resource trading:} To facilitate resource provisioning with proper incentives, spot trading among different parties has been widely adopted which allows resource buying and selling among servers and requestors (collectively known as players) under an onsite manner. Specifically, players in spot trading can reach a consensus on factors such as the amount of trading resources and the relevant price based on the current network/platform/application-related conditions, e.g., the present resource supply/demand and wireless channel quality. However, spot trading may lead to undesirable performance degradation, which are detailed below:

%\begin{itemize}

\noindent
$\bullet$ \textit{Latency on decision-making}: During each trading, spot players often have to spend extra latency to reach the final trading consensus, which may dramatically reduce the available time for dynamic resource sharing, especially for networks with moving devices, short trading period, as well as time-varying resources. 
For example, considering 2 seconds as the connection duration between a smart device and the nearby AP, and 1.5 seconds as the decision-making latency, there are only 0.5 seconds left for the actual computing service. Besides, resource condition (e.g., supply and demand) may also change during a long decision-making procedure.
Thus, timely resource provisioning presents a major challenge under dynamic network conditions when applying spot trading mechanism.

\noindent
$\bullet$ \textit{Energy consumption on decision-making}:
 During each trading, spot players may suffer from extra energy consumption to reach the trading consensus, bringing difficulties to power- and battery-constrained mobile devices. For example, a long decision-making procedure can cause heavy battery loss, that directly decreases the endurance of a smart device. Consequently, designing an energy-efficient trading mechanism is considered to be urgent and critical.
%\end{itemize}

Motivated by the abovementioned challenges, \textit{futures} trading~\cite{5} is applied as an
effective paradigm, which enables a forward contract between server and requestor with contract term such as the amount of trading resources, the relevant resource price, and default clause, etc., via analyzing historical data. Specifically, the pre-signed contract will be fulfilled during each practical trading in the future, without any further onsite discussion. 
Nevertheless, although futures can achieve low decision-making latency and energy
consumption, it may also be risky due to the insufficient and inaccurate
knowledge of historical statistics. Thus, this work investigates a novel hybrid
market via unifying both futures trading and spot trading in mobile edge networks.

%\subsubsection{Motivation of Overbooking}

\noindent
\textbf{Motivation of overbooking:} One common resource presale mode refers to equal-booking, where the amount of resources for booking (presale) generally does not exceed the resource owner's maximal resource supply. However, there exist several challenges, i.e. variation of resource demands and unreliable wireless communication links could prevent utilization of confirmed resources. Thus, \textit{overbooking}~\cite{7} has been encouraged where the total available capacity of resource owner is less than the theoretical maximal
required capacity of resource requestors, mainly
motivated by ``no shows''~\cite{8}. For example, a mobile device
has no task execution requirement during a trading may no longer need to use the required resources in accordance with the pre-signed contract. Consequently, overbooking efficiently offers substantial utilization and profit advantages in handling dynamic and unpredictable resource demands. Although overbooking may seem risky, it is a common technique widely adopted in commercial domains such as airlines~\cite{9}, hotels~\cite{10}, bandwidth reservation~\cite{11,12}, etc. For instance,
aiming to maximize the occupancy (and thus revenue), airlines routinely
overbook tickets by ensuring the maximum number of passengers on a flight;
otherwise, flights often depart with up to 15{\%} seats empty (without considering
overbooking), and thus incurs unsatisfying resource utilization and
economic~losses~\cite{8}. Thus, utilization inefficiency incurred by the dynamic and unpredictable resource demands of smart mobile devices greatly motivates us to study the overbooking-enabled resource trading mechanism.

Driven by the abovementioned motivations, this paper investigates a hybrid
market via unifying both futures and spot trading, under mobile edge networks, and proposes a novel overbooking-enabled resource trading
mechanism. Specifically, we consider an edge server with limited resource supply as
resource seller, and multiple smart devices with computation-intensive tasks as
resource buyers, each of which may purchase computing service from the seller
by offloading certain amount of task data through wireless communication.
The proposed trading mechanism effectively alleviates the unexpected latency
and cost (e.g., energy and battery consumption) on trading decision-making,
and greatly improves the resource utilization and time efficiency.
\vspace{-0.4cm}
\subsection{Related Work}

\noindent
\textbf{Resource trading mechanism:} Existing works devoted to resource
trading roughly fall into three categories: i) spot trading (also known as
onsite trading), where players reach a trading agreement relying on current
conditions (e.g., the current resource demand and supply, channel quality,
etc.), such as online game~\cite{13,14}, auction~\cite{15,16,17},
and bilateral negotiation~\cite{18,19}; ii) futures trading, where
players sign a forward contract over buying or selling a certain amount of
resources at a predetermined price in advance, that will be fulfilled during
each trading in the future, where existing studies mainly investigate
electricity market~\cite{20,21,22}, spectrum resource trading~\cite{23,24},
 and edge computing-assisted networks~\cite{5,25};
and iii) resource trading in hybrid market where both futures and spot
trading are allowed~\cite{26,27}.
In~\cite{13}, \textit{Messous et al}. investigated the computation offloading
problem in an edge computing-enabled UAV network by establishing a
non-cooperative theoretical game involving multiple players and three pure
strategies. A multi-user non-cooperative offloading game was investigated by
\textit{Wang et al}. \cite{14}, intended to maximize the utility of each vehicle
via a distributed best response algorithm. Considering resource auctions,
the VM allocation among edge clouds and mobile users was studied as an
n-to-one weighted bipartite graph matching problem by \textit{Gao et al}. in~\cite{15},
based on a greedy approximation algorithm. In~\cite{16},
\textit{Liwang et al}. studied a Vickrey--Clarke--Groves-based reverse auction
mechanism of vehicle-to-vehicle resource trading and suggested a unilateral
matching-based mechanism. In~\cite{17}, \textit{Gao et al}. developed a truthful
auction under computing resource trading market via considering graph tasks,
while providing both the optimal and an efficient sub-optimal algorithms.
\textit{Shojaiemehr et al}. in~\cite{18} proposed a novel
negotiation strategy to enhance the satisfaction of both trading parties
while supporting negotiation of composite cloud service. In~\cite{19},
\textit{Wang et al}. presented a smart contract-based negotiation framework while
providing a Bayesian Nash equilibrium of service providers which offer
flexible QoS.
However, the procedure to reach a trade-related decision usually results in
excessive latency and energy consumption~\cite{5,23,28}, which further
pose challenges to spot trading players. Take online auction as an example,
the winners gain the eventual auction contract while there is no such
compensation for the losers who have also spent extra time and energy during
decision making. Moreover, the latency from bidding to practical computing
service delivery can greatly impact the quality of experience and the
utilization if resources are reserved while waiting for the auction results
~\cite{5,8,23}.

Therefore, futures has been emerged as a practical paradigm and extensively
adopted in financial and commodity exchange markets. Benefitted from the
pre-signed forward contract, the unexpected time and energy consumption on
decision-making can be efficiently decreased. \textit{Khatib et al}. in~\cite{20}
proposed a systematic negotiation scheme, through which, a generator and
load can reach a mutually beneficial forward bilateral
contract in electricity markets. In~\cite{21}, \textit{Conejo
et al}. addressed the power producer's optimal involvement problem in a
futures electricity market, aiming to hedge against the risk of pool price
volatility. In~\cite{22}, \textit{Morales et al}. investigated scenario reduction
techniques to accurately convey the uncertainties in futures market trading
in electricity markets. In spectrum resource trading market, \textit{Sheng et al}. in~\cite{23}
 proposed a futures-based spectrum trading mechanism to
alleviate trading failures, and trading unfairness caused by price
fluctuation. In~\cite{24}, \textit{Li et al.} introduced a futures market to
manage the financial risk in spectrum trade and discovering future price.
Topics associated with futures-based resource trading have
rarely been studied in mobile edge networks, where factors such
as unpredictable nature of resource supply and demand, as well as the
ever-changing channel quality between resource provider and requestor caused
by mobile users' mobility, pose great difficulties to trading mechanism
design. We were among the first to address such challenges~\cite{5,25}.
In our previous work~\cite{25}, we investigated a futures-based resource
trading approach in edge computing-enabled internet of vehicles, where a
risk tolerable and mutually beneficial forward contract was designed through
estimating the historical statistics of future resource supply and network
condition. In~\cite{5}, we proposed an energy-aware resource trading mechanism under
edge computing-assisted UAV networks, where both the forward
contract design problem and power optimization problem were carefully analyzed.

Although futures brings benefits, it may also be risky due to factors such
as the lacking and inaccurate knowledge of historical data. Motivated by
which, several works also consider the integration of both the futures and
spot market. In~\cite{26}, \textit{Gao et al}. focused on the optimal spectrum
allocation among unlicensed secondary users in a hybrid market, which
maximized the secondary spectrum utilization efficiency. \textit{Vanmechelen et al}.
in~\cite{27} proposed a hybrid market in which a low-latency spot market
coexists with a higher latency futures market, to deal with the significant
delay of the allocation decision procedure of grid resources.
%\subsubsection{}

\noindent
\textbf{Overbooking:} ``Booking'' refers to a presale manner (rather than
spot trading), where ``overbooking'' presents the presale of a volatile
commodity or service in excess of actual supply, which has been shown to
provide substantial utilization and profit advantage under ``no shows''
(some consumers will cancel the trading of requested service)~\cite{8},
which, however, has been neglected in most previous mentioned works 
(namely, these works mainly consider equal-booking where the amount of
resources for sale equals to the actual supply). The widespread adoption of
overbooking techniques focus on many fields such as airlines and hotels~\cite{9,10}, spectrum reservation~\cite{11,12}, storage market~\cite{29}, network slicing~\cite{30,31,32},
 cloud computing~\cite{7,33,34,35,36}, and fog computing~\cite{37}. Specifically,
\textit{Liu et al}. in~\cite{11} proposed an opportunistic link overbooking
scheme for an edge gateway to improve its link efficiency, and developed an
integrated analytical framework for determining the suitable link
overbooking factor. In~\cite{12}, \textit{Adebayo et al}. proposed a spectrum
reservation prediction algorithm for wireless infrastructure providers to
reduce the probability of overbooking since it costs certain penalties. \textit{Gao
et a}l. in~\cite{29} proved that overbooking strategy plays an
important role in improving storage renting efficiency. \textit{Zanzi et al}.
in~\cite{30} deployed an overbooking network slices solution and a 5G
network slice broker as an entity in charge of mediating between vertical
network slice requests and physical network resources availability.
Additionally, in~\cite{31}, \textit{Zanzi et al}. proposed an orchestration
through a dashboard, allowing requesting network slices on-demand, monitored
their performance once deployed and displayed the achieved multiplexing gain
through overbooking. In~\cite{32}, \textit{Sexton et al}. employed the practice
of overbooking to increase resource utilization when offering auxiliary
resources in network slicing.
Among existing works considered overbooking, the most similar studies with
this work fall into cloud computing environment such as~\cite{7,33,34,35,36},
 and fog computing~\cite{37}.
\textit{Tomas et al}. in~\cite{7} focused on implementing an autonomic risk-aware
overbooking architecture capable of increasing the resource utilization of
cloud data centers by accepting more virtual machines than physical
available resources. In~\cite{33}, \textit{Son et al}. proposed a service level
agreement (SLA)-aware dynamic overbooking strategy in software defined
networking (SDN)-based cloud data centers, which jointly leveraged
virtualization capabilities and SDN for virtual machine (VM) and traffic
consolidation. \textit{Alanazi et al}. in~\cite{34} introduced an integrated
resource allocation framework for data centers that minimizes the number of
active physical machines through dynamic VM placement while ensuring that
SLAs of admitted VMs are not violated. \textit{Rahimzadeh et al}. proposed a cloud
resource management system that overbooks backup VMs by optimizing the
overbooking rate tradeoff in~\cite{35}. In~\cite{36}, \textit{Yao et al}.
presented an optimal overbooking policy to maximize resource providers'
profits in cloud federation and enhance cloud users' experiences.
\textit{Zhang et al}. \cite{37} studied a dynamic
resource allocation model through overbooking mechanism to maximize the
total welfare of fog servers.

Although ``resource overbooking'' has been applied in several fields and
achieves good performance, few of them paid attention to computing resource
trading problem in edge computing-enabled mobile networks. Besides, characteristic features in wireless communication environment, e.g., varying channel qualities, etc., also bring difficulties to trading mechanism design.

%Notably, this paper has made the following major differences, as compared with existing
%works in similar field:

%\begin{enumerate}
%\noindent
%$\bullet$ \textbf{Market and mechanism:}
% This work investigates a novel integration of both futures market and spot market, where an overbooking-enabled resource trading mechanism is studied, which, however, has rarely been considered in existing works under edge computing-assisted mobile network architecture.
%
%\noindent
%$\bullet$ \textbf{Random nature of the trading system:}
%This work analyzes the impacts of the unpredictable random nature of wireless communication environment (e.g., varying channel qualities and changing resource demands), which has been somehow neglected in most works in similar field;
%
%\noindent
%$\bullet$ \textbf{Major goals:} Comparing with existing works, this paper has different goals. For example, we aim to achieve mutually beneficial utilities for both resource owners and requestors, rather than those concerning single objective (e.g., the revenue of service providers). Then, this work puts emphasis on reducing the unexpected latency and energy consumption of trading decision-making, which, however, lacks consideration in existing works (e.g., online auction may result in excessive latency to reach the final trading decision~\cite{8}). Additionally, overbooking has been applied in this paper that efficiently improves the utilization of computing resource and time during each trading, which has rarely been investigated in related works.

%\end{enumerate}
\vspace{-0.5cm}
\subsection{Novelty and Contribution}

\noindent
To the best of our knowledge, this paper is among the first to study 
overbooking-enabled resource trading among an edge server (resource seller)
and multiple smart devices (resource buyers), via considering a hybrid
market integrating futures and spot. Specifically, in futures market, two
major issues are considered: i) overbooking rate design: players
determine a feasible overbooking rate, indicating the number of buyers that
can sign the forward contract with the seller (we name these buyers as
members); and ii) forward contract design: players determine a reasonable
forward contract, including the price of resource, the penalty that a member
has to pay to the seller if it breaks the forward contract, and the
compensation that a member with task can receive if the seller cannot offer
computing service due to overbooking. In spot market, the remaining buyers
who have not signed forward contract (we name these buyers as non-members)
can compete for available resources (if any) based on the current network
conditions. Major contributions are summarized as follows:

%\begin{itemize}

\noindent
$\bullet$ This paper introduces a novel hybrid resource trading market via integrating both futures and spot under mobile edge network architecture, which effectively alleviates extra latency and cost (e.g., energy consumption, etc.) on trading decision-making. Besides, overbooking is adopted which allows a larger number of members than the seller's capacity, that greatly achieves the improvements on both resource utilization and time efficiency.

\noindent
$\bullet$ To capture the unpredictable random nature of the resource trading market, two key uncertainties are considered: buyer's task arrival, which directly affects the resource demand; and the varying wireless channel quality, which reflects the unstable network condition caused by factors such as the mobility of each buyer. Specifically, buyers are divided into members and non-members, where members can sign a forward contract with the seller in advance, which will be fulfilled during each future practical trading; while non-members with tasks have to compete for available resources under a spot trading manner.

\noindent
$\bullet$ The proposed mechanism considers solving two key problems associated with different markets. The resource trading problem in futures market mainly relies on designing the feasible forward contract and overbooking rate, which is formulated as a multi-objective optimization (MOO) problem aiming to maximize both the seller's and the members' expected utilities, via analyzing historical statistics of the abovementioned key uncertainties (task arrival condition and wireless channel quality). Moreover, possible risks that players may face with during each practical trading are evaluated as constraints. To tackle this problem, an efficient bilateral negotiation scheme is proposed that facilitates the players reaching a consensus on futures trading.

\noindent
$\bullet$ In spot market, resource trading is defined as a MOO problem via maximizing seller's and each non-member's utilities based on the current resource supply and demand, as well as wireless channel qualities, under uniform pricing and differential pricing rules. To address the spot trading problem, for each pricing rule, we propose a bilateral negotiation-based scheme through solving a non-convex optimization problem and a knapsack~problem, within polynomial time.

\noindent
$\bullet$ Comprehensive simulation results demonstrate that the proposed overbooking-enabled resource trading mechanism in hybrid market achieves mutually beneficial players' utilities, while outperforming baseline methods on significant indicators such as decision-making latency and cost, task completion time, as well as time and resource utilization.
%\end{itemize}

%The rest of this paper is organized as follows. In Section~II, we introduce
%system models. Problem formulation and the relevant solution in futures
%market, and spot market is detailed in Section~III, and Section~IV,
%respectively. Simulation results are analyzed in Section~V before drawing
%the conclusion in Section~VI.

\section{System Model}
\subsection{Key Definition and System Overview}

\noindent
Considering a futures and spot integrated resource trading market containing multiple buyers (smart devices) $\bm{\mathcal{B}}=\left\{{\bm{b_1}},\ldots\,{\bm{b_m}},\ldots ,\bm{b}_{|\bm{\mathcal{B}}|}\right\}$, where each buyer may have a computation-intensive task needed to be processed in a trading; and one seller (edge server) with limited computing resources denoted by $Sd^{comp}$ (e.g., CPU cycles, $S$ presents a positive integer, $d^{comp}$ indicates the required amount of computing resources per task). Specifically, we consider a trading market where resource demand may exceed resource supply, $S<|\bm{\mathcal{B}}|$. Key definitions are listed below:

\noindent
\textbf{Definition 1} (Futures market and member). \textit{In futures market, some
of the buyers can sign a forward contract with the seller in advance, which
will be fulfilled with no further negotiation during each trading in the future. Correspondingly, we call a buyer with forward contract as a member, and the number of members is denoted by $\kappa $}.

\noindent
\textbf{Definition 2} (Spot market and non-member). \textit{In spot market, each
buyer without forward contract can purchase computing resources from the
seller under a real-time and on-demand manner through spot trading, we
regard these buyers as non-members.}

\noindent
\textbf{Definition 3} (Performer). \textit{A performer indicates a member who has 
task execution requirement during a trading}.

\noindent
\textbf{Definition 4} (Practical performer). \textit{A practical performer indicates
a performer who can practically obtain the required resources and service from seller during a trading}.

\noindent
\textbf{Definition 5} (Defaulter). \textit{A defaulter indicates a member who has no
task execution requirement during a trading although it has signed
the forward contract. Each defaulter has to pay penalty to the seller for
breaking the forward contract.}

\noindent
\textbf{Definition 6} (Volunteer). \textit{A volunteer indicates a performer who has
to process its task locally since the seller fails to afford performers'
task execution requirements due to ``overbooking''; correspondingly, each
volunteer will receive compensation from the seller.}

\noindent
\textbf{Definition 7} (Forward contract and contract term). \textit{A forward contract represents a trading consensus between the seller and members (prior to each practical trading), which will be fulfilled during each trading in the future. We consider three key contract terms: $p,q$, and $r$, where $p$ indicates the agreed unit price of resources, $q$ denotes the unit penalty that a defaulter has to pay to the seller ($p>q$), and $r$ refers to the unit compensation for each volunteer from the seller ($r>0$).}

\noindent
\textbf{Definition 8} (Overbooking rate). \textit{The overbooking rate ${\kappa }^o$ denotes the ratio of overbooked resources to the total available resources of the seller, which is calculated by ${\kappa}^{o}=\left(\kappa -S\right)/S$.}

\noindent
This paper studies an efficient resource trading mechanism considering two key problems: i) futures market focuses on designing a risk-aware and mutually beneficial forward contract, as well as the feasible overbooking rate ${\kappa}^{o}$ (namely, the number of members $\kappa$) by analyzing historical statistics (e.g., the task arrival of each buyer and channel quality between each buyer and the seller), to maximize the expected utilities of the seller and members; and ii) spot market concerns the design of spot trading mechanism among the seller and non-members, helping players obtain better utilities under real-time and on-demand manner. Fig. 1 provides the framework of the proposed resoure trading and some examples.

%f1
\begin{figure*}[!t]
\centering
\subfigtopskip=0.8pt
\subfigbottomskip=0.8pt
\includegraphics[width=\linewidth]{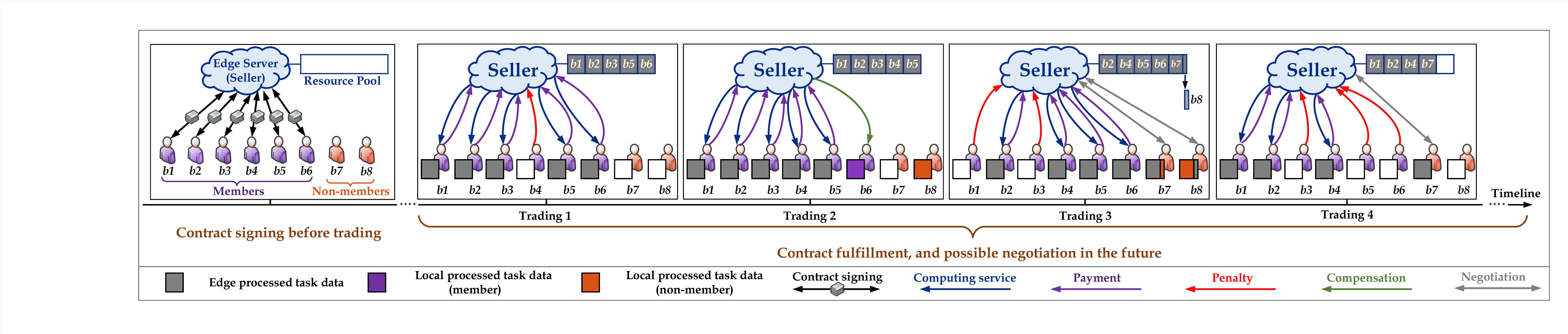}
%\vspace{-0.5cm}
\caption{Proposed overbooking-enabled resource trading framework and relevant examples ($S=5, \left|\bm{\mathcal{B}}\right|=8$, $\kappa =6$).}
\label{fig1}
\end{figure*}
\vspace{-0.5cm}
\subsection{Modeling of Buyers }

\noindent
 Considering buyers set $\bm{\mathcal{B}}$, where each buyer ${\bm{b_m}}\in \bm{\mathcal{B}}$ may have a task that needs to be processed during a trading, denoted by a 7-tuple ${\bm{b_m}}=\{d^{size},d^{comp},f^{b},e^{loc},e^{tran},{\alpha_m},\gamma_m\}$. Specifically, $d^{size}$ and $d^{comp}$ indicate the data size (e.g., bits), and the required amount of computing resources (e.g., CPU cycles) of the buyer's task, respectively\footnote{Suppose that tasks of buyers have the same data size and thus require the same amount of computing resources, for analytical simplicity.}. 
 $f^{b}$ denotes the local computing capability (CPU cycles/s) of each buyer, $e^{loc}$ and $e^{tran}$ describe the local computing power consumption (Watt), and the transmission power (Watt) of $\bm{b_m}$, respectively.
 Two key uncertainties are considered to describe the unpredictable nature of the trading process: ${\alpha}_{m}$ and ${\gamma}_{m}$\footnote{In this paper, trading statistics of uncertainties $\alpha_m$ and $\gamma_m$ are assumed to be
known based on the historical records\cite{5,23}.}. Specifically, ${\alpha}_{m}\in \left\{0,1\right\}$ represents the each buyer's task arrival during each trading (namely, if the buyer has a task execution requirement), which is a discrete random variable obeying the Bernoulli distribution denoted by ${\alpha}_{m}\sim\text{B}\left(\left\{1,0\right\},\{a,1-a\}\right)$. Thus, the relevant probability mass function (PMF) of ${\alpha}_{m}$ is given in~\eqref{eq1}.
{\setlength\abovedisplayskip{5pt}
\setlength\belowdisplayskip{5pt}
 \begin{equation}
 \label{eq1}
\mathrm{Pr}\left({\alpha}_{m}=i\right)=
\begin{cases}
a,&i=1\vspace{-1.5ex}\\
1-a, &i=0\\
\end{cases}
 \end{equation}
 }To better capture the uncertainty of the wireless communication environment, ${\gamma}_{m}$ is applied to describe the varying channel quality between buyer $\bm{b_m}$ and the seller, which represents a continuous random variable obeying an uniform distribution~\cite{5,23} in interval $[{\varepsilon }_1,{\varepsilon}_2]$, denoted by ${\gamma}_{m}\sim {\text{U}}({\varepsilon}_1,{\varepsilon}_2)$. Notably, we assume that all the buyers are independent and identically distributed (i.i.d). For notational simplicity, let $ \bm{\mathcal{A}}=\{{\alpha }_1,\ldots,{\alpha}_{m},\ldots,{\alpha}_{|\bm{\mathcal{B}}|}\}$ and $\bm{\mathcal{Y}}=\{{\gamma }_1,\ldots,{\gamma }_{m},\ldots,{\gamma}_{|\bm{\mathcal{B}}|}\}$ denote the vector of random variables ${\alpha}_{m}$ and ${\gamma}_{m}$, respectively.

%\subsubsection{Task completion time and energy consumption}

\noindent
\textbf{1) Task completion time and energy consumption:}
For each buyer, the local task completion time is calculated as $t^{loc}=\frac{d^{comp}}{f^{b}}$, and the relevant local energy consumption is thus given by $c^{loc}=e^{loc}t^{loc}=\frac{e^{loc}d^{comp}}{f^{b}}$ ~\cite{1,3,4,38}. Additionally, the task completion time of buyer ${\bm{b_m}}$ when it offloads a certain amount of task data to the seller is defined by~\eqref{eq2}, where $e^{tran}{\gamma}_{m}$ indicates the received SNR~\cite{39} of the seller from buyer ${\bm{b_m}}$.
{\setlength\abovedisplayskip{5pt}
\setlength\belowdisplayskip{5pt}
 \begin{equation}
 \label{eq2}
 t^{edge}_{m}\!=\!{\left(\frac{{\lambda}_{m}d^{size}}{W{{\log}}_{\text{2}}\left(1\!+\!e^{tran}{\gamma}_{m}\right)}+
 \frac{{\lambda}_{m}d^{comp}}{f^{s}},\frac{(1\!-\!{\lambda}_{m})d^{comp}}{f^{b}}\right)}^+,
 \end{equation}
 }where ${\lambda}_{m}$ $(0\le {\lambda}_{m}\le 1$) denotes the offloading rate of $\bm{b_m}$; symbol ${(i,j)}^+$ refers to the larger value between $i$ and $j$; ${\lambda}_{m}d^{size}$ and ${\lambda}_{m}d^{comp}$ denote the amount of data offloaded to the seller, and the relevant required resources, respectively. $W$ represents the bandwidth of the wireless channel\footnote{For analytical simplicity, we do not consider interference of wireless communication, as supported by existing works~\cite{1,40}, and innovative techniques such as OFDMA~\cite{41}.} between each buyer and the seller, and $W{{\log}}_{{2}}\left(1+e^{tran}{\gamma}_{m}\right)$ indicates the relevant data transmission rate. Moreover, $f^{s}$ depicts the seller's computing capability (e.g., CPU cycles/s), which is considered as a stable value (e.g., the value of $f^{s}$ doesn't change with seller's workloads), as illustrated in existing works~\cite{1,3}. Correspondingly, the relevant energy consumption $c^{edge}_{m}$ of $\bm{b_m}$ is defined by the following~\eqref{eq3}.
%{\setlength\abovedisplayskip{5pt}
%\setlength\belowdisplayskip{5pt}
\begin{equation}
\label{eq3}
c^{edge}_{m}=\frac{e^{tran}{\lambda}_{m}d^{size}}{W{\log}_{2}
\left(1+e^{tran}{\gamma}_{m}\right)}+\frac{e^{loc}\times \left(1-{\lambda}_{m}\right)\times d^{comp}}{f^{b}}
\end{equation}
%}
%\subsubsection{Utility, expected utility, and risks of members in futures market}

\noindent
\textbf{2) Utility, expected utility, and risks of member in futures market: }
 To avoid the notational redundancy, we use $m$ to represent the index of members hereafter. For analytical simplicity, the first $\kappa $ buyers $\bm{b_1}$, ${\bm{b_2}},\ldots,{\bm{b_\kappa}}$ are considered as members\footnote{ In the proposed resource trading market, sign the forward contract with any $\kappa$  of the buyers has no impact on the solution design since all the buyers are i.i.d. For example, considering   $\left|\mathcal{B}\right|=6$ and $\kappa =3$, the seller contracts with buyers $\bm{b_1}$, $\bm{b_2}$, and $\bm{b_3}$ makes no difference with that with buyers $\bm{b_4}$, $\bm{b_5}$, and $\bm{b_6}$.}, which are encouraged to offload the whole task\footnote{Namely, the proposed resource trading market encourages members to buy more resources (than non-members) from signing the contract in advance, which is close the real-life commodity exchange market.} to the seller (e.g., ${\lambda}_{m}=1$). Correspondingly, the utility $U^{PP}_{m}$ of a member ${\bm{b_m}}$ who is a practical performer is defined as the weighted sum of the time and energy saved from enjoying the computing service, minus payment of the required resources, as given by~\eqref{eq4}:
% \begin{align}
% \label{eq4}
% \begin{aligned}
%   U^{PP}_{m} &\!=\!{\omega}_{{1}}\left(t^{loc}-t^{edge}_{m}\right)+{\omega}_{\text{2}}\left(c^{loc}-c^{edge}_{m}\right)-pd^{comp}\\
%   &\!=\!{\omega}_{{1}}\left(\frac{d^{comp}}{f^{b}}-
%\frac{d^{size}}{W{\log}_{2}\left(1+e^{tran}{\gamma }_{m}\right)}-\frac{d^{comp}}{f^{s}}\right)+\\
%&\quad {\omega}_{\text{2}}\left(\frac{e^{loc}d^{comp}}{f^{b}}\!-\!
%\frac{e^{tran}d^{size}}{W{\log}_{2}
%\left(1+e^{tran}{\gamma}_{m}\right)}\right)\!-\!pd^{comp},
% \end{aligned}
% \end{align}
% 
%double column
\begin{small}
  \begin{align}
 \label{eq4}
 \begin{aligned}
   U^{PP}_{m} &\!=\!{\omega}_{{1}}\left(t^{loc}-t^{edge}_{m}\right)+{\omega}_{\text{2}}\left(c^{loc}-c^{edge}_{m}\right)-pd^{comp}\\
   &\!=\!{\omega}_{{1}}\left(\frac{d^{comp}}{f^{b}}-
\frac{d^{size}}{W{\log}_{2}\left(1+e^{tran}{\gamma }_{m}\right)}-\frac{d^{comp}}{f^{s}}\right)+
{\omega}_{\text{2}}\left(\frac{e^{loc}d^{comp}}{f^{b}}\!-\!
\frac{e^{tran}d^{size}}{W{\log}_{2}
\left(1+e^{tran}{\gamma}_{m}\right)}\right)\!-\!pd^{comp},
 \end{aligned}
 \end{align}
\end{small}where ${\omega}_1$ and ${\omega}_2$ are positive weight coefficients. Correspondingly, the utility $U^{DE}$ of a defaulter can be calculated as $U^{DE}=-qd^{comp}$, indicating that a member has to pay a penalty when it breaks the forward contract. Owing to overbooking, some of the performers with the worst channel qualities\footnote{In this paper, we consider a fair volunteer selection scheme where the performers with the worst channel qualities will be selected as volunteers during each trading, since all the buyers are i.i.d.} may be selected as volunteers when the seller fails to support task execution requirements of members during a trading (e.g., $\sum\nolimits^{m=\kappa}_{m=1}{{\alpha}_{m}}>S$). Thus, we define the utility of each volunteer as $U^{VO}=rd^{comp}$, indicating the compensation from the seller. Correspondingly, the number of volunteers $V$ can be expressed by~\eqref{eq5}.
{\setlength\abovedisplayskip{5pt}
\setlength\belowdisplayskip{5pt}
\begin{equation}
\label{eq5}
V=\sum\nolimits^{m=\kappa }_{m=1}{{\alpha }_{m}}-{\left(\sum\nolimits^{m=\kappa }_{m=1}{{\alpha }_{m}},S\right)}^-
\end{equation}}Let $v_m$ be the volunteer selection indicator, where $v_m=1$ denotes $\bm{b_m}$ is chosen as a volunteer; $v_m=0$, otherwise. Then, the utility of member ${\mathcal{U}}^{Mem}(p,q,r,\kappa , \bm{\mathcal{A}},\bm{\mathcal{Y}})$ is formulated by the following~\eqref{eq6}.
{\setlength\abovedisplayskip{5pt}
\setlength\belowdisplayskip{5pt}
\begin{align}
%\label{eq6}
%&{\mathcal{U}}^{Mem}(p,q,r,\kappa,\bm{\mathcal{A}},\bm{\mathcal{Y}})\notag\\
%&=\sum\nolimits^{m=\kappa}_{m=1}\left({\alpha }_{m}U^{PP}_{m}\!+\!\left.\left(1\!-\!{\alpha }_{m}\right)U^{DE}\right)+\right (pd^{comp}+U^{VO})V
%\end{align}
\label{eq6}
{\mathcal{U}}^{Mem}(p,q,r,\kappa,\bm{\mathcal{A}},\bm{\mathcal{Y}})=\sum\nolimits^{m=\kappa}_{m=1}\left({\alpha }_{m}U^{PP}_{m}+\left(1-\alpha_{m}\right)U^{DE}\right)-\sum\nolimits^{m=\kappa}_{m=1}v_mU^{PP}_{m}+U^{VO}V
\end{align}}Since the random nature of the resource trading market poses great challenges to maximize the members' utility directly, we consider the expected value of ${\mathcal{U}}^{Mem}(p,q,r,\kappa ,\bm{\mathcal{A}},\bm{\mathcal{Y}})$ as~\eqref{eq7}. Notably, (7) does not consider which specific member will be chosen as a volunteer. 
%\begin{align}
% \label{eq7}
%&\overline{{\mathcal{U}}^{Mem}}\left(p,q,r,\kappa,\bm{\mathcal{A}},\bm{\mathcal{Y}}\right)=\kappa \text{E}\left[{\alpha}_{m}\right]\times \text{E}\left[U^{PP}_{m}\right]-\notag\\
%&\kappa qd^{comp}+\kappa qd^{comp}\text{E}\left[{\alpha }_{m}\right]+\left(pd^{comp}+r\right)\text{E}\left[V\right],
%\end{align}
{\setlength\abovedisplayskip{5pt}
\setlength\belowdisplayskip{5pt}
\begin{align}
 \label{eq7}
 &\overline{{\mathcal{U}}^{Mem}}\left(p,q,r,\kappa,\bm{\mathcal{A}},\bm{\mathcal{Y}}\right)=\text{E}\left[\sum\nolimits^{m=\kappa}_{m=1}\left(\alpha_mU^{PP}_{m}+U^{DE}-\alpha_mU^{DE}\right)\right]+\text{E}\left[V\right]\left(\text{E}\left[U^{VO}\right]-\text{E}\left[U^{PP}_m\right]\right)
 \notag\\
 &=\kappa\text{E}\left[{\alpha}_{m}\right]\times \text{E}\left[U^{PP}_{m}\right]-\kappa qd^{comp}+\kappa qd^{comp}\text{E}\left[{\alpha }_{m}\right]+rd^{comp}\text{E}\left[V\right]-\text{E}\left[U^{PP}_m\right]\text{E}\left[V\right],
\end{align}}where $\text{E}\left[\cdot\right]$ denotes the mathematical expectation, and we can simply have $\text{E}\left[{\alpha}_{m}\right]=a$. Specifically, $\text{E}\left[V\right]$ is given by~\eqref{eq8},
%\begin{equation}
%\label{eq8}
%\text{E}\left[V\right]=\begin{cases}
%0,\qquad \kappa \le S \\
%\kappa a-\left(\sum\nolimits^{i=S-1}_{i=0}{iC^i_{\kappa }a^i{\left(1-a\right)}^{\kappa -i}}+\right.\\
%\quad~\left.S\sum\nolimits^{i=\kappa}_{i=S}{C^i_{\kappa}a^i{\left(1-a\right)}^{\kappa -i}}\right),\quad\kappa >S,
%\end{cases}
%\end{equation}
{\setlength\abovedisplayskip{5pt}
\setlength\belowdisplayskip{5pt}
\begin{equation}
\label{eq8}
\text{E}\left[V\right]=\begin{cases}
0,&\kappa \le S  \vspace{-1.5ex} \\
\kappa a-\left(\sum\nolimits^{i=S-1}_{i=0}{iC^i_{\kappa }a^i{\left(1-a\right)}^{\kappa -i}}+S\sum\nolimits^{i=\kappa}_{i=S}{C^i_{\kappa}a^i{\left(1-a\right)}^{\kappa -i}}\right),&\kappa >S
\end{cases},
\end{equation}}and $\text{E}\left[U^{PP}_{m}\right]$ is calculated by~\eqref{eq9},
%\begin{align}
%\label{eq9}
%\text{E}\left[U^{PP}_{m}\right]&=\left(\frac{{\omega}_1+{{\omega}_2e}^{loc}}{f^{b}}-\frac{{\omega}_1}{f^{s}}-p\right)d^{comp}-\notag\\
%&\quad~\frac{\ln 2d^{size}\left({\omega}_1+{\omega}_2e^{tran}\right)\times \int^{{\mathbb{C}}_2}_{{\mathbb{C}}_1}{\left(\frac{e^y}{y}\right)}{d}y}{We^{tran}\left({{\varepsilon }_2-\varepsilon }_1\right)},
%\end{align}
{\setlength\abovedisplayskip{5pt}
\setlength\belowdisplayskip{5pt}
\begin{align}
\label{eq9}
\text{E}\left[U^{PP}_{m}\right]&=\left(\frac{{\omega}_1+{{\omega}_2e}^{loc}}{f^{b}}-\frac{{\omega}_1}{f^{s}}-p\right)d^{comp}-\frac{\ln 2d^{size}\left({\omega}_1+{\omega}_2e^{tran}\right)\times \int^{{\mathbb{C}}_2}_{{\mathbb{C}}_1}{\left(\frac{e^y}{y}\right)}{d}y}{We^{tran}\left({{\varepsilon }_2-\varepsilon }_1\right)},
\end{align}}where ${\mathbb{C}}_1=\text{ln}2\times {\text{log}}_2\left(1+e^{tran}{\varepsilon }_{{1}}\right)$ and ${\mathbb{C}}_2= \ln2\times {\log}_{2}\left(1+e^{tran}{\varepsilon }_{{2}}\right)$, for notational simplicity. Detailed derivations of~\eqref{eq8} and~\eqref{eq9} are given in~Appendix A. Combine~\eqref{eq7}, \eqref{eq8}, and~\eqref{eq9}, we rewrite~\eqref{eq7} as~\eqref{eq10}
and (11) considering $\kappa \le S$ and $\kappa >S$, respectively.
%\begin{align}
%\label{eq10}
%&\overline{{\mathcal{U}}^{Mem}}\left(p,q,r,\kappa \bm{\le }S,\bm{\mathcal{A}}, \bm{\mathcal{Y}}\right)\notag\\
%&=\kappa a\text{E}\left[U^{PP}_{m}\right]-\kappa qd^{comp}+\kappa aqd^{comp}\\
%&\overline{{\mathcal{U}}^{Mem}}\left(p,q,r,\kappa >S,\bm{\mathcal{A}},\bm{\mathcal{Y}}\right)=\kappa a\text{E}\left[U^{PP}_{m}\right]-\notag\\
%&\quad~\kappa qd^{comp}+\kappa aqd^{comp}+\left(pd^{comp}+rd^{comp}\right)\text{E}\left[V\right]
%\end{align}
{\setlength\abovedisplayskip{5pt}
\setlength\belowdisplayskip{5pt}
\begin{align}
\label{eq10}
&\overline{{\mathcal{U}}^{Mem}}\left(p,q,r,\kappa \bm{\le }S,\bm{\mathcal{A}}, \bm{\mathcal{Y}}\right)=\kappa a\text{E}\left[U^{PP}_{m}\right]-\kappa qd^{comp}+\kappa aqd^{comp}\\
&\overline{{\mathcal{U}}^{Mem}}\left(p,q,r,\kappa >S,\bm{\mathcal{A}},\bm{\mathcal{Y}}\right)=\left(\kappa a-\text E[V]\right)\text{E}\left[U^{PP}_{m}\right]-\kappa qd^{comp}+\kappa aqd^{comp}+rd^{comp}\text{E}\left[V\right]
\end{align}}In this paper, we consider two key risks for members. First, the risk of a member ${\bm{b_m}}$ (not a volunteer) suffering from a non-positive utility (abbreviate to ``MRisk'') is defined as the probability that its utility is too close to or less than $U_{min}$ ($U_{min}$ denotes a value approaching to zero),
 expressed by the following~\eqref{eq12}.
%\begin{align}
%\label{eq12}
%&{\mathcal{R}}^{MRisk}\left(p,q,\bm{\mathcal{A}},\bm{\mathcal{Y}}\right)
%=\text{Pr}\left(\frac{{\alpha }_{m}U^{PP}_{m}+(1-{\alpha }_{m})U^{DE}}{U_{min}}\le {\xi }_1\right)\notag\\
%&=\begin{cases}
%\displaystyle{0,\quad {\mathbb{C}}_5<0} \\
%\displaystyle{1-a,\quad 0\le {\mathbb{C}}_5<{\mathbb{C}}_3-
%\frac{{\mathbb{C}}_4}{{\log}_{2}\left(1+e^{tran}{\varepsilon }_1\right)}} \\
%\displaystyle{1-a+a\left(\frac{2^{\frac{{\mathbb{C}}_4}{{\mathbb{C}}_3-{\mathbb{C}}_5}}-1-e^{tran}{\varepsilon }_1}{e^{tran}\left({{\varepsilon }_2-\varepsilon }_1\right)}\right),}\\[4pt] \qquad~\displaystyle{{\mathbb{C}}_3-\frac{{\mathbb{C}}_4}
%{{\log}_{2}\left(1+e^{tran}{\varepsilon }_1\right)}\le {\mathbb{C}}_5\le {\mathbb{C}}_3-}\\[4pt]
%\qquad~\displaystyle{\frac{{\mathbb{C}}_4}{{\log}_{2}\left({1+e}^{tran}{\varepsilon }_2\right)}} \\
%\displaystyle{1,\quad{\mathbb{C}}_5>{\mathbb{C}}_3-
%\frac{{\mathbb{C}}_4}{{\log}_{2}
%\left(1+e^{tran}{\varepsilon }_2\right)}}
% \end{cases}
%\end{align}
{\setlength\abovedisplayskip{5pt}
\setlength\belowdisplayskip{5pt}
\begin{align}
\label{eq12}
&{\mathcal{R}}^{MRisk}\left(p,q,\bm{\mathcal{A}},\bm{\mathcal{Y}}\right)
=\text{Pr}\left(\frac{{\alpha }_{m}U^{PP}_{m}+(1-{\alpha }_{m})U^{DE}}{U_{min}}\le {\xi }_1\right)\notag\\
&=\begin{cases}
0,&{\mathbb{C}}_5<0 \vspace{-1.5ex}\\
1-a,&0\le {\mathbb{C}}_5<{\mathbb{C}}_3-
\frac{{\mathbb{C}}_4}{{\log}_{2}\left(1+e^{tran}{\varepsilon }_1\right)} \vspace{-1.5ex}\\
1-a+a\left(\frac{2^{\frac{{\mathbb{C}}_4}{{\mathbb{C}}_3-{\mathbb{C}}_5}}-1-e^{tran}{\varepsilon }_1}{e^{tran}\left({{\varepsilon }_2-\varepsilon }_1\right)}\right),&{\mathbb{C}}_3-\frac{{\mathbb{C}}_4}
{{\log}_{2}\left(1+e^{tran}{\varepsilon }_1\right)}\le {\mathbb{C}}_5\le {\mathbb{C}}_3-
\frac{{\mathbb{C}}_4}{{\log}_{2}\left({1+e}^{tran}{\varepsilon }_2\right)} \vspace{-1.5ex}\\
1,&{\mathbb{C}}_5>{\mathbb{C}}_3-
\frac{{\mathbb{C}}_4}{{\log}_{2}
\left(1+e^{tran}{\varepsilon }_2\right)}
\end{cases}
\end{align}}Particularly, ${\xi}_1$ denotes a positive threshold coefficient; ${\mathbb{C}}_3=\frac{{{\omega}_{1}d}^{comp}+
{\omega}_{2}e^{loc}d^{comp}}{f^{b}}-\frac{{\omega}_{1}d^{comp}}{f^{s}}+qd^{comp}-pd^{comp}, {\mathbb{C}}_4=\frac{{\omega}_{2}e^{tran}d^{size}+{\omega}_{1}d^{size}}{W}$, and ${\mathbb{C}}_5={\xi}_1U_{min}+qd^{comp}$, which are constants under any given $p$ and $q$, for notational simplicity.
Then, the risk of a performer being selected as a volunteer\footnote{Although each volunteer will receive a compensation from the seller, this will always lead to bad trading experiences for those members who cannot enjoy computing service even they had signed the forward contract.} (abbreviate to ``VRisk'') is given by~\eqref{eq13}. Apparently, a larger value of ${\mathcal{R}}^{VRisk}(\kappa, \bm{\mathcal{A}})$ leads to a higher risk of being selected as a volunteer, which greatly impacts the members' trading experience. Derivations of~\eqref{eq12} and~\eqref{eq13} are provided in~Appendix B.
%\begin{align}
%\label{eq13}
%&{\mathcal{R}}^{VRisk}(\kappa ,\bm{\mathcal{A}})
%=\notag\\
%&\begin{cases}
%   0,\qquad 0\le \kappa \le S \\
%a-\sum\nolimits^{i=S-1}_{i=0}{C^i_{\kappa -1}a^{i+1}{\left(1-a\right)}^{\kappa -1-i}},~\kappa >S \end{cases}
%\end{align}
\begin{align}
\label{eq13}
{\mathcal{R}}^{VRisk}(\kappa ,\bm{\mathcal{A}})=
\begin{cases}
   0,&0\le \kappa \le S \vspace{-1.5ex}\\
a-\sum\nolimits^{i=S-1}_{i=0}{C^i_{\kappa -1}a^{i+1}{\left(1-a\right)}^{\kappa -1-i}},&\kappa >S \end{cases}
\end{align}

%\subsubsection{Utility of non-members in spot market}

\noindent
\textbf{3) Utility of non-members in spot market:}
For analytical simplicity, let $n$ be the index of non-members, where $n\in \{\kappa +1,\ldots,|\bm{\mathcal{B}}|\}$ (notably, there are no non-members when $\kappa =|\bm{\mathcal{B}}|$). During each trading, if the seller's resources are not fully occupied by tasks of members, each non-member with task execution requirement (${\alpha}_{n}=1$) can compete for the remaining resources based on the current channel quality, where partial offloading is allowed. Correspondingly, we define the utility ${\mathcal{U}}^{NonM}_{n}$ of each non-member ${\bm{b_n}}$ as~\eqref{eq14}, where $g_{n}$ denotes the unit price of resources that ${\bm{b_n}}$ has to pay during each trading.
%\begin{align}
%\label{eq14}
%&{\mathcal{U}}^{NonM}_{n}\left(g_{n},{\lambda}_{n},{\alpha }_{n},{\gamma }_{n}\right)={\alpha }_{n}\left({\omega}_1\left(t^{loc}-t^{edge}_{n}\right)+\right.\notag\\
%&\left.{\omega}_2\left(c^{loc}-c^{edge}_{n}\right)-g_{n}{\lambda}_{n}d^{comp}\right)
%\end{align}
%{\setlength\abovedisplayskip{5pt}
\begin{align}
\label{eq14}
{\mathcal{U}}^{NonM}_{n}\left(g_{n},{\lambda}_{n},{\alpha }_{n},{\gamma }_{n}\right)={\alpha }_{n}\left({\omega}_1\left(t^{loc}-t^{edge}_{n}\right)+{\omega}_2\left(c^{loc}-c^{edge}_{n}\right)-g_{n}{\lambda}_{n}d^{comp}\right)
\end{align}

\vspace{-0.6cm}
\subsection{Modeling of Seller}

%\subsubsection{Utility, expected utility and risk of seller in futures market}

\noindent
\textbf{1) Utility, expected utility and risk of seller in futures market:}
%\noindent
%$\bullet$ \textbf{Utility of the seller:} 
Suppose that an edge server owns $Sd^{comp}$ resources (e.g., CPU cycles). Utility of the seller contains two key factors: i) the revenue $U^{IN}$ obtained from practical performers and defaulters, and ii) the total refunds and compensations $U^{OUT}$ the seller has to pay for volunteers when the available resources fails to afford the members' task execution requirements owing to overbooking.  Correspondingly, $U^{IN}$ is defined as the following~\eqref{eq15}.
{\setlength\abovedisplayskip{5pt}
\setlength\belowdisplayskip{5pt}
\begin{equation}
\label{eq15}
U^{IN}=pd^{comp}\sum\nolimits^{m=\kappa }_{m=1}{{\alpha}_{m}}+qd^{comp}\sum\nolimits^{m=\kappa}_{m=1}{\left(1-{\alpha}_{m}\right)}
\end{equation}}Moreover, $U^{OUT}$ is calculated by~\eqref{eq16}.
{\setlength\abovedisplayskip{3pt}
\setlength\belowdisplayskip{3pt}
\begin{equation}
\label{eq16}
U^{OUT}=(p+r)d^{comp}V
\end{equation}}Correspondingly, utility of seller is considered as the difference between $U^{IN}$ and $U^{OUT}$.
{\setlength\abovedisplayskip{4pt}
\setlength\belowdisplayskip{4pt}
\begin{equation}
\label{eq17}
{\mathcal{U}}^{SelF}\left(p,q,r,\kappa,\bm{\mathcal{A}}\right)=U^{IN}-U^{OUT}
\end{equation}}Expected utilities of seller are given by~\eqref{eq18} and~\eqref{eq19} considering $ \kappa \le S$ and $\kappa >S$, respectively.
%\begin{align}
%\label{eq18}
%&\overline{{\mathcal{U}}^{SelF}}\left(p,q,r,\kappa \bm{\le }S,\bm{\mathcal{A}}\right)=\kappa apd^{comp}+\kappa qd^{comp}-\notag\\
%&\kappa aqd^{comp}=\kappa d^{comp}\left(q-aq+ap\right)
%\\
%\label{eq19}
%&\overline{{\mathcal{U}}^{SelF}}\left(p,q,r,\kappa >S,\bm{\mathcal{A}}\right)=\kappa d^{comp}\left(q-aq-ar\right)+\notag\\
%&\left(p+r\right)d^{comp}\left(\sum\nolimits^{i=S-1}_{i=0}{iC^i_{\kappa}a^i{\left(1-a\right)}^{\kappa -i}}+\right.\notag\\
%&\left.S\sum\nolimits^{i=\kappa}_{i=S}{C^i_{\kappa}a^i{\left(1-a\right)}^{\kappa -i}}\right)
%\end{align}
{\setlength\abovedisplayskip{4pt}
\setlength\belowdisplayskip{4pt}
\begin{align}
\label{eq18}
&\overline{{\mathcal{U}}^{SelF}}\left(p,q,r,\kappa \bm{\le }S,\bm{\mathcal{A}}\right)=\kappa apd^{comp}+\kappa qd^{comp}-\kappa aqd^{comp}=\kappa d^{comp}\left(q-aq+ap\right)
\\
\label{eq19}
&\overline{{\mathcal{U}}^{SelF}}\left(p,q,r,\kappa >S,\bm{\mathcal{A}}\right)\notag\\
&=\kappa d^{comp}\left(q-aq-ar\right)+\left(p+r\right)d^{comp}\left(\sum\nolimits^{i=S-1}_{i=0}{iC^i_{\kappa}a^i{\left(1-a\right)}^{\kappa -i}}+S\sum\nolimits^{i=\kappa}_{i=S}{C^i_{\kappa}a^i{\left(1-a\right)}^{\kappa -i}}\right)
\end{align}}In the proposed resource trading market, the seller always prefers to achieve a larger utility than expected. Thus, we define the risk of the seller as the probability that ${\mathcal{U}}^{SelF}\left(p,q,r,\kappa,\bm{\mathcal{A}}\right)$ is too close to or less than $\overline{{\mathcal{U}}^{SelF}}\left(p,q,r,\kappa ,\bm{\mathcal{A}}\right)$,
which is given by~\eqref{eq20}.
{\setlength\abovedisplayskip{5pt}
\setlength\belowdisplayskip{5pt}
\begin{equation}
\label{eq20}
{\mathcal{R}}^{SRisk}(p,q,r,\kappa ,\bm{\mathcal{A}})=\text{Pr}\left(\frac{{\mathcal{U}}^{SelF}\left(p,q,r,\kappa ,\bm{\mathcal{A}}\right)}{\overline{{\mathcal{U}}^{SelF}}\left(p,q,r,\kappa ,\bm{\mathcal{A}}\right)}\le {\xi }_2\right),
\end{equation}}where ${\xi}_2$ represents a positive threshold coefficient. According to~\eqref{eq18}, risk of the seller under $\kappa <S$ is calculated by~\eqref{eq21}, where ${\mathbb{C}}_6=\frac{{\xi }_2\overline{{\mathcal{U}}^{SelF}}\left(p,q,r,\kappa \le S,\mathcal{A}\right)}{d^{comp}\left(p-q\right)}-\frac{q\kappa }{\left(p-q\right)}$ for notational simplicity:
%\begin{align}
%\label{eq21}
%&{\mathcal{R}}^{SRisk}\left(p,q,r,\kappa \le S,\bm{\mathcal{A}}\right)
%=\notag\\
%&\begin{cases}
%0,&\mathbb{C}_6<0\\
%\sum\nolimits^{i={\lfloor\mathbb{C}}_6\rfloor}_{i=0}{C^i_{\kappa}
%a^i{\left(1-a\right)}^{\kappa-i}},&0\le \mathbb{C}_6 \le \kappa\\
%1,&\mathbb{C}_6> \kappa\\
%\end{cases}
%\end{align}
{\setlength\abovedisplayskip{5pt}
\setlength\belowdisplayskip{5pt}
\begin{align}
\label{eq21}
&{\mathcal{R}}^{SRisk}\left(p,q,r,\kappa \le S,\bm{\mathcal{A}}\right)=
\begin{cases}
0,&\mathbb{C}_6<0\vspace{-1.5ex}\\
\sum\nolimits^{i={\lfloor\mathbb{C}}_6\rfloor}_{i=0}{C^i_{\kappa}
a^i{\left(1-a\right)}^{\kappa-i}},&0\le \mathbb{C}_6 \le \kappa\vspace{-1.5ex}\\
1,&\mathbb{C}_6> \kappa\\
\end{cases}
\end{align}}Considering $\kappa>S$, risk of the seller is given by~\eqref{eq22}, based on~\eqref{eq19}.
%\begin{align}
%\label{eq22}
%&{\mathcal{R}}^{SRisk}\left(p,q,r,\kappa >S,\bm{\mathcal{A}}\right)=\notag\\
%&\begin{cases}
%0,\qquad {\mathbb{C}}_7<{\left(0,S\left(p-q\right)-\left(\kappa -S\right)(q+r)\right)}^- \\
%\sum\nolimits^{i=\left\lfloor \frac{{\mathbb{C}}_7}{p-q}\right\rfloor }_{i=0}{C^i_{\kappa }a^i{\left(1-a\right)}^{\kappa -i}}+\\
%\qquad\sum\nolimits^{i=\kappa }_{i=\left\lceil \frac{S\left(p-q\right)-{\mathbb{C}}_7}{q+r}+S\ \right\rceil }{C^i_{\kappa }a^i{\left(1-a\right)}^{\kappa -i}},\\
%\qquad{\left(0,S\left(p-q\right)-\left(\kappa -S\right)(q+r)\right)}^-\le\\
%\qquad {\mathbb{C}}_7\le S\left(p-q\right) \\
%1,\qquad {\mathbb{C}}_7>S\left(p-q\right)
%\end{cases}
%\end{align}
%\begin{small}
{\setlength\abovedisplayskip{5pt}
\setlength\belowdisplayskip{5pt}
\begin{align}
\label{eq22}
&{\mathcal{R}}^{SRisk}\left(p,q,r,\kappa >S,\bm{\mathcal{A}}\right)=
\begin{cases}
0,\qquad {\mathbb{C}}_7<{\left(0,S\left(p-q\right)-\left(\kappa-S\right)(q+r)\right)}^- \vspace{-0.6ex}\\
\sum\nolimits^{i=\left\lfloor \frac{{\mathbb{C}}_7}{p-q}\right\rfloor}_{i=0}{C^i_{\kappa}a^i{\left(1-a\right)}^{\kappa -i}}+\sum\nolimits^{i=\kappa }_{i=\left\lceil \frac{S\left(p-q\right)-{\mathbb{C}}_7}{q+r}+S\right\rceil }{C^i_{\kappa }a^i{\left(1-a\right)}^{\kappa -i}},\\
{\left(0,S\left(p-q\right)-\left(\kappa -S\right)(q+r)\right)}^-\le
{\mathbb{C}}_7\le S\left(p-q\right) \vspace{-1.5ex}\\
1,\qquad {\mathbb{C}}_7>S\left(p-q\right)
\end{cases}
\end{align}}where ${\mathbb{C}}_7=\frac{{\xi }_2\overline{{\mathcal{U}}^{Sel}}\left(p,q,r,\kappa >S,\bm{\mathcal{A}}\right)}{d^{comp}}-q\kappa $ for notational simplicity. Notably, let $\sum\nolimits^{i=\left\lfloor \frac{{\mathbb{C}}_7}{p-q}\right\rfloor}_{i=0}{C^i_{\kappa }a^i{\left(1-a\right)}^{\kappa -i}}\\=0$ when $\frac{{\mathbb{C}}_7}{p-q}<0$, and $\sum\nolimits^{i=\kappa}_{i=\lceil \frac{S\left(p-q\right)-{\mathbb{C}}_7}{q+r}+S\rceil}{C^i_{\kappa}a^i{\left(1-a\right)}^{\kappa -i}}=0$ when $\lceil \frac{S\left(p-q\right)-{\mathbb{C}}_7}{q+r}+S \rceil>\kappa $.
Derivations associated with~\eqref{eq21} and~\eqref{eq22} are detailed by~Appendix~C.
%\end{small}

%\subsubsection{Utility of seller in spot market}
\noindent
\textbf{2) Utility of seller in spot market:} Note that non-members with task execution requirements get chances to compete for available resources due to the possible ``no shows'' of members. Let binary indicator $x_{n}=1$ denote that the seller decides to trade with non-member ${\bm{b_n}}$, and $x_{n}=0$ otherwise; while $\bm{\mathcal{X}}=\{x_{n}|n\in \{\kappa+1,\ldots,|\bm{\mathcal{B}}|\}\}$ depicts the relevant trading decision vector. Moreover, let $\bm{\mathcal{G}}=\{g_{n}|n\in \{\kappa +1,\ldots,|\bm{\mathcal{B}}|\}\}$ present the price vector, and $\bm{\mathit{\Lambda}}=\{{\lambda}_{n}|n\in \{\kappa +1,
 \ldots,|\bm{\mathcal{B}}|\}\}$ indicate the offloading rate vector of non-members.
 Correspondingly, the seller's utility in spot market is defined by~\eqref{eq23}.
\begin{equation}
\label{eq23}
{\mathcal{U}}^{SelS}\left(\bm{\mathcal{ X}},\bm{\mathcal{G}},\bm{\mathit{\Lambda}},\bm{\mathcal{A}}\right)=
d^{comp}\sum\nolimits^{n=|\bm{\mathcal{B}}|}_{n=\kappa +1}{x_{n}g_{n}{\lambda}_{n}{\alpha }_{n}}
\end{equation}

\section{Problem Formulation and Solution Design in Futures Market}

\noindent
The proposed futures market mainly considers designing both the forward contract (e.g., $p$, $q$, and $r$) and overbooking rate (e.g., ${\kappa}^{o}$, which is equivalent to the design of $\kappa $), where the relevant problem is formulated by $\bm{\mathcal{F}_1}$ with two objectives (24a) and (24b), under constraints C1-C7. Specifically, (24a) describes that the seller aims to maximize its expected utility while meeting the tolerable risk (constraint C1); (24b) indicates the maximization of the expected utility of members under acceptable tolerant risks (constraints C2 and C3).
 %\textcolor{red}{subequations}
 %\vspace{-1.2ex}
 \begin{spacing}{0.6}
 \begin{align*}
 \hspace{0.5cm}\bm{\mathcal{F}_1}:
 \begin{cases}
\argmax\limits_{p,q,r,\kappa} \overline{{\mathcal{U}}^{SelF}}\left(p,q,r,\kappa ,\bm{\mathcal{A}}\right)\hspace{2.24cm}\text{(24a)} \\
\argmax\limits_{p,q,r,\kappa} \overline{{\mathcal{U}}^{Mem}}(p,q,r,\kappa,\bm{\mathcal{A}}, \bm{\mathcal{Y}})
\hspace{1.8cm}\text{(24b)}
 \end{cases}
 \end{align*}
\setcounter{equation}{24}
\begin{align*}
&s.t.\\
&C1:~{\mathcal{R}}^{SRisk}\left(p,q,r,\kappa ,\bm{\mathcal{A}}\right)\le {\xi}^{S}, \\
&C2:~{\mathcal{R}}^{MRisk}\left(p,q,\bm{\mathcal{A}},\bm{\mathcal{Y}}\right)\le {\xi}^{M},\\
&C3:~{\mathcal{R}}^{VRisk}(\kappa ,\bm{\mathcal{A}})\le {\xi }^V,\\
&C4:~1\le\kappa \le|\bm{\mathcal{B}}|,\\
&C5:~U^{PP}_{m}>0, \forall~m\in \{1,\ldots,\kappa \},\\
&C6:~p \ge p^{Sel}_{min},\\
&C7:~p>q,\ r>0.
\end{align*}
\end{spacing}

\vspace{0.5cm}
\noindent
 Specifically, ${\xi}^{S}$, ${\xi}^B$, and ${\xi}^V$ are positive threshold coefficients, constraints C1-C3 denote the acceptable tolerant risks of seller and members. Constraint C4 limits the practicable number of members (particularly, if the players fail to sign forward contract, let $\kappa =0$). Constraint C5 represents the individual rationality of members in this market, describing that each practical performer will receive at least non-negative utility from a trading even under a poor channel quality (e.g., ${\gamma}_{m}={\varepsilon}_1$). Additionally, constraint C6 indicates that $p$ should be larger than the seller's tolerable minimum price $p^{Sel}_{min}$ (e.g., the minimum price reflects the seller's cost for processing a task such as energy consumption, etc.); while C7 describes the relationships among $p$, $q$, and $r$. To facilitate the analysis, we integrate C5 and C6 as C8, where
 $p^{Mem}_{max}=\frac{{\omega}_{1}+{\omega}_{2}e^{loc}}{f^{b}}-
 \frac{{\omega}_{1}}{f^{s}}-\left(\frac{{\omega}_{\text{1}}d^{size}+{\omega}_{2}e^{tran}d^{size}}
 {Wd^{comp}{\log}_{2}\left(1+e^{tran}
 {\varepsilon}_1\right)}\right)$ denotes the maximum tolerable price of each member.
 {\setlength\abovedisplayskip{5pt}
\setlength\belowdisplayskip{5pt}
\begin{equation}
\label{eq25}
\text{C8:}~p^{Sel}_{min}\le p<p^{Mem}_{max}
\end{equation}}Note that $\bm{\mathcal{F}_1}$ represents a MOO problem, which, however, is difficult to be solved by the state-of-the-art methods (e.g., weighted sum method~\cite{2,42}, weighted metric method~\cite{43}, and $\epsilon$-constrained method~\cite{44}),  owing to the information privacy among players. For example,  seller is unaware of factors such as buyer's local capability $f^{b}$, local consumption $e^{loc}$, weight coefficients
 ${\omega}_{1}$ and ${\omega}_{2}$. Additionally, each objective in $\bm{\mathcal{F}_1}$ ((24a) and (24b)) refers to a mixed integer non-linear programing (MINLP) problem~\cite{45}, which considers determining both continuous (e.g., $p$, $q$ and $r$) and integer variables (e.g., $\kappa$), that further complicates the solution design.
 
%Alg1
%\setlength{\belowcaptionskip}{-10pt}
% {\setlength\abovedisplayskip{0pt}
%\setlength\belowdisplayskip{0pt}
\begin{algorithm*}[!t]
\setstretch{0.85} 
{\footnotesize
%\small
  \AlgoDisplayBlockMarkers
  %\SetAlgoNoLine
  \SetKwData{Left}{left}
  \SetKwData{This}{this}
  \SetKwData{Up}{up}
  \SetKwFunction{Union}{Union}
  \SetKwFunction{FindCompress}{FindCompress}
  \SetKwInOut{Input}{Input}
  \SetKwInOut{Output}{Output}

  %\SetKw{Init}{Initialization}
  \SetAlgoBlockMarkers{}{}%
  %\SetAlgoNoEnd
  \Input{$a,~{{\varepsilon}_1,~\varepsilon}_2,~d^{comp},~d^{size},~f^{b},~f^{s},~e^{loc},~e^{tran},~W,~|\bm{\mathcal{B}}|,~S,~{\omega}_1,~{\omega}_2,~{\xi}^{S},~{\xi}^{M},~\xi^{V},~\Delta p,~\Delta q,~\Delta r$}
  \Output{$p^*,~q^*,~r^*,~{\kappa}^*$}
  
\textbf{Initialization:} {${FuturesP}_1\leftarrow p^{Sel}_{min}$, ${FuturesQ}_1\leftarrow \Delta q$, ${FuturesR}_1\leftarrow \Delta r$, $i=j=l\leftarrow 1$,
  $\bm{CTerm}\leftarrow\emptyset$, $Count\leftarrow 0$},
  
  %\Init{${FuturesP}_1\leftarrow p^\text{Sel}_{\min}$, ${FuturesQ}_1\leftarrow \Delta q$, ${FuturesR}_1\leftarrow \Delta r$,$i=j=l\leftarrow 1$,
%  $\bm{CTerm}\bm{\leftarrow}\bm{\emptyset}$,$Count\leftarrow 0$\;}
The agent first determines the acceptable range of $\kappa $ denoted by $\bm{K^{Mem}}$ while meeting constraints C3 and C4,

  \While{${FuturesP}_i\le p^{Mem}_{max}$}{
  \While{${FuturesQ}_j\le constant_1\times \Delta q$}{
  \While{${FuturesR}_l\le constant_2 \times\Delta r$}{
The seller determines its acceptable range of $\kappa $ denoted by $\bm{K^{Sel}_{i,j,l}}$, while meeting C1 and C4,

The agent checks if the current price and penalty meets constraint C2, if yes, continue the negotiation; otherwise, go to step 17,

  \If{$\bm{K^{Sel}_{i,j,l}}\cap\bm{K^{Mem}}{\neq }\emptyset$}{
  ${\kappa}_{i,j,l}\leftarrow \argmax\limits_{\kappa} \overline{{\mathcal{U}}^{Mem}}\left({FuturesP}_i,
{FuturesQ}_j,{FuturesR}_l,\bm{\mathcal{A}}, \bm{\mathcal{Y}},\kappa \right)$, $\kappa \in \bm{K^{Sel}_{i,j,l}}\cap\bm{K^{Mem}}$ \% the agent chooses the value of $\kappa$ that maximizes the members' expected utility,\\
$\bm{CTerm}\leftarrow \bm{CTerm}\bigcup{\left\{{FuturesP}_i,{FuturesQ}_j,{FuturesR}_l,{\kappa }_{i,j,l}\right\}}$,

\ElseIf{${\bm{K^{Sel}_{i,j,l}}}=\emptyset$}{
$l\leftarrow 1$, $j=j+1$, ${FuturesQ}_j\leftarrow{FuturesQ}_{j-1}+\Delta q$, $Count\leftarrow Count+1$,

\Else{
 $Count\leftarrow Count+1$, break,  \% jump out of the current while loop
}
}
}$l\leftarrow l+1$,~${FuturesR}_l\leftarrow {FuturesR}_{l-1}+\Delta r$,~$Count\leftarrow Count+1$,

}$l\leftarrow 1$, $j\leftarrow j+1$,~${FuturesQ}_j\leftarrow {FuturesQ}_{j-1}+\Delta q$, $Count\leftarrow Count+1$,

}
$l\leftarrow 1$, $j\leftarrow 1$, $i\leftarrow i+1$,  ${FuturesP}_i\leftarrow {FuturesP}_{i-1}+\Delta p$, $Count\leftarrow Count+1$,}

 \If {$\bm{CTerm}\neq \emptyset$}{
\{$p^*,q^*,r^*,{\kappa }^*\}\leftarrow \argmax\limits_{p,q,r,\kappa} \overline{{\mathcal{U}}^{Sel}}\left(p,q,r,\kappa,\bm{\mathcal{A}}\right)$, $\{$$p,q,r,\kappa \}\in \bm{CTerm}$, \% the seller chooses a set of $p,q,r$ and $\kappa $ from $\bm{CTerm}$ that maximizes its expected utility\;
  \Else{Players fail to sign the forward contract,}
  }
  \textbf{end algorithm}
  
\caption{Proposed bilateral negotiation in futures market (solving problem $ \bm{\mathcal{F}_1}$)}
\label{algo1}
}
\end{algorithm*}
%\vspace{-0.8cm}

Consequently, bilateral negotiation is considered as an efficient approach which facilitates the negotiation among players with conflicting objectives~\cite{18} to reach the trading consensus on the forward contract (e.g., $p$, $q$, $r$, and $\kappa$). Since all the buyers are i.i.d, a trusted agent\footnote{The seller does not have to negotiate with every buyer since all the buyers are i.i.d. Thus, a trusted agent (e.g., access point, etc.) is supposed to be a representative of members, negotiates with the seller on forward contract and overbooking rate. Once the trading consensus has been reached, the relevant buyers who get the membership can sign the forward contract with seller.} is applied as the representative of members to negotiate with the seller. To facilitate analysis, let $\Delta p$, $\Delta q$, and $\Delta r$ denote the granularities of price, penalty, and refund, respectively. Specifically, we propose an alternative optimization-based bilateral negotiation mechanism to solve problem $\bm{\mathcal{F}_1}$, which is detailed by Algorithm~1, where $p^*,q^*,r^*$ and ${\kappa }^*$ denote the final trading consensus on forward contract, and the number of members, respectively.
As can be seen from Algorithm 1, the representative agent of buyers determines the acceptable range of $\kappa$ (line 2) via meeting constraint C3 to avoid too many possible volunteers. Under given price ${FuturesP}_i$, penalty ${FuturesQ}_j$ and compensation ${FuturesR}_l$, the seller first determines its acceptable range of $\kappa $ (e.g., $\bm{K^{Sel}_{i,j,l}}$) while meeting its tolerable risk (line 6); while the agent checkes of the current price and penalty meets MRisk (line 7). If $\bm{K^{Sel}_{i,j,l}}\cap\bm{K^{Mem}}\neq\emptyset$, the agent chooses a value of $\kappa $ from set $\bm{K^{Sel}_{i,j,l}}\cap\bm{K^{Mem}}$ that maximizes the expected utility of members (lines 8-9), where the relevant solution will be saved into a candidate set $\bm{CTerm}$ (line 10). Specifically, if $\bm{K^{Sel}_{i,j,l}}=\emptyset$, seller can directly raise the value of penalty (line 12) since the current expected utility may be unsatisfying; otherwise, the seller will adjust either the value of compensation, penalty, or price to start another quotation, as shown by lines 15-17. After all the quotations, the seller chooses the optimal forward contract terms and the number of available members from $\bm{CTerm}$ that maximize its expected utility, as the final trading consensus (lines 18-20); otherwise the players fail to sign the forward contract if there is no candidate terms.

\section{Problem Formulation and Solution Design in Spot Market}

\noindent
A spot trading may occur among the seller and non-members when the following two conditions happen concurrently: i): $\sum\nolimits^{m=\kappa}_{m=1}{{\alpha}_{m}<S}$, where seller has available resources after meeting the task execution requirements of members; ii) $\sum\nolimits^{n=|\bm{\mathcal{B}}|}_{n=\kappa +1}{{\alpha}_{n}}>0$, where at least one non-member has task execution requirement. Correspondingly, resource trading in spot market is formulated by problem $\bm{\mathcal{F}_2}$, where seller, and each non-member with task execution requirement is aiming to maximize its own utility, as shown by the following (26a), and (26b).
%\begin{align*}
%\hspace{0.3cm}\bm{\mathcal{F}_2}:
%\begin{cases}
%\argmax\limits_{\bm{\mathcal{X}},\bm{\mathcal{G}}} {\mathcal{U}}^{SelS}\left(\bm{\mathcal{X}},\bm{\mathcal{G}},\bm{\mathit{\Lambda}},\bm{\mathcal{A}}\right) \hspace{2.18cm}\text{(26a)}\\
%\argmax\limits_{{\lambda}_{n}} {\mathcal{U}}^{NonM}_{n}\left(g_{n},{\lambda}_{n}, {\alpha }_{n},{\gamma }_{n}\right),\\
%\qquad\forall~{\alpha}_{n}=1,n\in \left\{\kappa +1,\dots ,\left|\bm{\mathcal{B}}\right|\right\}\hspace{1.4cm}\text{(26b)}
%\end{cases}
%\end{align*}
%\setcounter{equation}{26}
%s.t.
%\begin{align*}
%&\text{\!~~C9:}~{\lambda}_{n}\triangleq 0,\quad \text{if}~
%U^{NonM}_{n}\left(g_{n},{\lambda}_{n},{\alpha}_{n},{\gamma }_{n}\right)\le 0,\\
%&\text{C10:}~0\le {\lambda}_{n}\le 1,\\
%&\text{C11:}~g_{n}\ge p^{Sel}_{min},\\
%&\text{C12:}~{\bm{\mathit{\Lambda}}}^{\text{T}}\bm{\mathcal{X}}\le S^{\prime},
%\end{align*}
\begin{spacing}{0.6}
\begin{align*}
\hspace{0.3cm}\bm{\mathcal{F}_2}:
\begin{cases}
\argmax\limits_{\bm{\mathcal{X}},\bm{\mathcal{G}}} {\mathcal{U}}^{SelS}\left(\bm{\mathcal{X}},\bm{\mathcal{G}},\bm{\mathit{\Lambda}},\bm{\mathcal{A}}\right) \hspace{8.15cm}\text{(26a)}\\
\argmax\limits_{{\lambda}_{n}} {\mathcal{U}}^{NonM}_{n}\left(g_{n},{\lambda}_{n}, {\alpha }_{n},{\gamma }_{n}\right),~\forall~{\alpha}_{n}=1,n\in \left\{\kappa +1,\dots ,\left|\bm{\mathcal{B}}\right|\right\}\hspace{1.8cm}\text{(26b)}
\end{cases}
\end{align*}
\setcounter{equation}{26}
\begin{align*}
&s.t.\\
&C9:~{\lambda}_{n}\triangleq 0,\quad\forall~
U^{NonM}_{n}\left(g_{n},{\lambda}_{n},{\alpha}_{n},{\gamma }_{n}\right)\le 0,\\
&C10:~0\le {\lambda}_{n}\le 1,\\
&C11:~g_{n}\ge p^{Sel}_{min}, \forall~{\alpha}_{n}=1,n\in \left\{\kappa +1,\dots ,\left|\bm{\mathcal{B}}\right|\right\},\\
&C12:~{\bm{\mathit{\Lambda}}}^{\text{T}}\bm{\mathcal{X}}\le S^{\prime},
\end{align*}
\end{spacing}
\vspace{0.3cm}
\noindent
where $S^{\prime}=S-\sum\nolimits^{m=\kappa}_{m=1}{{\alpha}_{m}}$ (apparently, $S^{\prime}d^{comp}$ indicates the remaining resources available for non-members). Constraints C9 ensures the non-negative utility of each non-member, C10 and C11 (similar with C6) limit the values of offloading rate and unit resource price, respectively. C12 restricts the limited seller's resources in spot market, where ${\bm{\mathit{\Lambda}}}^{\text{T}}$ denotes the transpose of vector $\bm{\mathit{\Lambda}}$. Similar with $\bm{\mathcal{F}_1}$, we consider bilateral negotiations among the seller and non-members to solve $\bm{\mathcal{F}_2}$, via considering two pricing rules~\cite{46}: uniform pricing and differential pricing.
%Alg2
\begin{algorithm*}[b!]
\setstretch{0.85} 
\footnotesize
%\small
  \AlgoDisplayBlockMarkers
  %\SetAlgoNoLine
  \SetKwData{Left}{left}\SetKwData{This}{this}\SetKwData{Up}{up}
  \SetKwFunction{Union}{Union}\SetKwFunction{FindCompress}{FindCompress}
  \SetKwInOut{Input}{Input}
  \SetKwInOut{Output}{Output}
  \SetAlgoBlockMarkers{}{}%
  %\SetAlgoNoEnd
  \Input{$\bm{\mathcal{A}},~\bm{\mathcal{Y}},~S^{\prime},~d^{comp},~d^{size},~f^{b},~f^{s},~e^{loc},~ e^{tran},~W,~{\omega}_1$,~${\omega}_2$,~$\Delta p$}
  \Output{$\bm{\mathcal{X}^*},~g^*,~\bm{\mathit{\Lambda}^*}$}
  %\BlankLine
  \textbf{Initialization}: ${SpotP}_i\leftarrow p^{Sel}_{min}$,$\ i\leftarrow 1,\ Count\leftarrow 0$, $\|\bm{\mathit{\Lambda}_0}\|_1>0,
  \mathbb{X}\leftarrow \emptyset$,

  \While{${SpotP}_i\ge p^{Sel}_{min}$}{
  \For{$n=\kappa +1$ and $n\le |\bm{\mathcal{B}}|$}{
  
  \If{${\alpha}_{n}=0$}{
  
  ${Count}_{n}\leftarrow 0,$ $n\leftarrow n+1$,

  \Else{${\lambda}^i_{n}\leftarrow \argmax\limits_{{\lambda}_{n}} {\mathcal{U}}^{NonM}_{n}({SpotP}_i,{\lambda}_{n},\ {\alpha}_{n},{\gamma }_{n})$, while meeting C10 and C13, \% this problem (27b) represents a non-convex optimization problem which is detailed in Appendix D,\\
$\bm{\mathit{\Lambda}_i}\leftarrow \bm{\mathit{\Lambda}_i}\bigcup{{\lambda}^i_{n}}$, $n\leftarrow n+1$, $Count\leftarrow Count+1$,
}
}
 
  \If{${\|\bm{\mathit{\Lambda}_i}\|}_1=0$}{
  %\% if all the non-members decide to process their tasks locally due to the excessive price\;
jump out of the current while loop, \% none of the non-members can accept a higher price, the seller stops quotation,
  
\Else{\bm{$\mathcal{X}_i}\leftarrow\argmax\limits_{\bm{\mathcal{X}}} {\bm{\mathit{\Lambda}_i}}^{\text{T}}\bm{\mathcal{X}}$, while meeting C12, \% this problem (27a) refers to a binary knapsack~problem which can be solved by dynamic programming\;
$u_i\leftarrow g_id^{comp}\bm{\mathit{\Lambda}_i}^{\text{T}}\bm{\mathcal{X}_i}$,
$\mathbb{X}\leftarrow \mathbb{X}\bigcup{\left\{u_i,{SpotP}_i, \bm{\mathcal{X}_i},\bm{\mathit{\Lambda}_i}\right\}}$, $i\leftarrow i+1$, \\$n\leftarrow n+1$,
${SpotP}_i\leftarrow {SpotP}_{i-1}+\Delta p$,
}  
}
}   
}
For all ${\alpha}_{n}=1$, ${Count}_{n}\leftarrow Count/\sum\nolimits^{n=|\bm{\mathcal{B}}|}_{n=\kappa +1}{{\alpha}_{n}}$,\\
The seller chooses the largest $u_i$ from set $\bm{\mathbb{X}}$, where the relevant ${SpotP}_i,\bm{\mathcal{X}_i}, \bm{\mathit{\Lambda}_i}$ stand for the final trading solution $g^*,\bm{\mathcal{X}^*}$ and $\bm{\mathit{\Lambda^*}}$,\\
\textbf{end algorithm}

  \caption{Proposed spot trading under uniform pricing (solving problem $ \bm{\mathcal{F}_3}$)}
  \label{algo2}
  \end{algorithm*}
  
  %Alg3
\begin{algorithm*}[t!]
\setstretch{0.85} 
\footnotesize
%\small
  \AlgoDisplayBlockMarkers
  %\SetAlgoNoLine
  \SetKwData{Left}{left}\SetKwData{This}{this}\SetKwData{Up}{up}
  \SetKwFunction{Union}{Union}\SetKwFunction{FindCompress}{FindCompress}
  \SetKwInOut{Input}{input}\SetKwInOut{Output}{output}
  \SetAlgoBlockMarkers{}{}%
  %\SetAlgoNoEnd
  %\SetKwInit{Initialization}{initialization}\SetKwInOut{Initialization}{initialization}
  \Input{ $\bm{\mathcal{A}},~\bm{\mathcal{Y}},~S^\prime,~d^{comp},~d^{size},~f^{b},~f^{s},~ e^{loc},~e^{tran},~W,~{\omega}_1$,~${\omega}_2$,~$\Delta p$}
  \Output{$\bm{\mathcal{X}^*},~\bm{\mathcal{G}^*},~\bm{\mathit{\Lambda}^*}$}

\textbf{Initialization}: ${SpotP}_1\leftarrow p^{Sel}_{min}$,$\ i\leftarrow 1,\mathbb{X}\leftarrow \emptyset $, ${\lambda}^0_{n}>0$,~${Count}_{n}\leftarrow 0$, ~$\forall n\in \left\{\kappa +1,\dots ,\left|\bm{\mathcal{B}}\right|\right\}$,\\

\For{$n=\kappa +1$ and $n\le |\bm{\mathcal{B}}|$}{
\If{${\alpha }_{n}=0$}{
$n\leftarrow n+1$,

\Else{
\While{${SpotP}_i\ge p^{Sel}_{min}$ and ${\lambda}^{i-1}_{n}>0$}{
${\lambda}^i_{n}\leftarrow \argmax\limits_{{\lambda}_{n}} {\mathcal{U}}^{NonM}_{n}({SpotP}_i,{\lambda}_{n},{\alpha }_{n},{\gamma }_{n})$, while meeting C9 and C10, \% similar with $\bm{\mathcal{F}_3}$, this problem (26b) represents a non-convex optimization problem which is detailed in Appendix D,

\If{${\lambda}^i_{n}>0$}{
$\bm{\mathit{\Lambda}_n}\leftarrow\bm{\mathit{\Lambda}_n}\bigcup{\{{SpotP}_i,{\lambda}^i_{n}\}}$, ${Count}_{n}\leftarrow {Count}_{n}+1$, \\$i\leftarrow i+1$,
${SpotP}_i\leftarrow {SpotP}_{i-1}+\Delta p$,

\Else{jump out of the current while loop, \% non-member $\bm{b_n}$ can not accept a higher price, the seller stops the current quotation,}
}}

$n\leftarrow n+1$, $i\leftarrow 1$,
}
}
}
The seller chooses $\bm{\mathcal{X}^*}$ that maximizes the value of ${\mathcal{U}}^{SelS}\left(\bm{\mathcal{X}},\bm{\mathcal{G}},\bm{\mathit{\Lambda}},\bm{\mathcal{A}}\right)$ based on set $\bigcup_{{\alpha}_{n}=1}{\bm{\mathit{\Lambda}_n}}$, where the relevant price set and offloading rate set will be the final solution $\bm{\mathcal{G}^*},\bm{\mathit{\Lambda}^*}$  \% this problem (26a) denotes a knapsack~problem with grouped items which can be solved by dynamic programming,\\
\textbf{end algorithm}
  \caption{Proposed spot trading under differential pricing (solving problem $\bm{\mathcal{F}_2}$)}\label{algo3}
  \end{algorithm*}

%\vspace{-0.5cm}
 \subsection{Spot Trading under Uniform Pricing}

\noindent
 Considering uniform pricing where the seller charges the same price for all the non-members, $\bm{\mathcal{F}_2}$ can be rewritten as $\bm{\mathcal{F}_3}$ by letting $g=g_{\kappa +1}=\ldots=g_{\left|\bm{\mathcal{B}}\right|}$.
% \begin{align*}
% \hspace{0.3cm}\bm{\mathcal{F}}:
% \begin{cases}
% \argmax\limits_{\bm{\mathcal{X}},g}\left(\bm{\mathcal{X}},g,\bm{\mathit{\Lambda}},\bm{\mathcal{A}}\right)
% \hspace{3.2cm}\text{(27a)}\\
% \argmax\limits_{{\lambda}_{n}}{\mathcal{U}}^{NonM}_{n}\left(g,{\lambda}_{n},\ {\alpha }_{n},{\gamma }_{n}\right),\\
% \qquad~\forall {\alpha}_{n}=1,n\in \left\{\kappa +1,\dots ,\left|\bm{\mathcal{B}}\right|\right\}
% \hspace{1.5cm}\text{(27b)}
% \end{cases}
% \end{align*}
%\setcounter{equation}{27}
%
%s.t.
%\begin{align*}
%&\text{C10, C12},\\
%&\text{C13:}~{\lambda}_{n}\triangleq 0,~\text{if}~U^{NonM}_{n}\left(g,{\lambda}_{n}, {\alpha}_{n},{\gamma}_{n}\right)\le 0,\\
%&\text{C14:}~g \ge p^{Sel}_{min}.
%\end{align*}
%\vspace{-1.2ex}
\begin{spacing}{0.6}
 \begin{align*}
 \hspace{0.3cm}\bm{\mathcal{F}_3}:
 \begin{cases}
 \argmax\limits_{\bm{\mathcal{X}},g}{\mathcal{U}^{SelS}}\left(\bm{\mathcal{X}},g,\bm{\mathit{\Lambda}},\bm{\mathcal{A}}\right)\hspace{8.2cm}\text{(27a)}\\
 \argmax\limits_{{\lambda}_{n}}{\mathcal{U}}^{NonM}_{n}\left(g,{\lambda}_{n},\ {\alpha }_{n},{\gamma }_{n}\right),~\forall~{\alpha}_{n}=1,n\in \left\{\kappa +1,\dots ,\left|\bm{\mathcal{B}}\right|\right\}
 \hspace{1.8cm}\text{(27b)}
 \end{cases}
 \end{align*}
\setcounter{equation}{27}
\begin{align*}
&s.t.~~~C10, C12,\\
&C13:~{\lambda}_{n}\triangleq 0,~\forall~U^{NonM}_{n}\left(g,{\lambda}_{n}, {\alpha}_{n},{\gamma}_{n}\right)\le 0,\\
&C14:~g \ge p^{Sel}_{min}.
\end{align*}
\end{spacing}

\vspace{0.3cm}
\noindent
Constraints C13 and C14 are similar with C9 and C11. Notably, (27a) depicts a binary knapsack problem with the weight ${\lambda}_{n}$, and the value $g{\lambda}_{n}$ for a non-member ${\bm{b_n}}$; while (27b) in $\bm{\mathcal{F}_3}$ represents a non-convex optimization problem under any given $g$. Apparently, (27a) is NP-complete which poses difficulty to find efficient algorithms, thus, we apply dynamic programming~\cite{46} to solve the binary knapsack problem in pseudo-polynomial time (e.g., by using the kp01 software package in MATLAB). Moreover, algorithm for obtaining the optimal offloading rate (solve (27b)) is detailed by Appendix~D. 
%Fig.~2(a) shows an example of spot trading under uniform pricing rule, where the seller determines to trade with non-members $\bm{b_{12}}$, $\bm{b_{13}}$ and $\bm{b_{15}}$ at \textit{price2}, on offloading rates 85\%, 59\%, and 48\%.
Pseudocodeof solving $\bm{\mathcal{F}_3}$ is given by Algorithm~2, where ${\bm{\mathcal{X}}}^{\bm{*}}$ and ${\bm{\mathit{\Lambda}}}^{\bm{*}}$ indicate the final trading decision vector and offloading rate vector, respectively; and $g^*$ denotes the relevant final agreed unit price of resource. Specifically, only non-members with tasks are considered in the proposed spot market, as depicted by lines 4-5. Lines 6-8 indicate that under a given price ${SpotP}_i$, each non-member $\bm{b_n}$ with task decides the optimal offloading rate $\bm{\mathit{\Lambda}^i_{n}}$ that maximizes its utility; while lines 9-10 shows that the seller will stop raising price if all the non-members decide to process their tasks locally, mainly owing to an excessive price. In line 12, the seller determines a trading decision vector that maximizes its utility under price ${SpotP}_i$ (by solving a knapsack problem), and saves the relevant utility $u_i$, price ${SpotP}_i$, trading vector $\bm{\mathcal{X}_i}$, and offloading vector $\bm{\mathit{\Lambda}_i}$ in to a candidate set $\mathbb{X}$. After the~quotation procedure is completed, the seller chooses a solution from $\mathbb{X}$ with the largest utility, through line 16.

\vspace{-0.5cm}
 \subsection{Spot Trading under Differential Pricing}

\noindent
Differential pricing rule considers a more general case where the seller charges different non-members with different prices, where problem $\bm{\mathcal{F}_2}$ in discussed. Specifically, (26a) in $\bm{\mathcal{F}_2}$ refers to a knapsack problem
  with grouped items~\cite{47}, for which the dynamic programming can also be applied, similar with (27a); while (26b) in $\bm{\mathcal{F}_2}$ represents a non-convex problem under any given $g_{n}$, for which the solution of obtaining the optimal offloading rate is given by Appendix~D. Specifically, each non-member with task execution requirement decides an optimal offloading rate based on each price, while the seller determines the trading vector by changing non-members with different prices, associated the relevant offloading rates, to maximize its utility. Pseudocode for solving $\bm{\mathcal{F}_2}$ is detailed by Algorithm~3, where $\bm{\mathcal{X}^*},\bm{\mathcal{G}^*}$, and $\bm{\mathit{\Lambda}^*}$ indicates the final trading decision vector, price vector, and offloading rate vector, respectively. The negotiation procedure between the seller and non-member $\bm{b_{n}}$ are mainly shown by lines 5-12, where the seller keeps raising price until $\bm{b_n}$ decides to process its task locally (lines 5 and 12). Specifically, under each given price, $\bm{b_n}$ determines the optimal offloading rate that maximizes its utility (line 7). After all the~quotation procedure are completed, in line 14, the seller decides the trading decision vector via charging different non-members with different price, to maximize its utility (by solving a knapsack problem with grouped items).

%Fig. 2(b) shows an example of spot trading under differential pricing, where the seller decides to trade with $\bm{b_{11}}$, $\bm{b_{13}}$ and $\bm{b_{15}}$ at \textit{price1}, \textit{price3}, and \textit{price2}, respectively. Correspondingly, non-members $\bm{b_{11}}$, $\bm{b_{13}}$ and $\bm{b_{15}}$ will offload 93\%, 59\%, and 48\% of their task data to the seller through wireless communications. 

%
%%f2
%\begin{figure}[!t]
%\centering
%\subfigtopskip=0.8pt
%\subfigbottomskip=0.8pt
%\subfigure[]{\includegraphics[width=.3\linewidth]{fig2a}}
%\subfigure[]{\includegraphics[width=.3\linewidth]{fig2b}}
%%\begin{minipage}{0.45\linewidth}
%%  \centering
%%  \includegraphics[width=1.5in]{fig2a}
%%\end{minipage}\quad
%%\begin{minipage}{0.45\linewidth}
%%  \centering
%%  \includegraphics[width=1.5in]{fig2b}
%%\end{minipage}
%\caption{Trading examples in spot market where $S=8$, $S^{\prime}=2,|\bm{\mathcal{B}}|=15$,
%$\kappa=10,{\alpha}_{11}={\alpha}_{12}={\alpha}_{13}={\alpha}_{15}=1$, ${\alpha}_{14}=0$: a) uniform pricing, where seller determines to trade with $\bm{b_{12}}$, $\bm{b_{13}}$, and $\bm{b_{15}}$ at \textit{price 2} based on the relevant offloading rate; b) differential pricing, where seller determines to trade with $\bm{b_{11}}$ at \textit{price 1}, $\bm{b_{13}}$ at \textit{price 3}, and $\bm{b_{15}}$ at \textit{price2}, based on the relevant offloading rate.}
%\label{fig2}
%\end{figure}

%\vspace{-0.5cm}

\section{Experimental Results}

\noindent
This section presents comprehensive simulation results and performance evaluations, illustrating the validity of the proposed overbooking-enabled computing resource trading mechanism. Specifically, simulations are implemented via MATLAB R2019b platform on desktop computer with Intel Core i7--4770 3.40 GHz CPU and 16.0 GB RAM. For notational simplicity, the proposed mechanisms under uniform and differential pricing are abbreviated to ``Futures\_Spot\_OverB\_UP'', and ``Futures\_Spot\_OverB\_DP'', respectively.
\vspace{-0.4cm}
\subsection{Baseline Method}

\noindent
To achieve better evaluation, key baseline methods in this simulation are considered:

%\begin{enumerate}
\noindent
$\bullet$ \textbf{Equal-booking-based trading under uniform pricing in futures and spot integrated market (Futures\_Spot\_EqualB\_UP):} In futures market, Algorithm 1 is performed considering $\kappa =S$; in spot market, seller trades with non-members under uniform pricing (Algorithm 2).

\noindent
$\bullet$ \textbf{Equal-booking-based trading under differential pricing in futures and spot integrated market (Futures\_Spot\_EqualB\_DP):} In futures market, Algorithm 1 is performed considering $\kappa =S$; in spot market, seller trades with non-members under differential pricing (Algorithm 3).

\noindent
$\bullet$ \textbf{Spot trading under uniform pricing (Spot\_UP):} Without considering futures market, all the trading are performed by following spot trading mode under uniform pricing. Namely, $\kappa =0$, and Algorithm 2 is performed during each trading.

\noindent
$\bullet$ \textbf{Spot trading under differential pricing (Spot\_DP):} Without considering futures market, all the trading are performed by following spot trading mode under differential pricing. Namely, $\kappa =0$, and Algorithm 3 is performed during each trading.
%\end{enumerate}
\vspace{-0.3cm}
 \subsection{Critical Indicator }

\noindent
 In addition to players' utilities, in this simulation, significant indicators are considered as follows:

%\begin{enumerate}
\noindent
$\bullet$ \textbf{Decision-making cost (DMC):} in each trading, DMC denotes the cost (e.g., energy and battery consumption) that players have spent on trading decision-making. Since DMC is difficult to be quantized by a numerical value (e.g., it is challenging to estimate the amount of battery capacity consumed in each quotation); in this simulation, DMC is described by the number of quotations (e.g., the value of ${Count}_{n}$ in Algorithm 2 and Algorithm 3). Apparently, larger DMC presents heavier energy and battery consumption of both players.

\noindent
$\bullet$ \textbf{Decision-making latency (DML): }DML denotes the time that players have spent on trading decision-making, which is estimated by considering the end-to-end delay $t^{E2E}$ of wireless communication channels. Note that members are no longer have to spend extra time on trading decision-making, as benefitted from the pre-signed forward contract, in this simulation, DML is mainly considered for non-members. Consequently, DML of a non-member $\bm{b_n}$ ($n\in \{\kappa +1,\ldots,|\bm{\mathcal{B}}|\}$) is calculated by $t^{DML}_{n}={\alpha }_{n} \times {Count}_{n}\times t^{E2E}$.

\noindent
$\bullet$ \textbf{Task completion time (TCT):} Since DML can directly affect the actual TCT of each non-member, for $\bm{b_n}$ ($n\in \{\kappa +1,\ldots,|\bm{\mathcal{B}}|\}$), TCT of which can be calculated by \eqref{eq28}.
%\begin{align}
%\label{eq28}
%t^{TCT}_{n}=x_{n}{\alpha }_{n}\left(\frac{{\lambda}_{n}d^{size}}{W{\log}_{2}\left(1+e^{tran}{\gamma }_{n}\right)}+\frac{{\lambda}_{n}d^{comp}}{f^{s}}, \right.\notag\\
%\left.
%\frac{\left(1-{\lambda}_{n}\right)d^{comp}}{f^{b}}\right)^{+}+
%\left(1-x_{n}\right)\frac{d^{comp}}{f^{b}}+t^{DML}_{n}
%\end{align}
{\setlength\abovedisplayskip{5pt}
\setlength\belowdisplayskip{5pt}
\begin{align}
\label{eq28}
t^{TCT}_{n}=x_{n}{\alpha }_{n}\left(\frac{{\lambda}_{n}d^{size}}{W{\log}_{2}\left(1+e^{tran}{\gamma }_{n}\right)}+\frac{{\lambda}_{n}d^{comp}}{f^{s}}, \frac{\left(1-{\lambda}_{n}\right)d^{comp}}{f^{b}}\right)^{+}+\left(1-x_{n}\right)\frac{d^{comp}}{f^{b}}+t^{DML}_{n}
\end{align}}Besides, TCT of a member $\bm{b_m}$ ($m\in \{1,2,\ldots,\kappa \}$) is computed by the following~\eqref{eq29}.
%\begin{equation}
%\label{eq29}
%t^{TCT}_{m}=
%\begin{cases}
%\displaystyle{{\alpha}_{m}\left(\frac{d^{size}}{W{\log}_{2}\left(1+e^{tran}{\gamma }_{m}\right)}+\frac{d^{comp}}{f^{s}}\right)},\\
%\qquad {\bm{b_m}}~\text{is not a volunteer} \\
%\dfrac{d^{comp}}{f^{b}},\qquad {\bm{b_m}}\bm{\ }\text{is\ a\ volunteer}
%\end{cases}
%\end{equation}
{\setlength\abovedisplayskip{5pt}
\setlength\belowdisplayskip{5pt}
\begin{equation}
\label{eq29}
t^{TCT}_{m}=
\begin{cases}
\displaystyle{{\alpha}_{m}\left(\frac{d^{size}}{W{\log}_{2}\left(1+e^{tran}{\gamma }_{m}\right)}+\frac{d^{comp}}{f^{s}}\right)},~{\bm{b_m}}~\text{is not a volunteer} \\
\dfrac{d^{comp}}{f^{b}},\qquad {\bm{b_m}}\bm{\ }\text{is\ a\ volunteer}
\end{cases}
\end{equation}}$\bullet$ \textbf{Time utilization rate (TUR):} TUR represents the time efficiency of each resource trading calculated by~\eqref{eq30}. Apparently, large TUR refers to better time efficiency of resource trading.
{\setlength\abovedisplayskip{5pt}
\setlength\belowdisplayskip{5pt}
\begin{equation}
\label{eq30}
\text{TUR}=1-\frac{\sum\nolimits^{n=|\bm{\mathcal{B}}|}_{n=\kappa +1}{t^{DML}_{n}}}{\sum^{m=\kappa }_{m=1}{t^{TCT}_{m}}+\sum\nolimits^{n=|\bm{\mathcal{B}}|}_{n=\kappa +1}{t^{TCT}_{n}}}
\end{equation}}$\bullet$ \textbf{Resource utilization rate (RUR):} RUR indicates the ratio of the amount of resources occupied by the buyers to the seller's total available resources in each trading. Apparently, a large value of RUR presents a better utilization of computing resources.
%\end{enumerate}

\noindent
 Major parameters in this simulation are set as follows: $S=15$, $|\bm{\mathcal{B}}|=30$, $a=0.76$, $d^{size}=0.5$Mb, $d^{comp}=600 \text{cycles}\text{/bit}\times d^{size}$, $f^{s}={10}^{11}$cycles/s, $f^{b}={10}^9$cycles/s~\cite{1,2,3,4}, $e^{loc}=500\text{mWatt}$, $e^{tran}=550\text{mWatt}$~\cite{5}, ${\varepsilon }_1=100$, ${\varepsilon }_2=500$, $W=6\text{MHz}$, ${\xi}^{S}={\xi}^B=0.33$, ${\xi}^V=0.45$, $t^{E2E}\in [2,10]\text{ms}$ \cite{48}.

%tab1
\begin{table*}[b!]
\setstretch{0.9} 
{\footnotesize
%\small
%\setstretch{0.9}
%\renewcommand{\arraystretch}{1.3}
\caption{Additional analysis on short-term performance 
(Algo 1: Futures\_Spot\_OverB\_UP, Algo 2: Futures\_Spot\_OverB\_DP, Algo 3: Futures\_Spot\_EqualB\_UP, Algo 4: Futures\_Spot\_EqualB\_DP, Algo 5: Spot\_UP, Algo 6: Spot\_DP)}
\label{table_example}
\centering
\setlength{\tabcolsep}{3.2mm}{
\begin{tabular}{|c|c|c|c|c|c|c|}
\hline
 & \textbf{Algo 1} & \textbf{Algo 2} & \textbf{Algo 3} & \textbf{Algo 4} & \textbf{Algo 5} & \textbf{Algo 6} \\ \hline
Sum utility of buyers (Figs.~2(a)-Fig. 2(b)) & 251.61 & 247.32 & 220.15 & 195.70 & 108.19 & 8.83 \\ \hline
Sum utility of seller (Figs.~2(a)-Fig. 2(b)) & 72.29 & 74.94 & 58.79 & 77.14 & 4.57 & 80.41 \\ \hline
Sum task completion time (Figs.~3(a)-3(b)) & 324.07 & 328.58 & 568.26 & 586.35 & 904.59 & 992.20 \\ \hline
Sum energy consumption (Figs.~3(a)-3(b)) & 131.39 & 135.21 & 131.47 & 146.98 & 131.47 & 190.77 \\ \hline
Sum decision-making cost (Figs.~4(a)-4(b)) & 13181 & 12792 & 53851 & 52318 & 109907 & 106793 \\ \hline
Sum decision-making latency (Figs.~4(a)-4(b)) & 79.09 & 76.75 & 323.11 & 313.91 & 659.44 & 640.76 \\ \hline
Average time utilization rate (Fig.~5(a)) & 79.51\% & 80.97\% & 42.42\% & 46.35\% & 26.39\% & 35.30\% \\ \hline
Average resource utilization rate (Fig.~5(b)) & 100\% & 98.41\% & 100\% & 93.60\% & 100\% & 75.13\% \\ \hline
\end{tabular}}
}
\end{table*}

\vspace{-0.3cm}
 \subsection{Performance Evaluation}

 %f2
\begin{figure*}[b!]
\centering
\subfigtopskip=0pt
\subfigbottomskip=0pt
\subfigure[]{\includegraphics[width=.252\linewidth]{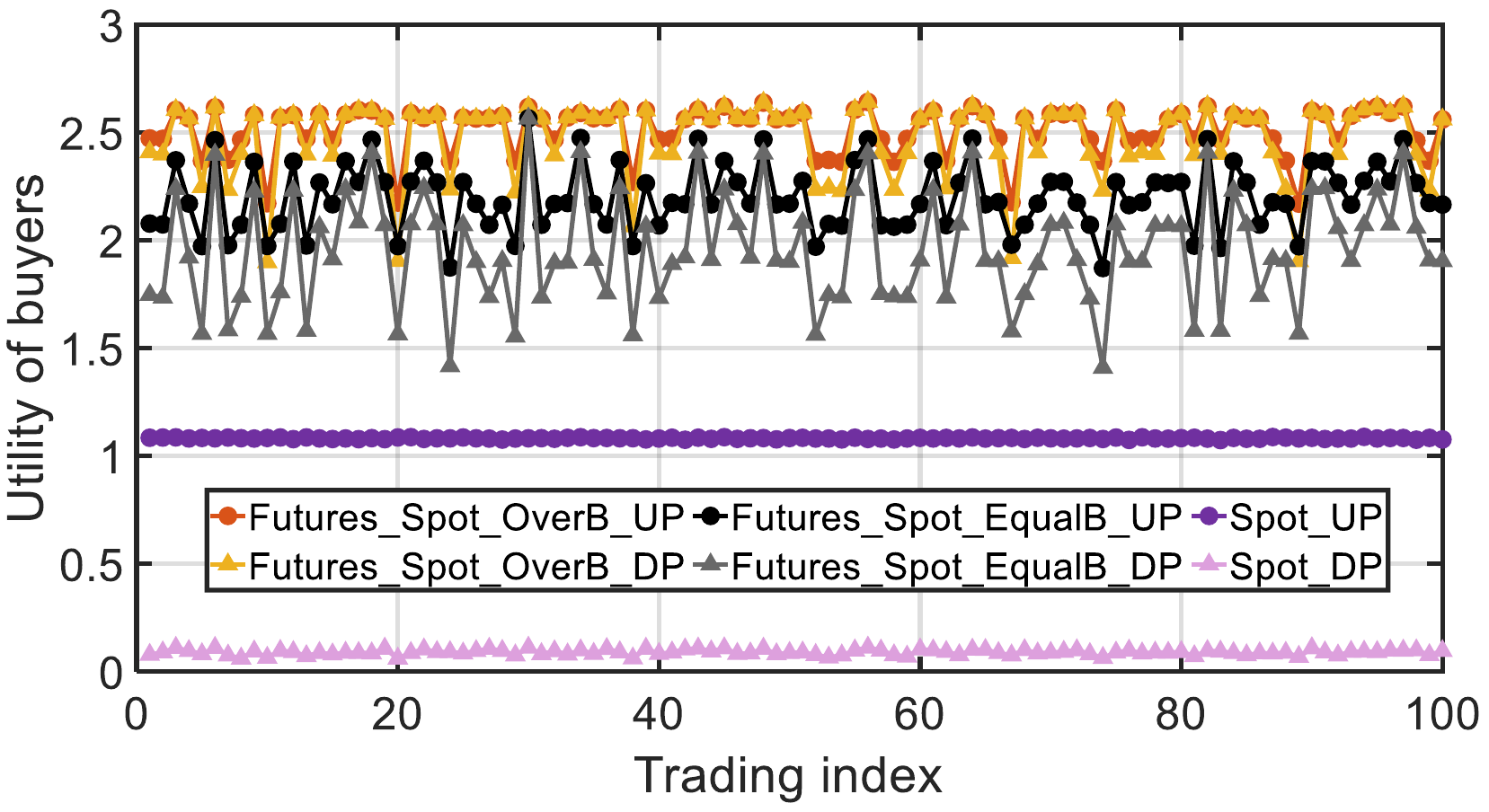}}\hfill
\subfigure[]{\includegraphics[width=.252\linewidth]{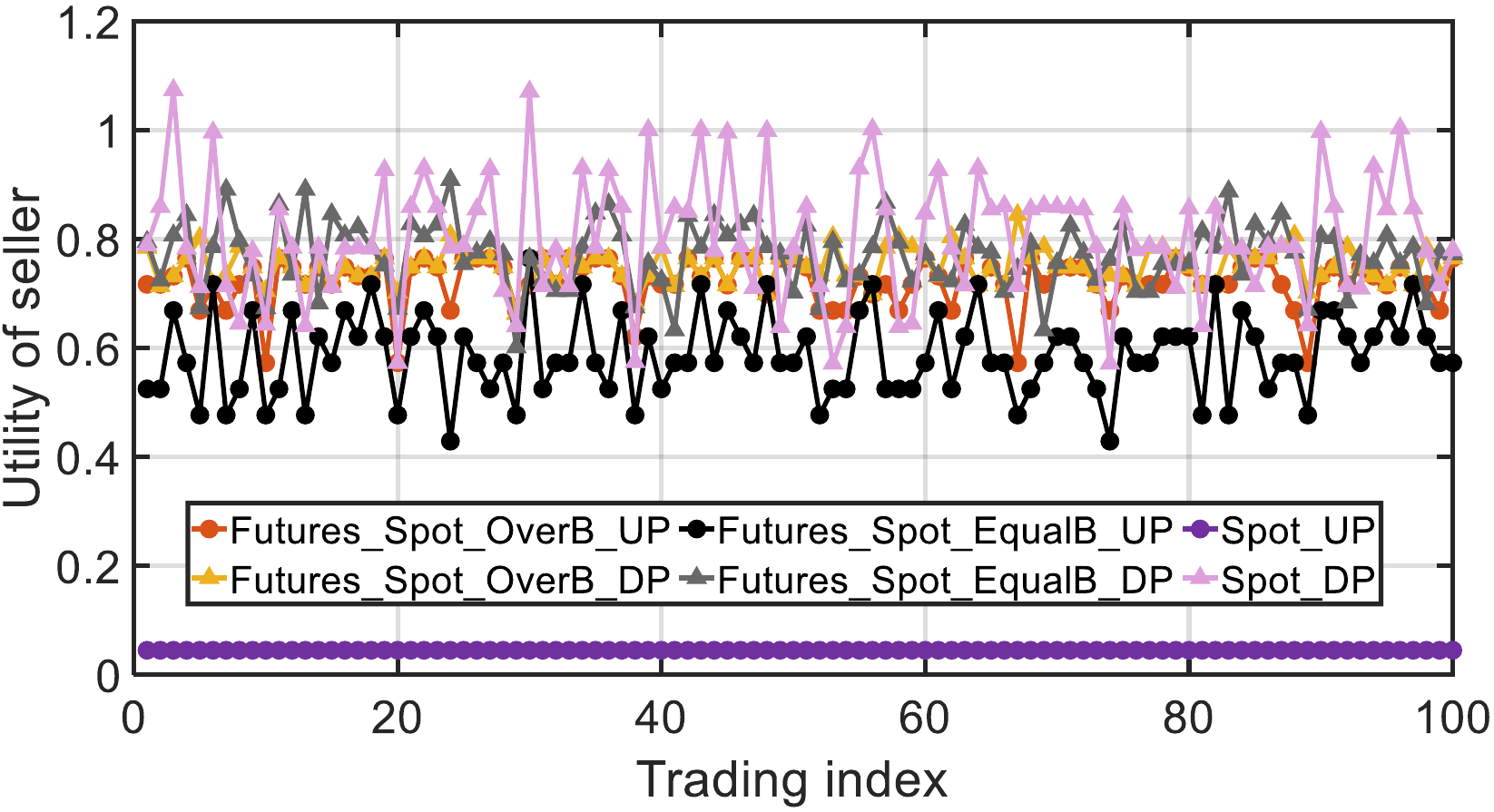}}\hfill
\subfigure[]{\includegraphics[width=.245\linewidth]{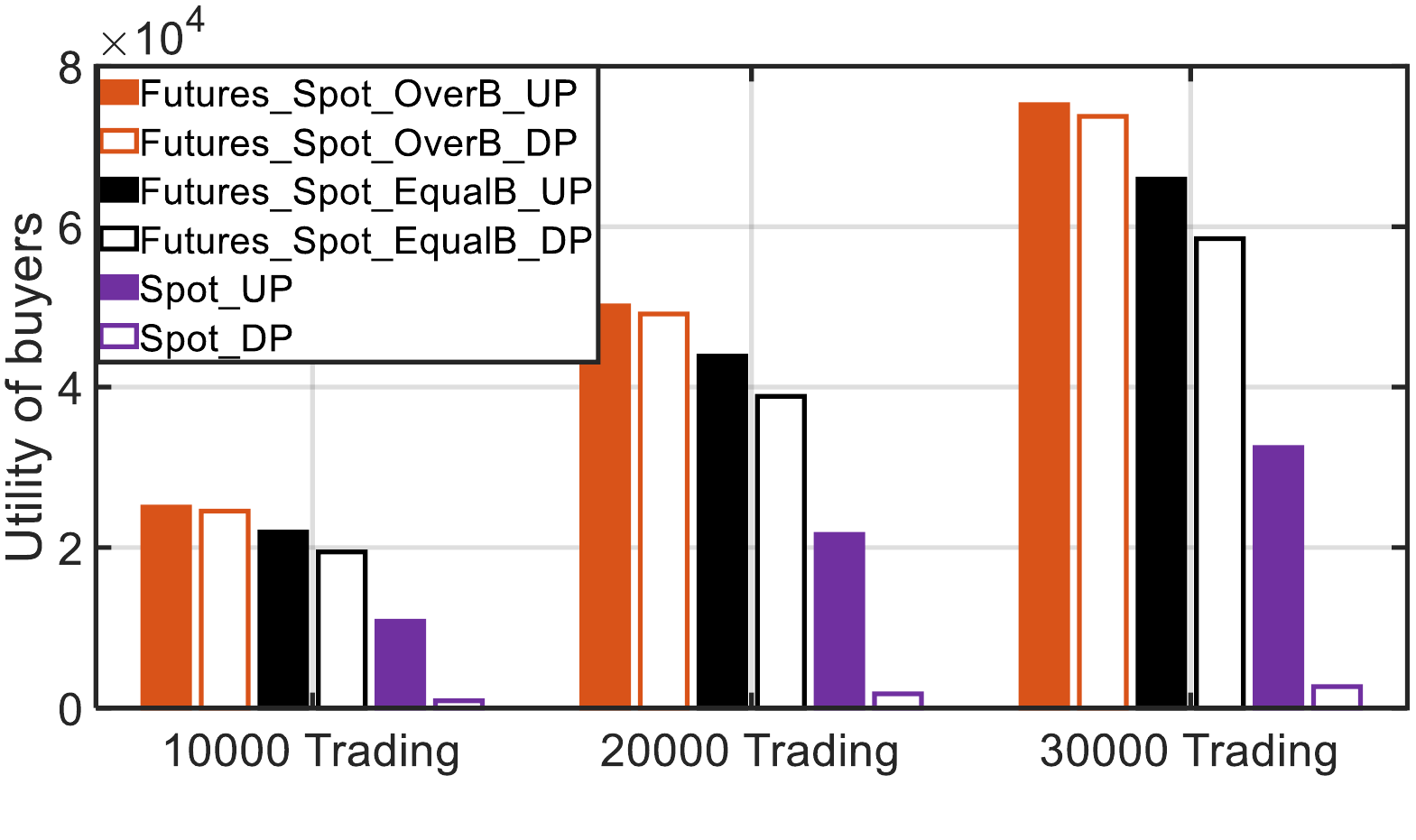}}\hfill
\subfigure[]{\includegraphics[width=.246\linewidth]{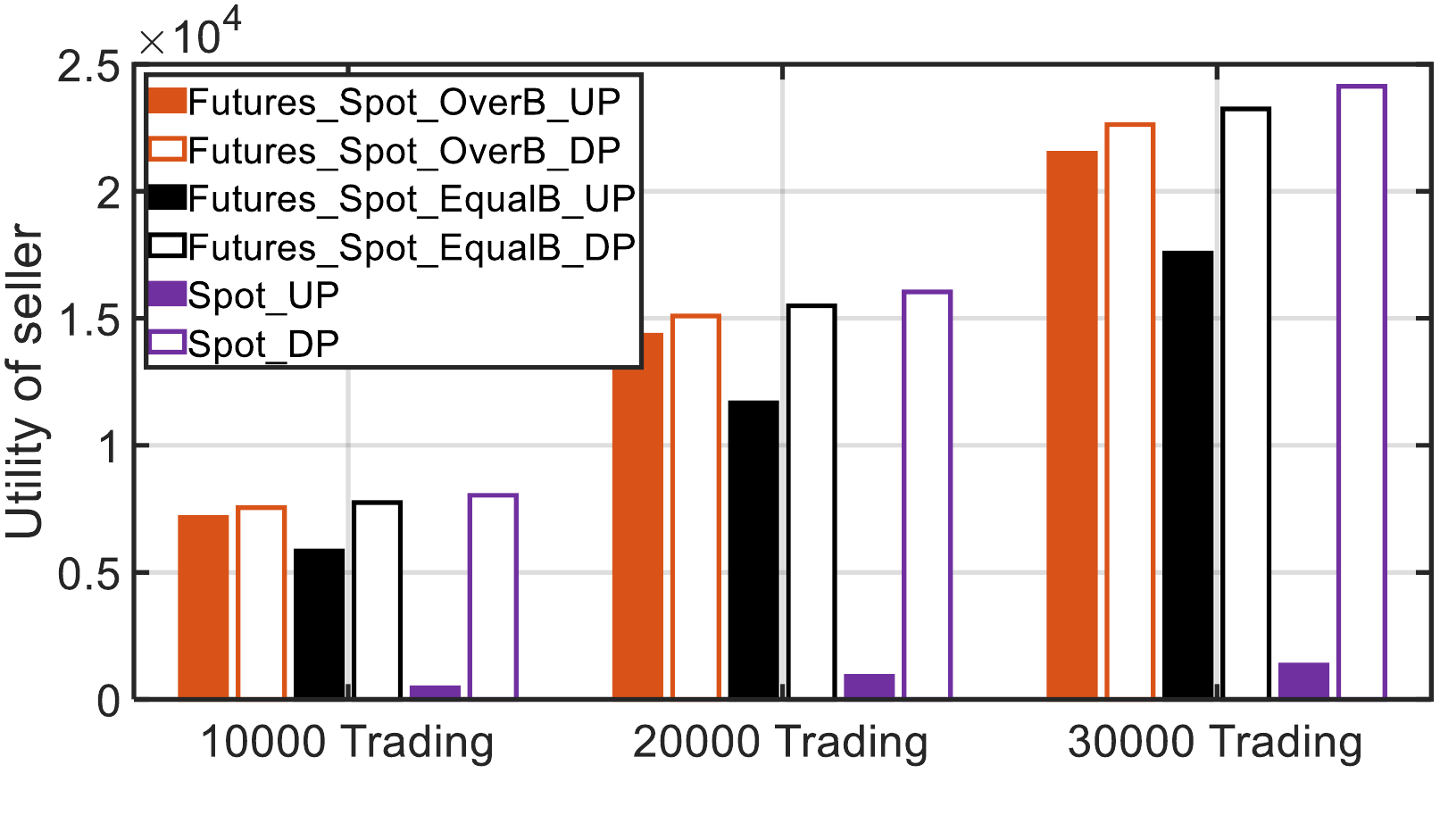}}
\vspace{-0.2cm}
\caption{Short-term and long-term performance on players' utilities.}
\label{fig3}
\end{figure*}
%\vspace{-1cm}
 
%f3
\begin{figure*}[b!]
\centering
\subfigtopskip=0.1pt
\subfigbottomskip=0.8pt
\subfigure[]{\includegraphics[width=.255\linewidth]{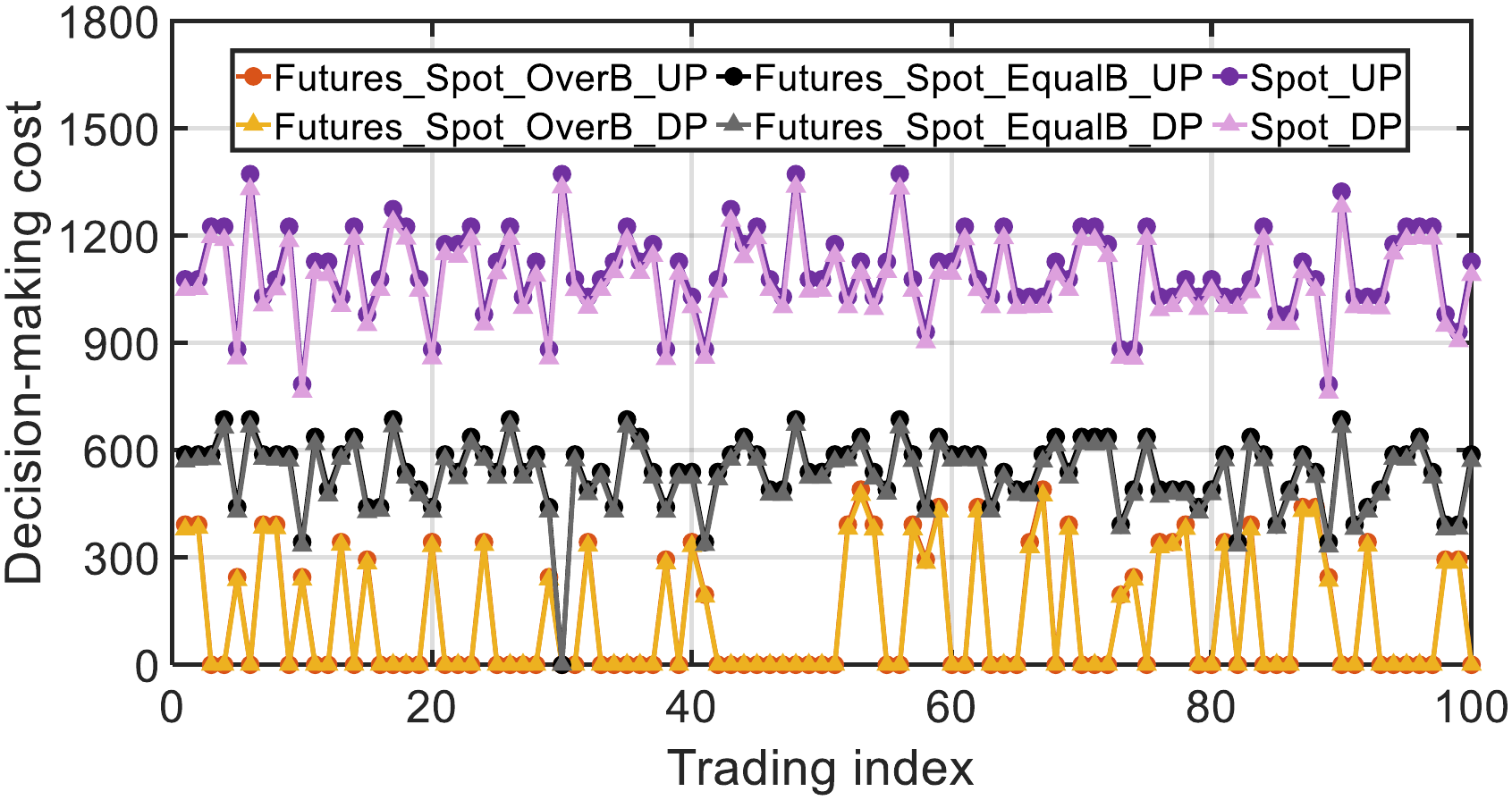}}\hfill
\subfigure[]{\includegraphics[width=.251\linewidth]{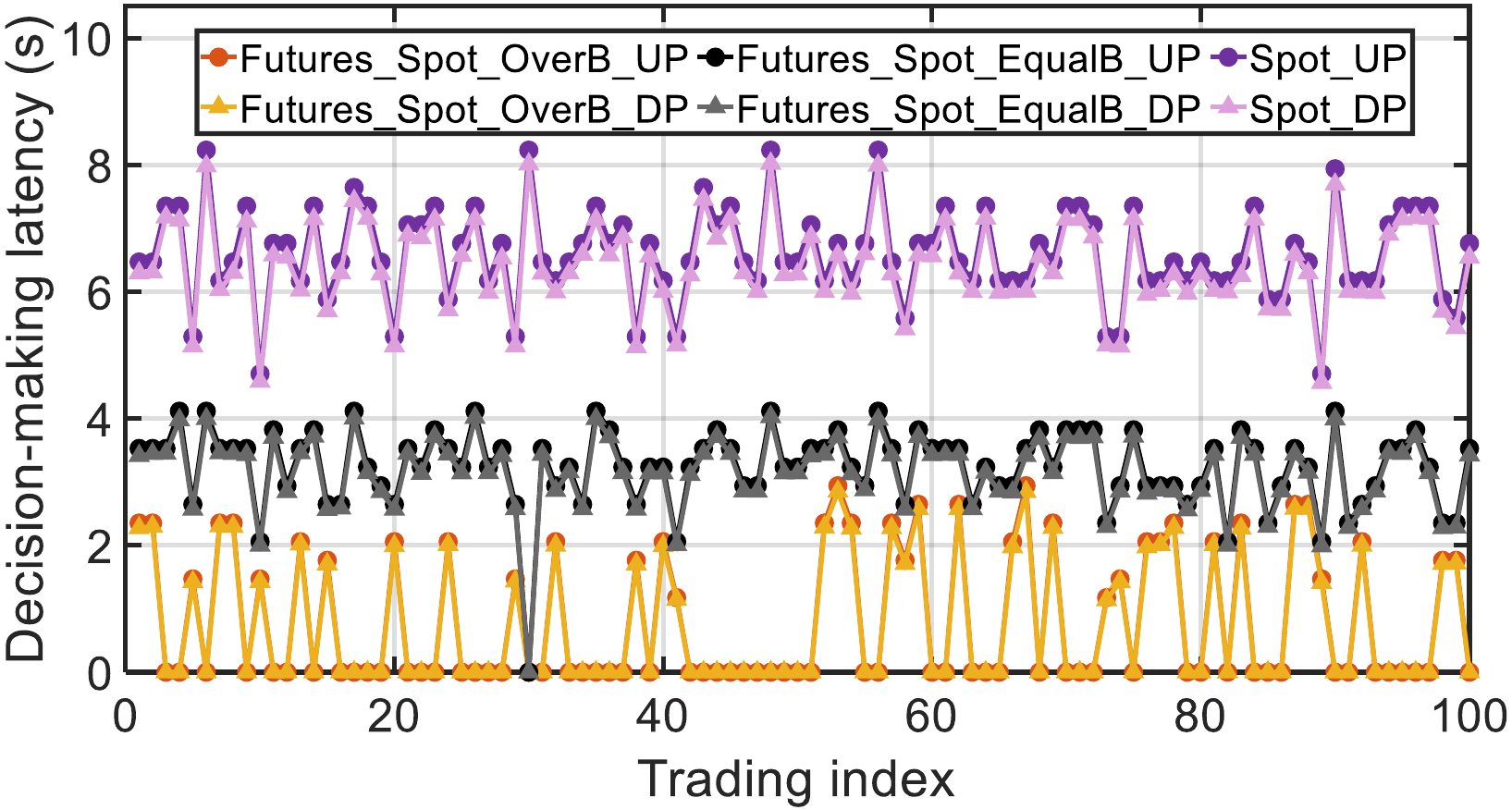}}\hfill
\subfigure[]{\includegraphics[width=.246\linewidth]{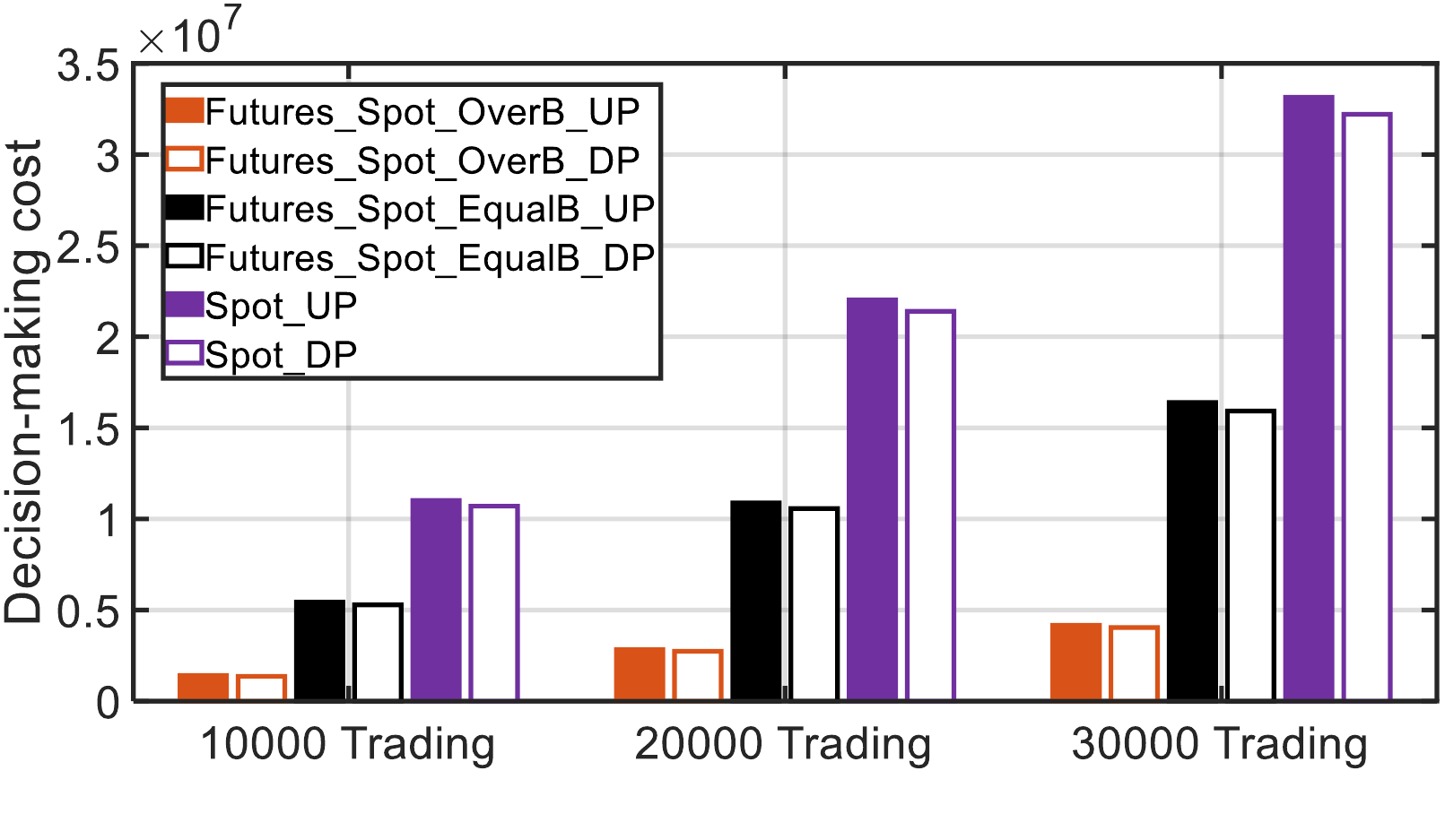}}\hfill
\subfigure[]{\includegraphics[width=.246\linewidth]{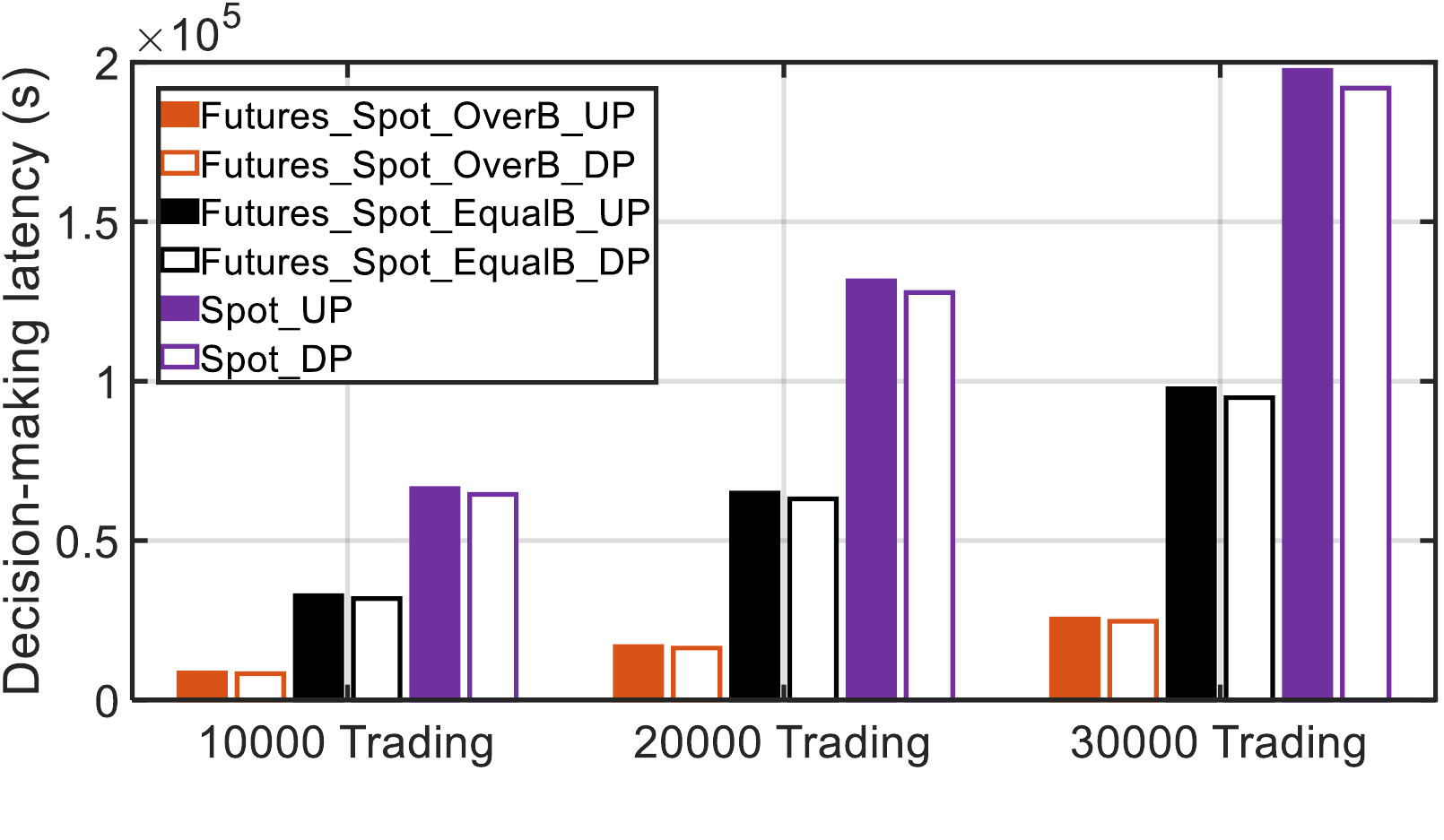}}
\vspace{-0.2cm}
\caption{Short-term and long-term performance on decision-making cost and decision-making latency.}
\label{fig4}
\end{figure*}
%\vspace{-1cm}
%f4
\begin{figure*}[h!t]
\centering
\subfigtopskip=0.1pt
\subfigbottomskip=0.8pt
\subfigure[]{\includegraphics[width=.254\linewidth]{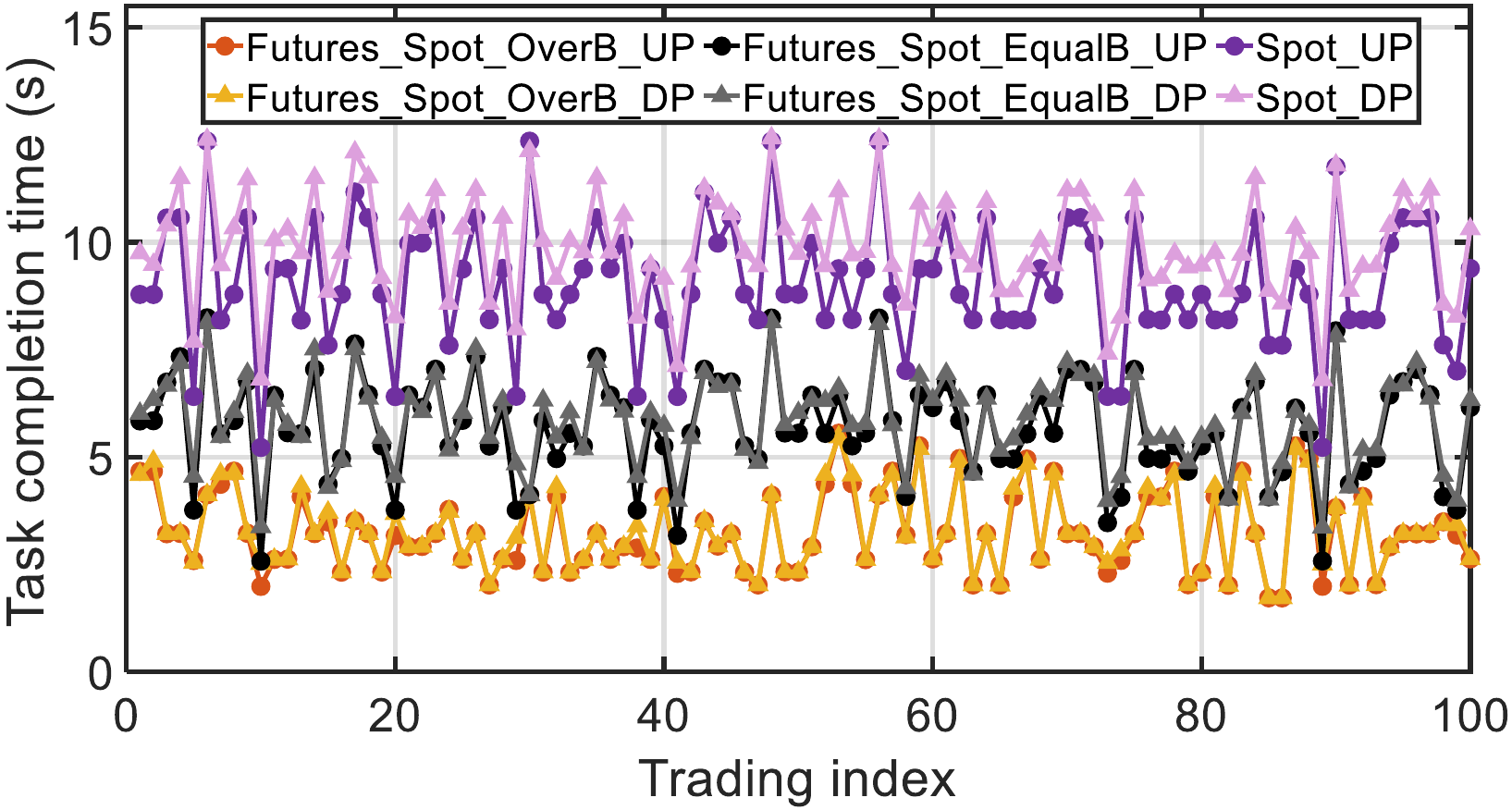}}\hfill
\subfigure[]{\includegraphics[width=.256\linewidth]{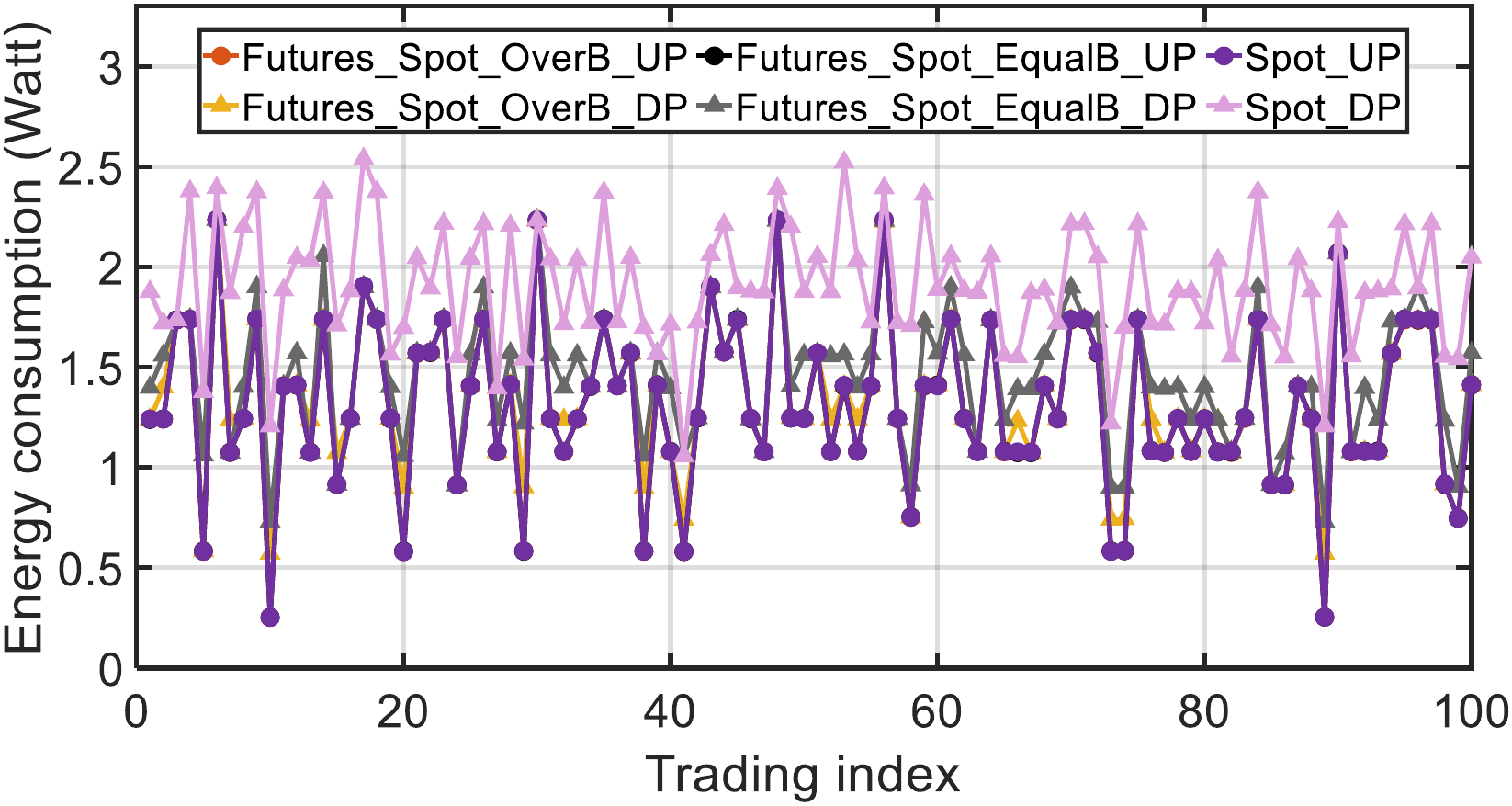}}\hfill
\subfigure[]{\includegraphics[width=.252\linewidth]{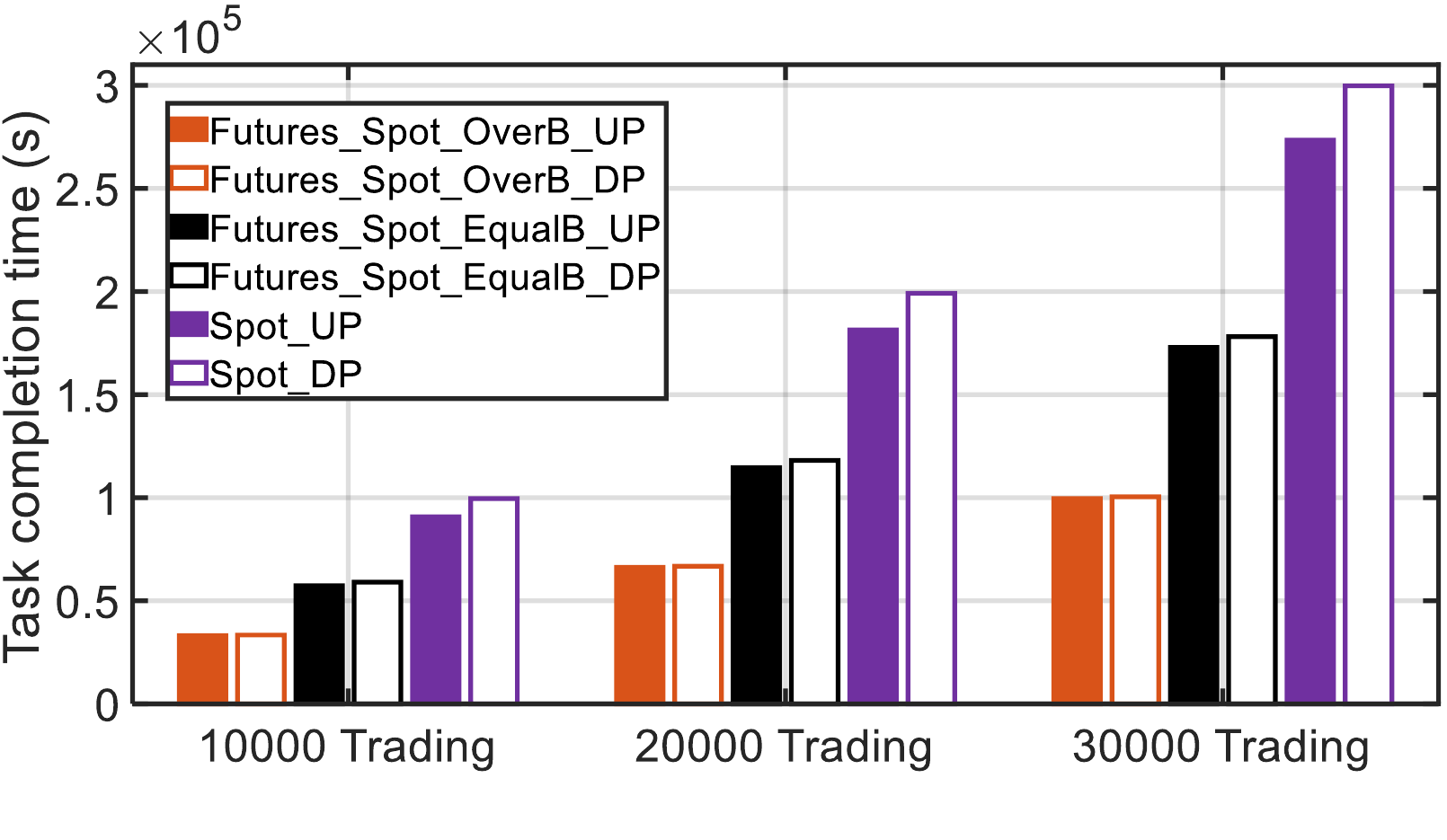}}\hfill
\subfigure[]{\includegraphics[width=.237\linewidth]{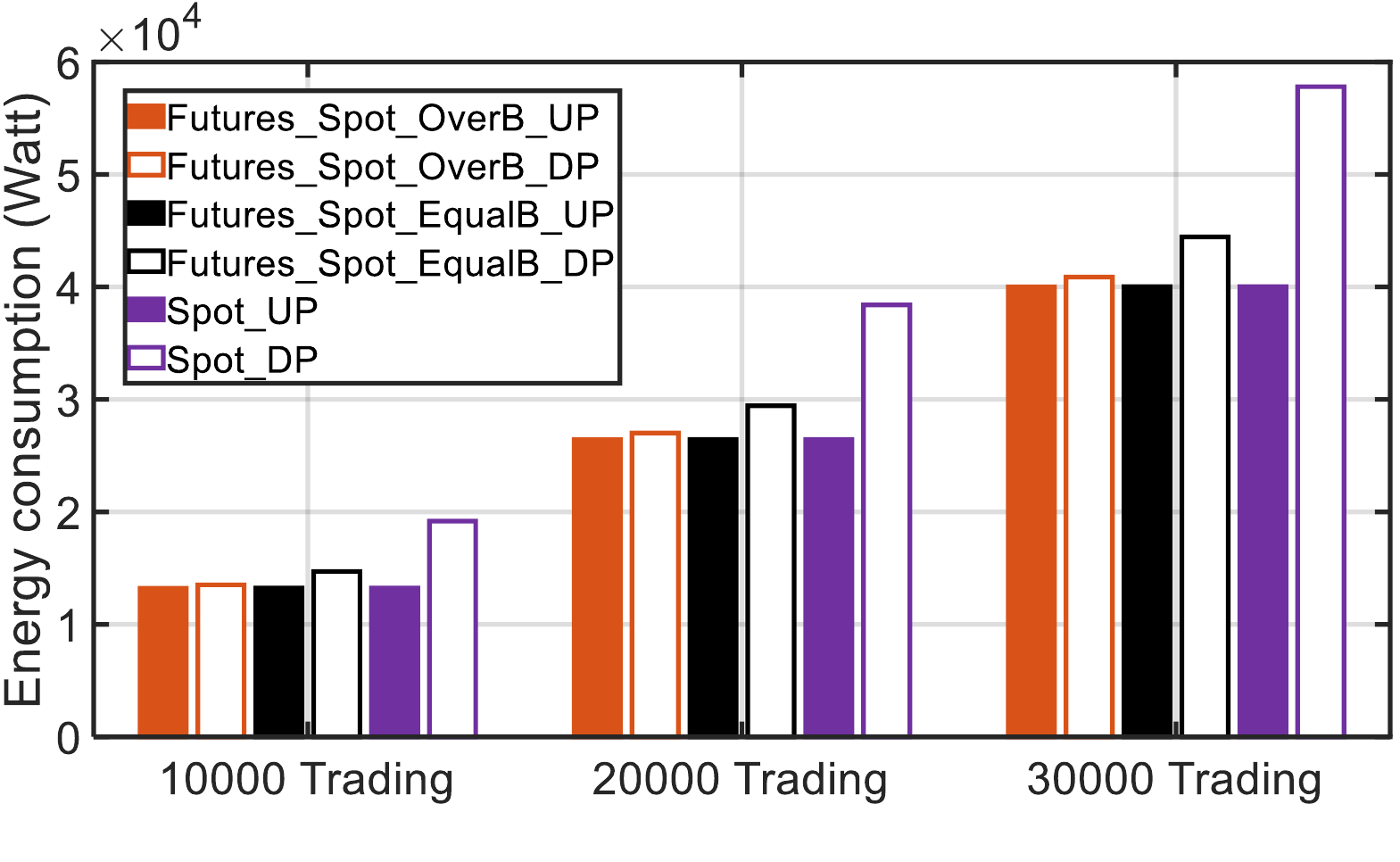}}
%\vspace{-0.2cm}
\caption{Short-term and long-term performance on task completion time and energy consumption.}
\label{fig_sim}
\end{figure*}
%f5
\begin{figure*}[t!]
\centering
\subfigtopskip=0.1pt
\subfigbottomskip=0.1pt
\subfigure[]{\includegraphics[width=.266\linewidth]{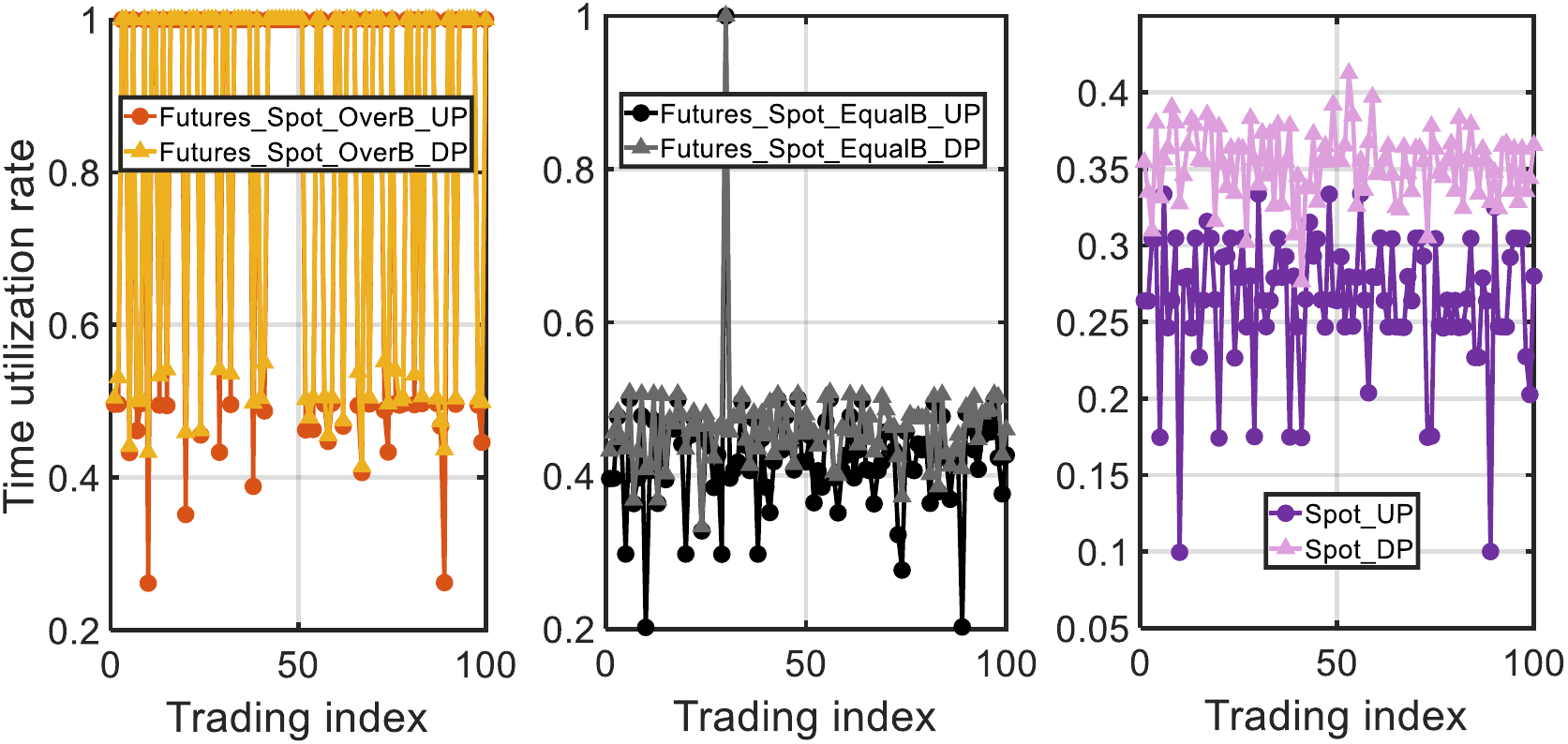}}\hfill
\subfigure[]{\includegraphics[width=.266\linewidth]{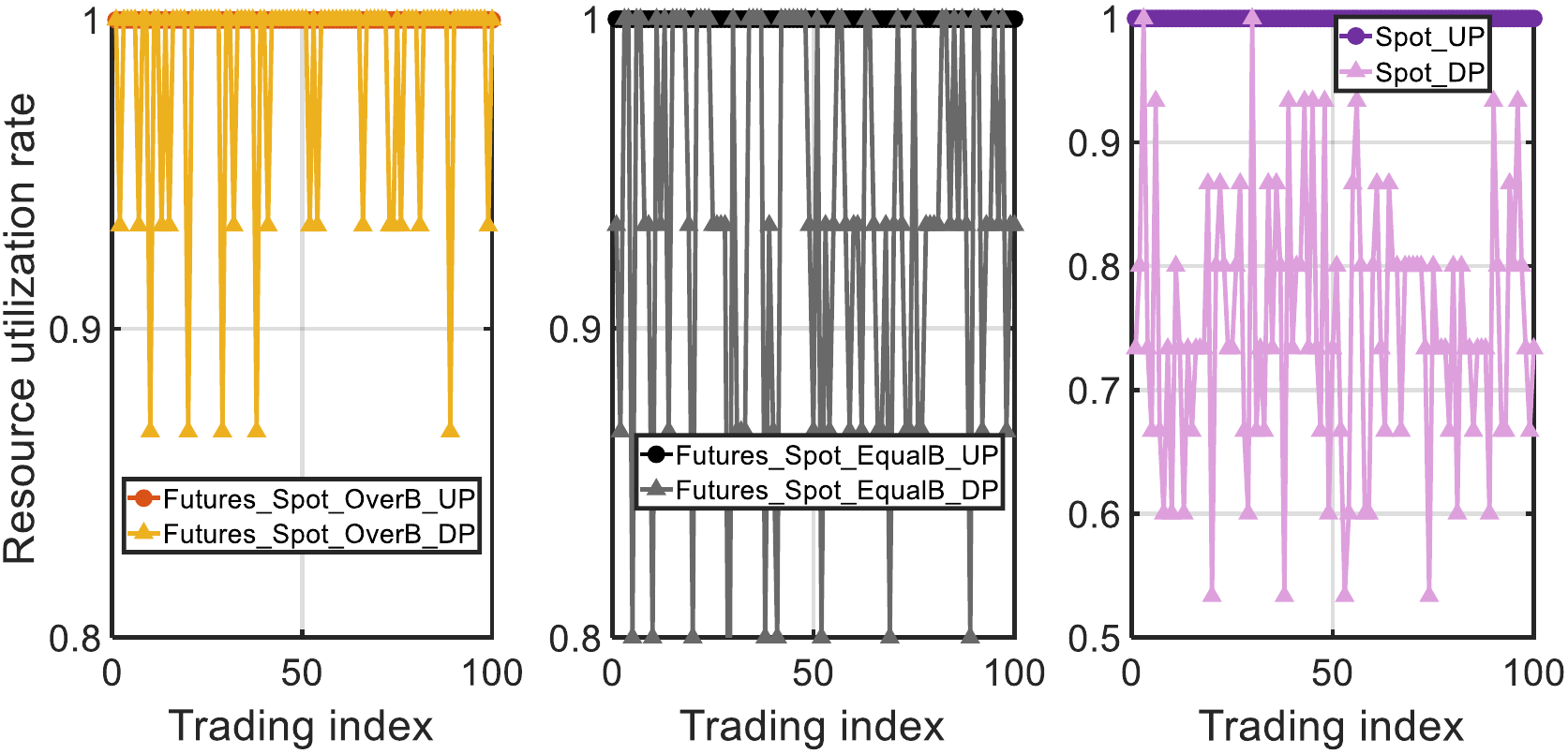}}\hfill
\subfigure[]{\includegraphics[width=.231\linewidth]{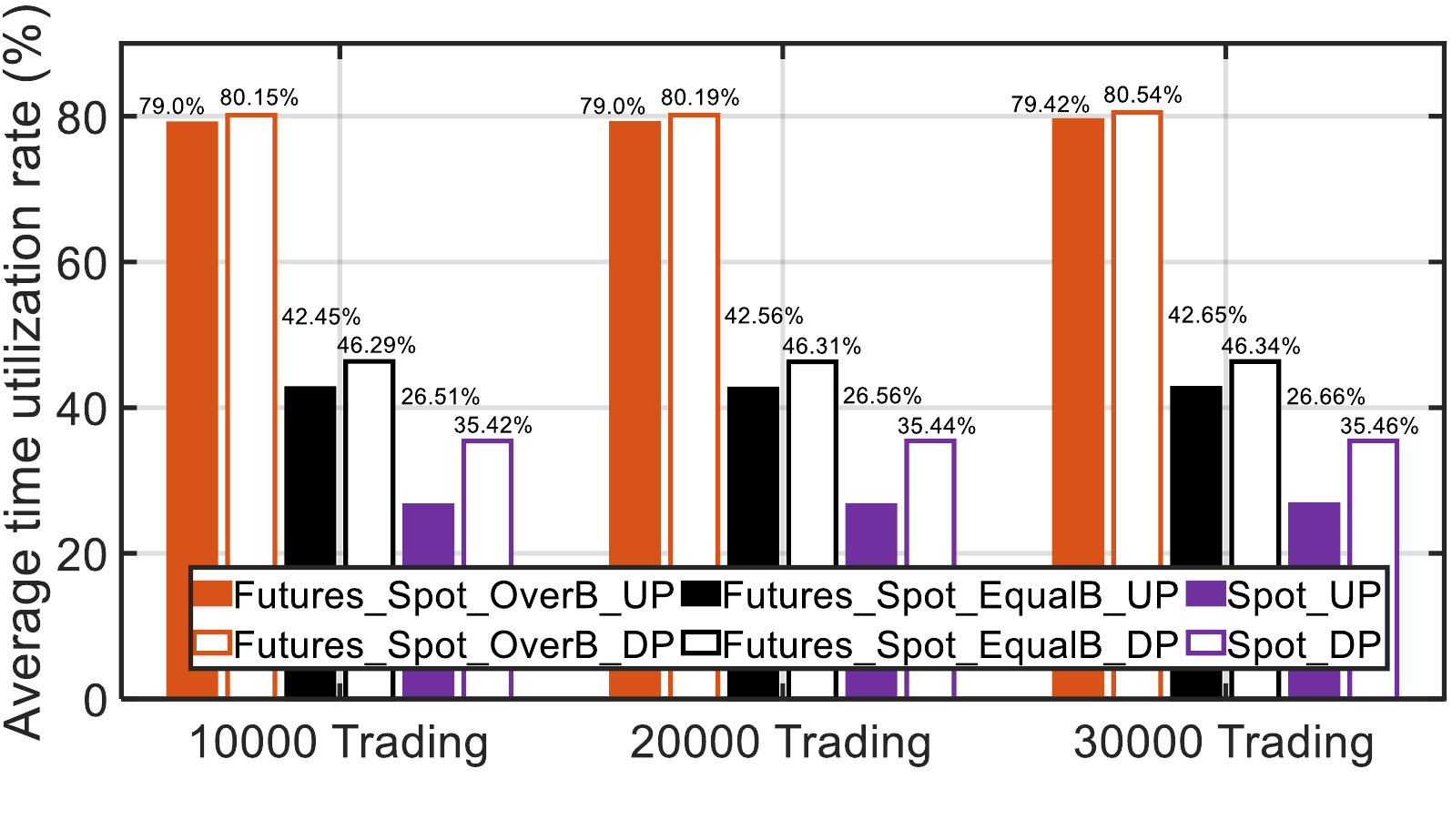}}\hfill
\subfigure[]{\includegraphics[width=.2365\linewidth]{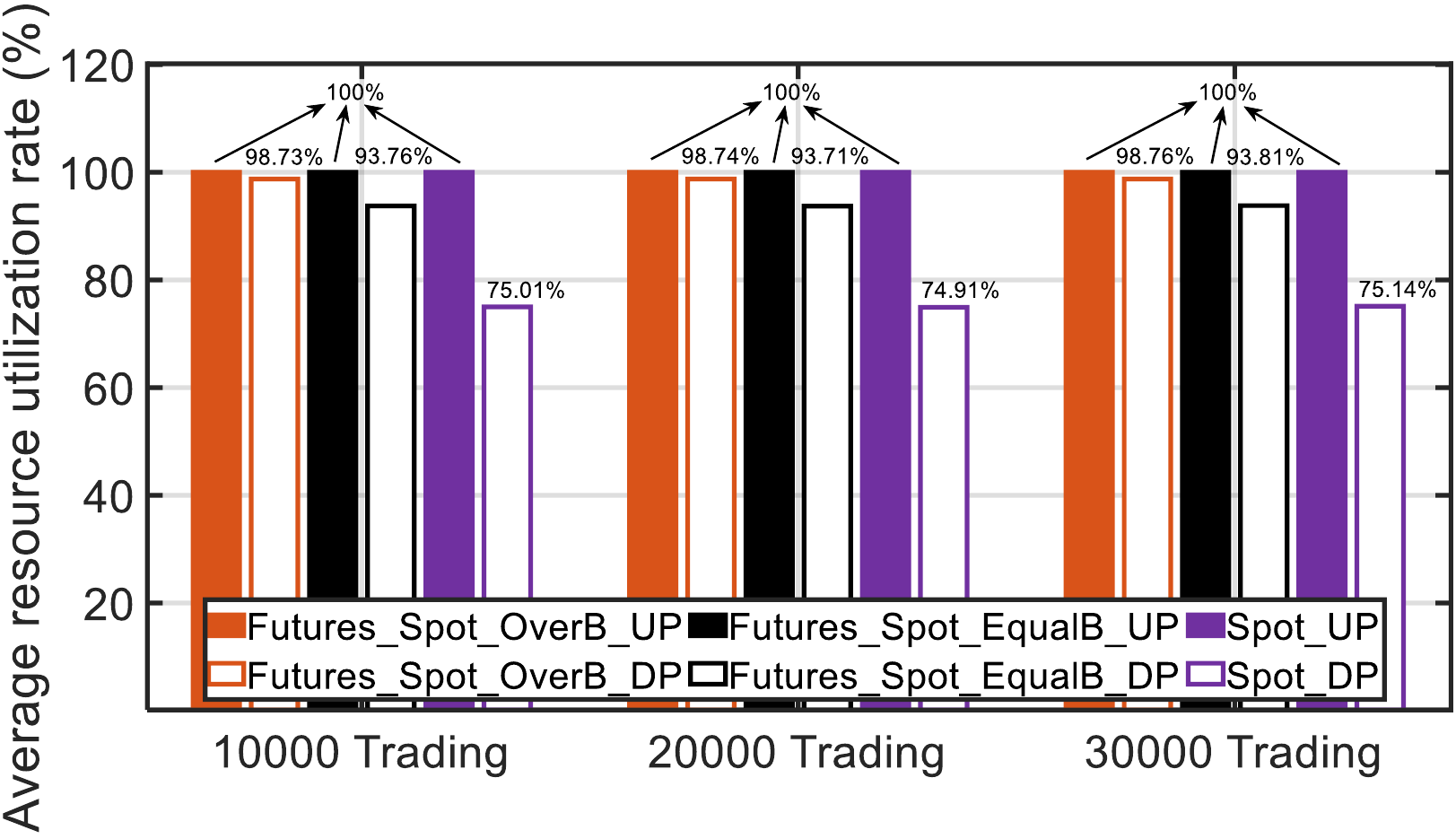}}
\vspace{-0.2cm}
\caption{Short-term and long-term performance on time and resource utilization.}
\label{fig_sim}
\end{figure*}
%\vspace{-0.5cm}
\noindent
In this simulation, we analyze both short-term performance via simulating 100 trading, and long-term performance via simulating large numbers of trading (e.g., 10000, 20000, and 30000), to evaluate the validity of the proposed overbooking-enabled resource trading mechanism. Fig.~2 and Table~I depict the short-term (Figs.~2(a)-2(b)) and long-term performance (Figs.~2(c)-2(d)) on utilities of seller and buyers. Specifically, Fig.~2(a) and Fig.~2(b) show the sum utility of 30 buyers, and utility of seller during each trading; where the proposed Futures\_Spot\_OverB\_UP achieves better buyers' (and also seller's) utility than Futures\_Spot\_EqualB\_UP in most trading, and Spot\_UP in all trading. Besides, the proposed Futures\_Spot\_OverB\_DP outperforms baseline methods on buyers' utility, although sometimes gets slightly lower seller's utility than Futures\_Spot\_EqualB\_DP and Spot\_DP, the total seller's utility (of 100 trading) of which has a small gap comparing with the two baseline methods under differential pricing, as given by Table~I. Figs.~2(c)-2(d) investigate long-term players' utilities (cumulative) via considering large numbers of trading, through monte carlo method. As can be seen from Fig. 2(c), the proposed overbooking-enabled mechanism obtains better buyers' utility than baseline methods under both uniform and differential pricing rules. In Fig.~2(d), although the proposed Futures\_Spot\_OverB\_DP gets slightly lower long-term seller's utility than Futures\_Spot\_EqualB\_DP, which, however, can achieve better performance on other factors as described by the following Figs.~3-5. Moreover, although Spot\_DP enables tinily higher seller's utility, it fails to provide mutually beneficial utilities to both players, owing to a low value of buyers' utility (see Fig.~2(c)).

Fig.~3 illustrates both the short-term (Fig.~3(a) and Fig.~3(b)) and long-term (Fig.~3(c) and Fig.~3(d)) performance on decision-making cost and latency. Fig.~3 demonstrates that the proposed overbooking-enabled mechanism greatly outperforms baseline methods on both DMC and DML (also in Table~I), as benefitted from the pre-signed forward contract and the relevant overbooking rate ($\kappa^*=20$). Specifically, all the 20 members will no longer have to spend extra time and energy on trading decision-making, significantly accelerating the service provision procedure. Although equal-booking-enabled trading mechanism may reduce DMC and DML to~some~extent ($\kappa^*=15$), which, however, faces challenges to handle ``no shows'' of buyers. Since the seller and buyers in Spot\_UP and Spot\_DP have to negotiate a consensus before every practical trading, they are suffering from excessive and unexpected DMC and DML, which pose great challenges to power/battery-constrained mobile devices. 

Performance evaluation on task completion time and energy consumption is analyzed by Fig.~4, where Fig.~4(a), Fig.~4(c) and Table~I depict that the proposed overbooking-enabled trading mechanism facilitates faster task completion especially comparing with spot trading, from both short-term and long-term perspectives. As shown by Fig.~3, the commendable performance on DML brought by the proposed mechanism enables an efficient trading mode, where the seller only has to negotiate with non-members about the trading consensus during each practical trading. Additionally, the proposed overbooking-enabled mechanism reaches similar energy consumption comparing with Futures\_Spot\_EqualB\_UP and Spot\_UP, while outperforming Futures\_Spot\_EqualB\_DP and Spot\_DP on both short- and long-term energy consumption, as demonstrated by Fig.~4(b), Fig.~4(d) and Table~I. For example, considering 10000 trading under uniform pricing in Fig. 4(c), the proposed mechanism (3.307s/trading) achieves 42.23\% and 63.55\% improvment on task completion time than equal-booking-based method (5.724s/trading) and spot trading (9.072s/trading). Namely, spot buyers may spend roughly 2.7 times longer to complete the same number of tasks, rather than the proposed overbooking-enabled mechanism.

Investigation on time and resource utilization are detailed in Fig. 5 from short- and long- term perspectives. Benefitted from the pre-determined forward contract and the overbooking policy, the proposed mechanism facilitates far better TUR than baseline methods since members do not have to spend extra time on negotiating a consensus during each practical trading, which significantly improves the time efficiency (see Fig.~5(a) and Table~I). As depicted by Fig.~5(c), the proposed Futures\_Spot\_OverB\_UP achieves averagely 85.98\% and 197.78\% improvement on time utilization, comparing with Futures\_Spot\_EqualB\_UP and Spot\_UP. Besides, Futures\_Spot\_OverB\_DP obtains averagely 73.37\% and 126.56\% increase in time utilization, comparied with Futures\_Spot\_EqualB\_DP and Spot\_DP. 
Fig. 5(b), Fig. 5(d), and Table~I illustrate that the proposed mechanism offers substantial resource utilization; namely, overbooking provides a commendable solution to handle the dynamic resource demand, e.g.,  ``no shows'', in trading market. Specifically, the proposed mechanism achieves 5.32\% and 31.62\% improvement on resource utilization under differential pricing, as compared with equal-booking-based and spot trading mechanisms (depicted by Fig. 5(d)).

In summary, the proposed overbooking-enabled resource trading mechanism under futures and spot integrated market offers mutually beneficial utilities to both seller and buyers, achieves commendable decision-making cost and latency, as well as faster task completion and lower energy consumption, while facilitating sufficient time and resource utilization, comparing with equal-booking-based and spot trading mechanisms.

\section{Conclusion}

\noindent
Motivated by challenges of excessive latency and cost incurred by onsite decision-making, as well as the possible ``no shows'' of smart devices, in this paper, an overbooking-enabled resource trading mechanism considering an edge server (seller) and multiple smart devices (buyers) is investigated under mobile edge network architecture, via integrating both futures and spot market. Specifically, in futures market, a mutually beneficial and risk tolerable forward contract as well as the relevant overbooking rate are studied, for which an effective bilateral negotiation scheme is proposed by alternatively optimizing the seller's and members' expected utilities. For spot trading problem, considering uniform pricing and differential pricing, we propose two bilateral negotiation schemes via addressing non-convex optimization and knapsack problems, based on the current network condition. Experiential results demonstrate that the proposed mechanism can achieve mutually beneficial utilities for both the seller and buyers, and outperform the baseline methods on critical indicators such as decision-making latency and cost, as well as time and resource utilization.

\appendices
%\begin{spacing}{1}
\section{Derivation of Expected Utility of Member}
%$\overline{{\mathcal{U}}^{Mem}}(p,q,r,\kappa ,\bm{\mathcal{A}},\bm{\mathcal{Y}})$
\noindent
%Note that random variables ${\alpha}_{m}$ and ${\gamma}_{m}$ are independent with each other, we discuss the derivations of $\text{E}\left[V\right]$ and $\text{E}\left[U^{PP}_{m}\right]$ hereafter.
%\noindent
%$\bullet$ \textbf{Derivation of $\text{E}\left[V\right]$:} 
Let random variable $X1=\sum\nolimits^{m=\kappa}_{m=1}{{\alpha}_{m}}$, and $X2={\left(X1,S\right)}^-$ for analytical simplicity, we first discuss the PMF of $X1$ as given by~\eqref{eq31}.
%\begin{align}
% \label{eq31}
% \begin{aligned}
% &\quad~\text{Pr}\left(X1=x\right)
%=
% \begin{cases}
%  0,& x<0 \\
%{\left(1-a\right)}^{\kappa },& x=0 \\
%{C^1_{\kappa }a^1\left(1-a\right)}^{\kappa -1},& x=1 \\
%{C^2_{\kappa }a^2\left(1-a\right)}^{\kappa -2},& x=2 \\
%\vdots & \\
%a^{\kappa},& x=\kappa\\
%0,& x>\kappa \\
% \end{cases}\\
% &=C^x_{\kappa }a^x{\left(1-a\right)}^{\kappa-x},x\in \{0,1,\dots ,\kappa \}
% \end{aligned}
%\end{align}
{\setlength\abovedisplayskip{5pt}
\setlength\belowdisplayskip{5pt}
\begin{align}
 \label{eq31}
 \begin{aligned}
 &\text{Pr}\left(X1=x\right)
%\begin{cases}
%  0,& x<0 \vspace{-1.5ex}\\
%{\left(1-a\right)}^{\kappa },& x=0 \vspace{-1.5ex}\\
%{C^1_{\kappa }a^1\left(1-a\right)}^{\kappa -1},& x=1 \vspace{-1.5ex}\\
%%{C^2_{\kappa }a^2\left(1-a\right)}^{\kappa -2},& x=2 \\
%\vdots & \vspace{-1.5ex}\\
%a^{\kappa},& x=\kappa \vspace{-1.5ex}\\
%0,& x>\kappa \\
% \end{cases}
=C^x_{\kappa }a^x{\left(1-a\right)}^{\kappa-x},x\in \{0,1,\dots ,\kappa \}
 \end{aligned}
\end{align}}Based on~\eqref{eq31}, we consider $\text{E}\left[X2\right]$ via the following two cases:

\noindent
$\bullet$ \textbf{Case 1 ($\kappa \le S$):} $X2=X1$, we have $\text{E}\left[X2\right]=\text{E}\left[X1\right]=\kappa a$.

\noindent
$\bullet$ \textbf{Case 2 ($\kappa > S$):} $X2=\begin{cases}
X1,& X1<S \vspace{-1.5ex} \\
S,& X1\ge S
 \end{cases}
$ and we have PMF of $X2$ as given by~\eqref{eq32}.
%\begin{align}
%\label{eq32}
%&\text{Pr}\left(X2=x\right)
% =
% \begin{cases}
%  0,&x<0 \\
%{\left(1-a\right)}^{\kappa },& x=0 \\
%{C^1_{\kappa }a^1\left(1-a\right)}^{\kappa -1},& x=1 \\
%{C^2_{\kappa }a^2\left(1-a\right)}^{\kappa -2},& x=2 \\
%\vdots & \\
%{C^{S-1}_{\kappa }a^{S-1}\left(1-a\right)}^{\kappa -S+1},& x=S-1\\
%\sum\nolimits^{i=\kappa}_{i=S}{C^i_{\kappa}a^i{\left(1-a\right)}^{\kappa -i}},& x=S \\
%0,& x>S \\
%   \end{cases}\notag\\
%&=
%\begin{cases}
%C^x_{\kappa }a^x{\left(1-a\right)}^{\kappa-x},& x\in \left\{0,1,\dots ,S-1\right\} \\
%\sum\nolimits^{i=\kappa}_{i=S}{C^i_{\kappa}a^i{\left(1-a\right)}^{\kappa -i}},& x=S\\
%\end{cases}
%\end{align}
{\setlength\abovedisplayskip{5pt}
\setlength\belowdisplayskip{5pt}
\begin{align}
\label{eq32}
&\text{Pr}\left(X2=x\right)
 =
% \begin{cases}
%  0,&x<0 \vspace{-1.5ex}\\
%{\left(1-a\right)}^{\kappa },& x=0 \vspace{-1.5ex}\\
%{C^1_{\kappa }a^1\left(1-a\right)}^{\kappa -1},& x=1 \vspace{-1.5ex}\\
%%{C^2_{\kappa }a^2\left(1-a\right)}^{\kappa -2},& x=2 \\
%\vdots & \\
%%{C^{S-1}_{\kappa }a^{S-1}\left(1-a\right)}^{\kappa -S+1},& x=S-1\\
%\sum\nolimits^{i=\kappa}_{i=S}{C^i_{\kappa}a^i{\left(1-a\right)}^{\kappa -i}},& x=S \vspace{-1.5ex}\\
%0,& x>S \\
%   \end{cases}
\begin{cases}
C^x_{\kappa}a^x{\left(1-a\right)}^{\kappa-x},&0 \leq x\leq S-1 \vspace{-1.5ex} \\
\sum\nolimits^{i=\kappa}_{i=S}{C^i_{\kappa}a^i{\left(1-a\right)}^{\kappa -i}},&x=S\\
\end{cases}
\end{align}}Accordingly, $\text{E}\left[X2\right]$ in \textbf{Case 2} can be
 calculated by~\eqref{eq33}.
%\begin{align}
%\label{eq33}
%\text{E}\left[X2\right]&=\sum\nolimits^{i=S-1}_{i=0}{iC^i_{\kappa }a^x{\left(1-a\right)}^{\kappa -i}}+\notag\\
%&\quad~S\sum\nolimits^{i=\kappa }_{i=S}{C^i_{\kappa }a^i{\left(1-a\right)}^{\kappa -i}}
%\end{align}
{\setlength\abovedisplayskip{5pt}
\setlength\belowdisplayskip{5pt}
\begin{align}
\label{eq33}
\text{E}\left[X2\right]&=\sum\nolimits^{i=S-1}_{i=0}{iC^i_{\kappa }a^x{\left(1-a\right)}^{\kappa -i}}+S\sum\nolimits^{i=\kappa }_{i=S}{C^i_{\kappa }a^i{\left(1-a\right)}^{\kappa -i}}
\end{align}}As a result, $\text{E}\left[V\right]$ is represented by~\eqref{eq34}.
%\begin{align}
%\label{eq34}
%\text{E}\left[V\right]&=\text{E}\left[X1\right]-\text{E}\left[X2\right]\notag\\
%&=\begin{cases}
%0,&\kappa \le S \\
%\kappa a-\left(\sum\nolimits^{i=S-1}_{i=0}{iC^i_{\kappa}a^i{\left(1-a\right)}^{\kappa -i}}+\right.\\
%\qquad \left.S\sum\nolimits^{i=\kappa }_{i=S}{C^i_{\kappa }a^i{\left(1-a\right)}^{\kappa -i}}\right),&\kappa >S\\
% \end{cases}
%\end{align}
{\setlength\abovedisplayskip{5pt}
\setlength\belowdisplayskip{5pt}
\begin{align}
\label{eq34}
\text{E}\left[V\right]&=\text{E}\left[X1-X2\right]
=
\begin{cases}
0,&\kappa \le S \vspace{-1.5ex} \\
\kappa a-\left(\sum\nolimits^{i=S-1}_{i=0}{iC^i_{\kappa}a^i{\left(1-a\right)}^{\kappa -i}}+S\sum\nolimits^{i=\kappa }_{i=S}{C^i_{\kappa }a^i{\left(1-a\right)}^{\kappa -i}}\right),&\kappa >S\\
 \end{cases}
\end{align}}Let random variable $Y1=\frac{{1}}{{\log}_{2}\left(1+e^{tran}{\gamma }_{m}\right)}$ for notational simplicity, the CDF of $Y1$ is given by~\eqref{eq35}, according to ${\gamma}_{m} \sim\bm{\text{U}}({\varepsilon}_{1},{\varepsilon}_{2})$.
% \begin{align}
% \label{eq35}
%   \text{F}_{Y1}\left(y\right)&=\text{Pr}\left(\frac{1}{{\log}_{2}
%\left(\text{1+}e^{tran}{\gamma}_{m}\right)}\le y\right)\notag\\
%&=\text{Pr}\left({\gamma}_{m}\ge \frac{2^{\frac{1}{y}}-1}{e^{tran}}\right)=1-\text{Pr}\left({\gamma}_{m}\le \frac{2^{\frac{1}{y}}-1}{e^{tran}}\right)\notag\\
%&=\begin{cases}
%\displaystyle{0,\quad y<\frac{1}{{log}_{2}\left(1+e^{tran}{\varepsilon }_{2}\right)}}\\
%\displaystyle{1-\frac{2^{\frac{1}{y}}-1-e^{tran}{\varepsilon}_{1}}{e^{tran}\left({{\varepsilon }_{2}-\varepsilon }_{1}\right)}},\\ \qquad~\displaystyle{\frac{1}{{\log}_{2}\left(1+e^{tran}{\varepsilon}_{2}\right)}\le y\le \frac{1}{{\log}_{2}\left(1+e^{tran}{\varepsilon }_{1}\right)} }\\
%\displaystyle{1,\quad y>\frac{1}{{\log}_2\left(1+e^{tran}{\varepsilon }_{1}\right)}}
%\end{cases}
% \end{align}
\begin{align}
 \label{eq35}
   \text{F}_{Y1}\left(y\right)&
%=\text{Pr}\left(\frac{1}{{\log}_{2}
%\left(1+e^{tran}{\gamma}_{m}\right)}\le y\right)=\text{Pr}\left({\gamma}_{m}\ge \frac{2^{\frac{1}{y}}-1}{e^{tran}}\right)
%=1-\text{Pr}\left({\gamma}_{m}\le \frac{2^{\frac{1}{y}}-1}{e^{tran}}\right)
=\begin{cases}
0,&y<\frac{1}{{log}_{2}\left(1+e^{tran}{\varepsilon }_{2}\right)}\vspace{-0.9ex} \\
1-\frac{2^{\frac{1}{y}}-1-e^{tran}{\varepsilon}_{1}}{e^{tran}\left({{\varepsilon }_{2}-\varepsilon }_{1}\right)},&\frac{1}{{\log}_{2}\left(1+e^{tran}{\varepsilon}_{2}\right)}\le y\le \frac{1}{{\log}_{2}\left(1+e^{tran}{\varepsilon }_{1}\right)}\vspace{-0.9ex} \\
1,&y>\frac{1}{{\log}_2\left(1+e^{tran}{\varepsilon }_{1}\right)}
\end{cases}
 \end{align}
 
\noindent
The PDF of $Y$ can be obtained as given in~\eqref{eq36}, based on~(35).
%\begin{align}
%\label{eq36}
%&\text{Pr}\left(Y1=y\right)=\frac{\partial F_{Y1}\left(y\right)}{\partial y}=\notag\\
%&\begin{cases}
%\displaystyle{\frac{\text{ln}2\times 2^{\frac{1}{y}}}{y^2e^{tran}\left({{\varepsilon }_{\text{2}}-\varepsilon }_{\text{1}}\right)}, \frac{1}{{\text{log}}_2\left(1+e^{tran}{\varepsilon }_{\text{2}}\right)}\le}\\
%\qquad \displaystyle{y\le \frac{1}{{\text{log}}_2\left(1+e^{tran}{\varepsilon }_{\text{1}}\right)}}\\
%0,\quad \text{otherwise}\\
%\end{cases}
%\end{align}
{\setlength\abovedisplayskip{5pt}
\setlength\belowdisplayskip{5pt}
\begin{align}
\label{eq36}
&\text{Pr}\left(Y1=y\right)=\frac{\partial F_{Y1}\left(y\right)}{\partial y}=
\begin{cases}
\frac{\text{ln}2\times 2^{\frac{1}{y}}}{y^2e^{tran}\left({{\varepsilon }_{\text{2}}-\varepsilon }_{\text{1}}\right)}, &\frac{1}{{\text{log}}_2\left(1+e^{tran}{\varepsilon }_{\text{2}}\right)}\le~y\le \frac{1}{{\text{log}}_2\left(1+e^{tran}{\varepsilon }_{\text{1}}\right)}\vspace{-0.9ex} \\
0,&\text{otherwise}\\
\end{cases}
\end{align}}Accordingly, $\text{E}\left[Y1\right]$ is thus calculated by~\eqref{eq37}, where ${\mathbb{C}}_1=\ln2\times {\log}_2\left(1+e^{tran}{\varepsilon }_{1}\right)$ and ${\mathbb{C}}_2=\ln2\times {\log}_{2}\left(1+e^{tran}{\varepsilon }_{2}\right)$, for notational simplicity.
%\begin{align}
%\label{eq37}
%\text{E}\left[Y1\right]&=\text{E}\left[\frac{1}{{\log}_{2}\left(1+e^{tran}{\gamma }_{m}\right)}\right]\notag\\
%&=\int^{\frac{1}{{\log}_{2}
%\left(1+e^{tran}{\varepsilon}_{1}\right)}}_{\frac{1}{{\text{log}}_{\text{2}}\left(1+e^{tran}{\varepsilon }_{\text{2}}\right)}}{y\text{Pr}\left(Y=y\right)dy}\notag\\
%&=\frac{ln2}{e^{tran}\left({{\varepsilon }_{2}-\varepsilon }_{1}\right)}\int^{\frac{1}{{\log}_{2}\left(1+e^{tran}{\varepsilon }_{1}\right)}}_{\frac{1}{{\log}_{2}\left(1+e^{tran}{\varepsilon }_{2}\right)}}{\left(\frac{2^{\frac{1}{y}}}{y}\right)dy}\notag\\
%&=\frac{\ln2\times \int^{{\mathbb{C}}_2}_{{\mathbb{C}}_1}{\left(\frac{e^y}{y}\right)}dy}{e^{tran}\left({{\varepsilon }_{2}-\varepsilon }_{1}\right)}
%\end{align}
{\setlength\abovedisplayskip{5pt}
\setlength\belowdisplayskip{5pt}
\begin{align}
\label{eq37}
\text{E}\left[Y1\right]=
%\text{E}\left[\frac{1}{{\log}_{2}\left(1+e^{tran}{\gamma }_{m}\right)}\right]
\int^{\frac{1}{{\log}_{2}
\left(1+e^{tran}{\varepsilon}_{1}\right)}}_{\frac{1}{{\text{log}}_{\text{2}}\left(1+e^{tran}{\varepsilon }_{\text{2}}\right)}}{y\text{Pr}\left(Y=y\right)dy}
=\frac{\ln2\times\int^{\frac{1}{{\log}_{2}\left(1+e^{tran}{\varepsilon }_{1}\right)}}_{\frac{1}{{\log}_{2}\left(1+e^{tran}{\varepsilon }_{2}\right)}}{\left(\frac{2^{\frac{1}{y}}}{y}\right)dy}}{e^{tran}({{\varepsilon }_{2}-\varepsilon }_{1})}=\frac{\ln2\times \int^{{\mathbb{C}}_2}_{{\mathbb{C}}_1}{\left(\frac{e^y}{y}\right)}dy}{e^{tran}\left({{\varepsilon }_{2}-\varepsilon }_{1}\right)}
\end{align}}According to~\eqref{eq37}, $\text{E}\left[U^{PP}_{m}\right]$ is expressed by~\eqref{eq38} and $\overline{{\mathcal{U}}^{Mem}}(p,q,r,\kappa ,\bm{\mathcal{A}},\bm{\mathcal{Y}})$ is thus obtained.
%\begin{align}
%\label{eq38}
% &\quad\;\text{E}\left[U^{PP}_{m}\right]\notag\\
% &=\text{E}\left[\left(\frac{{\omega}_{1}d^{comp}}{f^{b}}-\frac{{\omega}_{1}d^{size}}{W{\log}_{2}\left(1+e^{tran}{\gamma }_{m}\right)}-\frac{{\omega}_{1}d^{comp}}{f^{s}}\right)+\right.\notag\\
% &\quad~\left.\left(\frac{{\omega}_{2}e^{loc}d^{comp}}{f^{b}}-\frac{{\omega}_{2}e^{tran}d^{size}}{W{\log}_{2}
% \left(1+e^{tran}\times {\gamma }_{m}\right)}\right)-pd^{comp}\right]\notag\\
% &=\frac{{\omega}_{1}d^{comp}}{f^{b}}-\frac{{\omega}_{1}d^{size}}{W}\text{E}\left[Y1\right]
% -\frac{{\omega}_{1}d^{comp}}{f^{s}}+\notag\\
% &\quad~\frac{{{\omega}_{2}e}^{loc}d^{comp}}{f^{b}}-\frac{{\omega}_{2}e^{tran}d^{size}}{W}\text{E}
% \left[Y1\right]-pd^{comp}\notag\\
% &=\frac{{{\omega}_{1}d}^{comp}+
% {{\omega}_{2}e}^{loc}d^{comp}}{f^{b}}-\frac{{\omega}_{1}d^{comp}}{f^{s}}-\notag\\
% &\quad~pd^{comp}-\left(\frac{{\omega}_{1}d^{size}+
% {\omega}_{2}e^{tran}d^{size}}{W}\right)\text{E}\left[Y1\right]\notag\\
% &=\left(\frac{{\omega}_{1}+{{\omega}_{2}e}^{loc}}{f^{b}}-\frac{{\omega}_{1}}{f^{s}}-p\right)d^{comp}-\notag\\
% &\quad\frac{\ln2d^{size}\left({\omega}_{1}+
% {\omega}_{2}e^{tran}\right)\times \int^{{\mathbb{C}}_2}_{{\mathbb{C}}_1}{\left(\frac{e^y}{y}\right)}{d}y}{We^{tran}\left({{\varepsilon }_{2}-\varepsilon }_{1}\right)}
%\end{align}
\begin{align}
\label{eq38}
 &\text{E}\left[U^{PP}_{m}\right]=
% \text{E}\left[\left(\frac{{\omega}_{1}d^{comp}}{f^{b}}-\frac{{\omega}_{1}d^{size}}{W{\log}_{2}\left(1+e^{tran}{\gamma }_{m}\right)}-\frac{{\omega}_{1}d^{comp}}{f^{s}}\right)+\left(\frac{{\omega}_{2}e^{loc}d^{comp}}{f^{b}}-\frac{{\omega}_{2}e^{tran}d^{size}}{W{\log}_{2}
% \left(1+e^{tran}\times {\gamma }_{m}\right)}\right)-pd^{comp}\right]\notag\\
%=\frac{{\omega}_{1}d^{comp}}{f^{b}}-\frac{{\omega}_{1}d^{size}}{W}\text{E}\left[Y1\right]
% -\frac{{\omega}_{1}d^{comp}}{f^{s}}+\frac{{{\omega}_{2}e}^{loc}d^{comp}}{f^{b}}-\frac{{\omega}_{2}e^{tran}d^{size}}{W}\text{E}
% \left[Y1\right]-pd^{comp}\notag\\
% &=\frac{{{\omega}_{1}d}^{comp}+
% {{\omega}_{2}e}^{loc}d^{comp}}{f^{b}}-\frac{{\omega}_{1}d^{comp}}{f^{s}}-pd^{comp}-\left(\frac{{\omega}_{1}d^{size}+
% {\omega}_{2}e^{tran}d^{size}}{W}\right)\text{E}\left[Y1\right]\notag\\
 \left(\frac{{\omega}_{1}+{{\omega}_{2}e}^{loc}}{f^{b}}-\frac{{\omega}_{1}}{f^{s}}-p\right)d^{comp}-\frac{\ln2d^{size}\left({\omega}_{1}+
 {\omega}_{2}e^{tran}\right)\times \int^{{\mathbb{C}}_2}_{{\mathbb{C}}_1}{\left(\frac{e^y}{y}\right)}{d}y}{We^{tran}\left({{\varepsilon }_{2}-\varepsilon }_{1}\right)}
\end{align}

\section{Derivation of Risks of Member}

\noindent
%$\bullet$ \textbf{Derivation of ${\mathcal{R}}^{MRisk}\left(p,q,\bm{\mathcal{A}},\bm{\mathcal{Y}}\right)$:} 
Based on the pre-determined $Y1=\frac{1}{{\log}_{2}\left(1+e^{tran}{\gamma }_{m}\right)}$, ${\mathcal{R}}^{MRisk}\left(p,q,\bm{\mathcal{A}},\bm{\mathcal{Y}}\right)$ is rewritten as~(39).
% \begin{gather}
% \label{eq39}
%  {\mathcal{R}}^{MRisk}\left(p,q,\bm{\mathcal{A}},\bm{\mathcal{Y}}\right)
%=\text{Pr}\left(\frac{{\alpha}_{m}U^{PP}_{m}+(1-{\alpha}_{m})U^{DE}}{U_{min}}\le {\xi}_1\right)\notag\\
%=\text{Pr}\left({\alpha}_{m}\left(\frac{{{\omega}_{1}d}^{comp}+
%{\omega}_{2}e^{loc}d^{comp}}{f^{b}}-\right.\right.\notag\\
%\frac{{{\omega}_{1}d}^{comp}}{f^{s}}+qd^{comp}-pd^{comp}-\notag\\
%\left.\frac{{\omega}_{2}e^{tran}d^{size}+
%{\omega}_{1}d^{size}}{W}Y1\right)\le \left.{\xi}_1U_{min}+qd^{comp}\right)\notag\\
%=\text{Pr}\left({\alpha}_{m}\left({\mathbb{C}}_3-{\mathbb{C}}_4Y1\right)\le {\mathbb{C}}_5\right)
% \end{gather}
{\setlength\abovedisplayskip{5pt}
\setlength\belowdisplayskip{5pt}
 \begin{gather}
 \label{eq39}
  {\mathcal{R}}^{MRisk}\left(p,q,\bm{\mathcal{A}},\bm{\mathcal{Y}}\right)
=\text{Pr}\left(\frac{{\alpha}_{m}U^{PP}_{m}+(1-{\alpha}_{m})U^{DE}}{U_{min}}\le {\xi}_1\right)
%=\text{Pr}\left({\alpha}_{m}\left(\frac{{{\omega}_{1}d}^{comp}+
%{\omega}_{2}e^{loc}d^{comp}}{f^{b}}-\right.\right.
%\frac{{{\omega}_{1}d}^{comp}}{f^{s}}+qd^{comp}-pd^{comp}-\notag\\
%\left.\frac{{\omega}_{2}e^{tran}d^{size}+{\omega}_{1}d^{size}}{W}Y1\right)\le \left.{\xi}_1U_{min}+qd^{comp}\right)\notag\\
=\text{Pr}\left({\alpha}_{m}\left({\mathbb{C}}_3-{\mathbb{C}}_4Y1\right)\le {\mathbb{C}}_5\right),
 \end{gather}}where ${\mathbb{C}}_3=\frac{{\omega}_{1}d^{comp}+{\omega}_{2}e^{loc}d^{comp}}{f^{b}}-\frac{{\omega}_{1}d^{comp}}{f^{s}}+qd^{comp}-pd^{comp}$, ${\mathbb{C}}_4=\frac{{\omega}_{2}e^{tran}d^{size}+{\omega}_{1}d^{size}}{W}$, and ${\mathbb{C}}_5={\xi }_1U_{min}+qd^{comp}$, which are constants under any given $p$ and $q$ for notational simplicity. Let random variable $Y2={\mathbb{C}}_3-{\mathbb{C}}_4Y1$, we discuss the CDF of $Y2$ which is given by~\eqref{eq40}.
%\begin{align}
%\label{eq40}
%{\text{F}}_{Y2}\left(y\right)&=\text{Pr}\left(Y2\le y\right)=\text{Pr}\left(Y1\ge \frac{{\mathbb{C}}_3-y}{{\mathbb{C}}_4}\right)\notag\\
%&=1-\text{Pr}\left(Y1<\frac{{\mathbb{C}}_3-y}{{\mathbb{C}}_4}\right)\notag\\
%&=1-{\text{F}}_{Y1}\left(\frac{{\mathbb{C}}_3-y}{{\mathbb{C}}_4}\right)\notag\\
%&=\begin{cases}
%\displaystyle{0, y<{\mathbb{C}}_3-\frac{{\mathbb{C}}_4}{{\log}_2
%\left(1+e^{tran}{\varepsilon}_{\text{1}}\right)}}\\
%\displaystyle{\frac{2^{\frac{{\mathbb{C}}_4}{{\mathbb{C}}_3-y}}-1-e^{tran}{\varepsilon }_{1}}{e^{tran}\left({{\varepsilon }_{2}-\varepsilon }_{1}\right)}},\\ \qquad~\displaystyle{{\mathbb{C}}_3-\frac{{\mathbb{C}}_4}{{\log}_{2}\left(1+e^{tran}{\varepsilon }_{1}\right)}\le y\le}\\
%\qquad~\displaystyle{{\mathbb{C}}_3-\frac{{\mathbb{C}}_4}{{\log}_2\left(1+e^{tran}{\varepsilon }_{2}\right)} }\\
%\displaystyle{1, y>{\mathbb{C}}_3-\frac{{\mathbb{C}}_4}{{\log}_2\left(1+e^{tran}{\varepsilon }_{2}\right)}}\\
% \end{cases}
% \end{align}
{\setlength\abovedisplayskip{5pt}
\setlength\belowdisplayskip{5pt}
\begin{align}
\label{eq40}
{\text{F}}_{Y2}\left(y\right)
%&=\text{Pr}\left(Y2\le y\right)=\text{Pr}\left(Y1\ge \frac{{\mathbb{C}}_3-y}{{\mathbb{C}}_4}\right)
%=1-\text{Pr}\left(Y1<\frac{{\mathbb{C}}_3-y}{{\mathbb{C}}_4}\right)
=1-{\text{F}}_{Y1}\left(\frac{{\mathbb{C}}_3-y}{{\mathbb{C}}_4}\right)
=\begin{cases}
0,&y<{\mathbb{C}}_3-\frac{{\mathbb{C}}_4}{{\log}_2\left(1+e^{tran}{\varepsilon}_{\text{1}}\right)}\vspace{-0.6ex} \\
\frac{2^{\frac{{\mathbb{C}}_4}{{\mathbb{C}}_3-y}}-1-e^{tran}{\varepsilon }_{1}}{e^{tran}\left({{\varepsilon }_{2}-\varepsilon }_{1}\right)},&{\mathbb{C}}_3-\frac{{\mathbb{C}}_4}{{\log}_{2}\left(1+e^{tran}{\varepsilon }_{1}\right)}\le y\le{\mathbb{C}}_3-\frac{{\mathbb{C}}_4}{{\log}_2\left(1+e^{tran}{\varepsilon}_{2}\right)}\vspace{-0.6ex} \\
1,&y>{\mathbb{C}}_3-\frac{{\mathbb{C}}_4}{{\log}_2\left(1+e^{tran}{\varepsilon }_{2}\right)}\\
 \end{cases}
 \end{align}}Let random variable $Y3={\alpha}_{m}Y2$, the CDF of $Y3$ is thus considered by~\eqref{eq41}.
% \begin{align}
% \label{eq41}
%  {\text{F}}_{Y3}\left(y\right)&=\text{Pr}\left(Y3\le y\right)
%  =\text{Pr}\left({\alpha }_{m}Y2\le y\right)\notag\\
%  &\begin{cases}
% 0,\quad y<0 \\
%\displaystyle{1-a,\quad 0\le y<{\mathbb{C}}_3-\frac{{\mathbb{C}}_4}{{\log}_{2}\left({1+e}^{tran}{\varepsilon }_1\right)} }\\
%\displaystyle{1-a+a\left(\frac{2^{\frac{{\mathbb{C}}_4}{{\mathbb{C}}_3-y}}-1-e^{tran}{\varepsilon }_1}{e^{tran}\left({{\varepsilon }_2-\varepsilon }_1\right)}\right)},\\ \qquad~\displaystyle{{\mathbb{C}}_3-\frac{{\mathbb{C}}_4}{{\log}_{2}\left(1+e^{tran}{\varepsilon }_1\right)}\le}\\
%\qquad \displaystyle{y\le {\mathbb{C}}_3-\frac{{\mathbb{C}}_4}{{log}_{2}\left(1+e^{tran}{\varepsilon }_2\right)}} \\
%\displaystyle{1,\quad y>{\mathbb{C}}_3-\frac{{\mathbb{C}}_4}{{\log}_{2}\left(1+e^{tran}{\varepsilon }_2\right)}} \\
%  \end{cases}
% \end{align}
{\setlength\abovedisplayskip{5pt}
\setlength\belowdisplayskip{5pt}
\begin{align}
 \label{eq41}
  {\text{F}}_{Y3}\left(y\right)
%  &=\text{Pr}\left(Y3\le y\right)
%  =\text{Pr}\left({\alpha}_{m}Y2\le y\right)\notag\\
 =
\begin{cases}
 0,&y<0\vspace{-0.8ex}  \\
1-a,&0\le y<{\mathbb{C}}_3-\frac{{\mathbb{C}}_4}{{\log}_{2}\left({1+e}^{tran}{\varepsilon }_1\right)}\vspace{-0.8ex} \\
1-a+a\left(\frac{2^{\frac{{\mathbb{C}}_4}{{\mathbb{C}}_3-y}}-1-e^{tran}{\varepsilon }_1}{e^{tran}\left({{\varepsilon}_2-\varepsilon }_1\right)}\right),&{\mathbb{C}}_3-\frac{{\mathbb{C}}_4}{{\log}_{2}\left(1+e^{tran}{\varepsilon }_1\right)}\le y\le {\mathbb{C}}_3-\frac{{\mathbb{C}}_4}{{log}_{2}\left(1+e^{tran}{\varepsilon }_2\right)} \vspace{-0.8ex} \\
1,&y>{\mathbb{C}}_3-\frac{{\mathbb{C}}_4}{{\log}_{2}\left(1+e^{tran}{\varepsilon }_2\right)}\vspace{-0.8ex}  \\
  \end{cases}
 \end{align}}Accordingly, we recalculate~(39) as~\eqref{eq42}, according to~\eqref{eq41}.
%\begin{align*}
%{\mathcal{R}}^{MRisk}\left(p,q,\bm{\mathcal{A}},\bm{\mathcal{Y}}\right)=\text{Pr}\left(Y3\le {\mathbb{C}}_5\right)
%\end{align*}
%\begin{align}
%\label{eq42}
%=&\begin{cases}
%0,\quad {\mathbb{C}}_5<0 \\
%\displaystyle{1-a,~ 0\le {\mathbb{C}}_5<{\mathbb{C}}_3-\frac{{\mathbb{C}}_4}{{\log}_{2}\left(1+e^{tran}{\varepsilon }_1\right)}} \\
%\displaystyle{1-a+a\left(\frac{2^{\frac{{\mathbb{C}}_4}{{\mathbb{C}}_3-{\mathbb{C}}_5}}-1-e^{tran}{\varepsilon }_1}{e^{tran}\left({{\varepsilon }_2-\varepsilon }_1\right)}\right)},\\ \qquad~\displaystyle{{\mathbb{C}}_3-\frac{{\mathbb{C}}_4}{{\log}_{2}\left(1+e^{tran}{\varepsilon }_1\right)}\le {\mathbb{C}}_5\le} \\ \qquad~\displaystyle{{\mathbb{C}}_3-\frac{{\mathbb{C}}_4}{{\log}_{2}\left(1+e^{tran}{\varepsilon }_2\right)}} \\
%1,\quad \displaystyle{{\mathbb{C}}_5>{\mathbb{C}}_3
%-\frac{{\mathbb{C}}_4}{{\log}_{2}\left(1+e^{tran}{\varepsilon }_2\right)}}
%\end{cases}
%\end{align}
{\setlength\abovedisplayskip{5pt}
\setlength\belowdisplayskip{5pt}
\begin{align}
\label{eq42}
&{\mathcal{R}}^{MRisk}\left(p,q,\bm{\mathcal{A}},\bm{\mathcal{Y}}\right)
=\text{Pr}\left(Y3\le {\mathbb{C}}_5\right)\notag\\
=
&\begin{cases}
0,&\mathbb{C}_5<0 \vspace{-0.8ex}  \\
1-a,&0\le {\mathbb{C}}_5<{\mathbb{C}}_3-\frac{{\mathbb{C}}_4}{{\log}_{2}\left(1+e^{tran}{\varepsilon }_1\right)} \vspace{-0.8ex}  \\
1-a+a\left(\frac{2^{\frac{{\mathbb{C}}_4}{{\mathbb{C}}_3-{\mathbb{C}}_5}}-1-e^{tran}{\varepsilon }_1}{e^{tran}\left({{\varepsilon }_2-\varepsilon }_1\right)}\right),&{\mathbb{C}}_3-\frac{{\mathbb{C}}_4}{{\log}_{2}\left(1+e^{tran}{\varepsilon }_1\right)}\le {\mathbb{C}}_5\le{\mathbb{C}}_3-\frac{{\mathbb{C}}_4}{{\log}_{2}\left(1+e^{tran}{\varepsilon }_2\right)} \vspace{-0.8ex}  \\
1,&{\mathbb{C}}_5>{\mathbb{C}}_3-\frac{{\mathbb{C}}_4}{{\log}_{2}\left(1+e^{tran}{\varepsilon }_2\right)}
\end{cases}
\end{align}}For VRsik, we first discuss the conditional probability~$\text{Pr}(\sum\nolimits^{m=\kappa -1}_{m=1}{{\alpha}_{m}}$ $>S-1|{\alpha}_{\kappa }=1)$, describing that a member $\bm{b_\kappa}$ is under the risk of being selected as a volunteer when it is a performer (e.g., ${\alpha}_{\kappa}=1$, here, we consider member  ${\bm{b_\kappa}}$ as an example, where the risk is universal for all the members in the proposed market due to that all the buyers are i.i.d.). Let random variable $X3=\sum\nolimits^{m=\kappa-1}_{m=1}{{\alpha }_{m}}$, the CDF of $X3$ is expressed by~\eqref{eq43}.
%\begin{align}
%\label{eq43}
%{\text{F}}_{X3}\left(x\right)&=\text{Pr}\left(X3\le x\right)\notag\\
%&=\begin{cases}
%0,\quad x<0 \\
%\sum\nolimits^{i=\left\lfloor x\right\rfloor}_{i=0}
%{C^i_{\kappa -1}a^i{\left(1-a\right)}^{\kappa -1-i}}, \\
%\qquad~0\le x\le \kappa -1\\
%1,\quad x>\kappa -1
%\end{cases}
%\end{align}
{\setlength\abovedisplayskip{5pt}
\setlength\belowdisplayskip{5pt}
\begin{align}
\label{eq43}
{\text{F}}_{X3}\left(x\right)&=\text{Pr}\left(X3\le x\right)
=\begin{cases}
0,&x<0   \vspace{-1.5ex}\\
\sum\nolimits^{i=\left\lfloor x\right\rfloor}_{i=0}
{C^i_{\kappa -1}a^i{\left(1-a\right)}^{\kappa -1-i}},&0\le x\le \kappa -1   \vspace{-1.5ex}\\
1,&x>\kappa -1
\end{cases}
\end{align}}Correspondingly, we have $\text{Pr}\left(\left.\sum^{m=\kappa-1}_{m=1}{{\alpha }_{m}}>S-1\right|{\alpha }_{\kappa}=1\right)$ calculated by~\eqref{eq44}.
%\begin{align}
%\label{eq44}
%&\quad\;\text{Pr}\left(\left.\sum\nolimits^{m=\kappa -1}_{m=1}{{\alpha }_{m}}>S-1\right|{\alpha }_{\kappa}=1\right)\notag\\
%&=\text{Pr}\left(\sum\nolimits^{m=\kappa -1}_{m=1}{{\alpha }_{m}}>S-1\right)=\text{Pr}\left(X3>S-1\right)\notag\\
%&=1-\text{Pr}\left(X3\le S-1\right)\notag\\
%&=\begin{cases}
%   0, & 0\le \kappa \le S \\
%1-\sum\nolimits^{i=S-1}_{i=0}{C^i_{\kappa -1}a^i{\left(1-a\right)}^{\kappa -1-i}},& \kappa >S\\
% \end{cases}
%\end{align}
{\setlength\abovedisplayskip{5pt}
\setlength\belowdisplayskip{5pt}
\begin{align}
\label{eq44}
\text{Pr}\left(\left.\sum\nolimits^{m=\kappa -1}_{m=1}{{\alpha }_{m}}>S-1\right|{\alpha }_{\kappa}=1\right)
%=\text{Pr}\left(\sum\nolimits^{m=\kappa -1}_{m=1}{{\alpha }_{m}}>S-1\right)=\text{Pr}\left(X3>S-1\right)\notag\\
%=1-\text{Pr}\left(X3\le S-1\right)
=\begin{cases}
0, &~0\le \kappa \le S   \vspace{-1.5ex}\\
1-\sum\nolimits^{i=S-1}_{i=0}{C^i_{\kappa -1}a^i{\left(1-a\right)}^{\kappa -1-i}},&~\kappa >S\\
 \end{cases}
\end{align}}Consequently, the probability of a performer who is undergoing the risk of being selected as a volunteer is given by~\eqref{eq45}.
%\begin{align}\label{eq45}
%  {\mathcal{R}}^{VRisk}(\bm{\mathcal{A}},\kappa)&\!=\!\text{Pr}\left(\left(\sum\nolimits^{m=\kappa -1}_{m=1}{{\alpha }_{m}}>S\!-\!1\right)\cap \left({\alpha}_{\kappa}=1\right)\right)\notag\\
%  &\!=\!\text{Pr}\left(\left.\sum\nolimits^{m=\kappa -1}_{m=1}{{\alpha }_{m}}>S-1\right|{\alpha}_{\kappa}=1\right)\times\notag\\
%  &\quad~ \text{Pr}\left({\alpha}_{\kappa }=1\right)\notag\\
%  &\!=\!\begin{cases}
%0,\quad 0\le \kappa \le S \\
%a-\sum\nolimits^{i=S-1}_{i=0}{C^i_{\kappa -1}a^{i+1}{\left(1-a\right)}^{\kappa -1-i}},\\
%\qquad \kappa >S
% \end{cases}
%\end{align}
\begin{align}\label{eq45}
  {\mathcal{R}}^{VRisk}(\bm{\mathcal{A}},\kappa)
%  &\!=\!\text{Pr}\left(\left(\sum\nolimits^{m=\kappa -1}_{m=1}{{\alpha }_{m}}>S\!-\!1\right)\cap \left({\alpha}_{\kappa}=1\right)\right)\notag\\
%&\!=\!\text{Pr}\left(\left.\sum\nolimits^{m=\kappa -1}_{m=1}{{\alpha }_{m}}>S-1\right|{\alpha}_{\kappa}=1\right)\times \text{Pr}\left({\alpha}_{\kappa }=1\right)\notag\\
&\!=\!\begin{cases}
0,&0\le \kappa \le S \vspace{-1.5ex}\\
a-\sum\nolimits^{i=S-1}_{i=0}{C^i_{\kappa -1}a^{i+1}{\left(1-a\right)}^{\kappa -1-i}},&\kappa >S
 \end{cases}
\end{align}

%\vspace{-1.7em}

 \section{Derivation of Risk of Seller}
\noindent
 We apply the previous defined $X1$ and $X2$ to describe ${\mathcal{R}}^{SRisk}(p,q,r,\kappa ,\bm{\mathcal{A}})$ as (46).
% \begin{gather}
% \label{eq46}
%   {\mathcal{R}}^{SRisk}(p,q,r,\kappa ,\bm{\mathcal{A}})=\text{Pr}\left(\frac{{\mathcal{U}}^{SelF}\left(p,q,r,\kappa ,\bm{\mathcal{A}}\right)}{\overline{{\mathcal{U}}^{SelF}}\left(p,q,r,\kappa ,\bm{\mathcal{A}}\right)}\le {\xi }_2\right)\notag\\
%   \begin{align}
%   &=\text{Pr}\Bigg({\mathcal{U}}^{SelF}\left(p,q,r,\kappa ,\bm{\mathcal{A}}\right)\notag\\
%   &\left.\le {\xi }_2\overline{{\mathcal{U}}^{SelF}}\left(p,q,r,\kappa ,\bm{\mathcal{A}}\right)\right)\notag\\
%   &=\text{Pr}\left((p+r)X2-(q+r)X1\right.\notag\\
%   &\left.\le \frac{{\xi}_2\overline{{\mathcal{U}}^{SelF}}\left(p,q,r,\kappa ,\bm{\mathcal{A}}\right)}{d^{comp}}-q\kappa \right)
%   \end{align}
% \end{gather}
{\setlength\abovedisplayskip{5pt}
\setlength\belowdisplayskip{5pt}
\begin{gather}
 \label{eq46}
   {\mathcal{R}}^{SRisk}(p,q,r,\kappa ,\bm{\mathcal{A}})
%   \text{Pr}\left(\frac{{\mathcal{U}}^{SelF}\left(p,q,r,\kappa ,\bm{\mathcal{A}}\right)}{\overline{{\mathcal{U}}^{SelF}}\left(p,q,r,\kappa ,\bm{\mathcal{A}}\right)}\le {\xi }_2\right)\notag\\
%   &=\text{Pr}\Bigg({\mathcal{U}}^{SelF}\left(p,q,r,\kappa ,\bm{\mathcal{A}}\right)\notag\\
%   &\left.\le {\xi }_2\overline{{\mathcal{U}}^{SelF}}\left(p,q,r,\kappa ,\bm{\mathcal{A}}\right)\right)\notag\\
=\text{Pr}\left((p+r)X2-(q+r)X1\le \frac{{\xi}_2\overline{{\mathcal{U}}^{SelF}}\left(p,q,r,\kappa ,\bm{\mathcal{A}}\right)}{d^{comp}}-q\kappa \right)
 \end{gather}}Consider $\kappa \le S$, we have$\ X2=X1$. Thus, ${\mathcal{R}}^{SelF}(p,q,r,\kappa$ $\le S,\bm{\mathcal{A}})$ can be calculated by~(47), where ${\mathbb{C}}_6=\frac{{\xi }_2\overline{{\mathcal{U}}^{SelF}}(p,q,r,\kappa \le S,\bm{\mathcal{A}})}{d^{comp}(p-q)}-\frac{q\kappa }{(p-q)}$ for notational simplicity.
%\begin{gather}
%\label{eq47}
%{\mathcal{R}}^{SRisk}\left(p,q,r,\kappa \le S,\bm{\mathcal{A}}\right)=\text{Pr}\left(X1\le {\mathbb{C}}_6\right)\notag\\
%=\begin{cases}
%  0,&{\mathbb{C}}_6<0\\
%\sum\nolimits^{i=\left\lfloor {\mathbb{C}}_6\right\rfloor }_{i=0}{C^i_{\kappa }a^i{\left(1-a\right)}^{\kappa -i}},&0\le {\mathbb{C}}_6\le \kappa \\
%1,&{\mathbb{C}}_6>\kappa
% \end{cases}
%\end{gather}
{\setlength\abovedisplayskip{5pt}
\setlength\belowdisplayskip{5pt}
\begin{gather}
\label{eq47}
{\mathcal{R}}^{SRisk}\left(p,q,r,\kappa \le S,\bm{\mathcal{A}}\right)=\text{Pr}\left(X1\le {\mathbb{C}}_6\right)
=
\begin{cases}
  0,&{\mathbb{C}}_6<0 \vspace{-1.5ex}\\
\sum\nolimits^{i=\left\lfloor {\mathbb{C}}_6\right\rfloor }_{i=0}{C^i_{\kappa }a^i{\left(1-a\right)}^{\kappa -i}},&0\le {\mathbb{C}}_6\le \kappa \vspace{-1.5ex}\\
1,&{\mathbb{C}}_6>\kappa
 \end{cases}
\end{gather}}For $\kappa>S$, we consider a random variable $Z$ given by~\eqref{eq48},
%\begin{align}
%\label{eq48}
% Z&=(p+r)X2-\left(q+r\right)X1\notag\\
% &=\begin{cases}
%   \left(p-q\right)X1,& X1<S \\
%\left(p+r\right)S-\left(q+r\right)X1,& X1\ge S\\
% \end{cases}
%\end{align}
{\setlength\abovedisplayskip{5pt}
\setlength\belowdisplayskip{5pt}
\begin{align}
\label{eq48}
 Z&=(p+r)X2-\left(q+r\right)X1
 =\begin{cases}
   \left(p-q\right)X1,& X1<S \vspace{-1.5ex}\\
\left(p+r\right)S-\left(q+r\right)X1,& X1\ge S \\
 \end{cases}
\end{align}}Correspondingly, the PMF of $Z$ can be calculated as~\eqref{eq49} based on~\eqref{eq31}.
%\begin{align}
%\label{eq49}
%&\text{Pr}\left(Z=z\right)=\notag\\
%&\begin{cases}
%{\left(1-a\right)}^{\kappa},& z=0\\
%{C^1_{\kappa }a^1\left(1-a\right)}^{\kappa -1},& z=\left(p-q\right)\\
%\vdots  \\
%{C^{S-1}_{\kappa }a^{S-1}\left(1-a\right)}^{\kappa -S+1},&z=(S-1)\left(p-q\right)\\
%{C^{\rm S}_{\kappa }a^{\rm S}\left(1-a\right)}^{\kappa -S},
%&z=S\left(p-q\right)\\
%{C^{S+1}_{\kappa }a^{S+1}\left(1\!-\!a\right)}^{\kappa\!-\!S\!-\!1},&z=S\left(p\!-\!q\right)-(q+r) \\
%\vdots  \\
%a^{\kappa},&z=S\left(p-q\right)-\\
%&\left(\kappa -S\right)(q+r)\\
%0,& \text{otherwise}
%\end{cases}
%\end{align}
{\setlength\abovedisplayskip{5pt}
\setlength\belowdisplayskip{5pt}
\begin{align}
\label{eq49}
&\text{Pr}\left(Z=z\right)=
\begin{cases}
{\left(1-a\right)}^{\kappa},& z=0 \vspace{-1.5ex}\\
%{C^1_{\kappa }a^1\left(1-a\right)}^{\kappa -1},& z=\left(p-q\right) \vspace{-1.5ex}\\
\vdots  \vspace{-1.5ex}\\
%{C^{S-1}_{\kappa }a^{S-1}\left(1-a\right)}^{\kappa -S+1},&z=(S-1)\left(p-q\right)\\
{C^{S}_{\kappa }a^{S}\left(1-a\right)}^{\kappa-S},
&z=S\left(p-q\right) \vspace{-1.5ex}\\
%{C^{S+1}_{\kappa }a^{S+1}\left(1\!-\!a\right)}^{\kappa\!-\!S\!-\!1},&z=S\left(p\!-\!q\right)-(q+r) \\
\vdots  \vspace{-1.5ex}\\
a^{\kappa},&z=S\left(p-q\right)-\left(\kappa-S\right)(q+r)  \vspace{-1.5ex}\\
0,& \text{otherwise}
\end{cases}
\end{align}}Correspondingly, the CDF ${\text{F}}_Z\left(z\right)$ of $Z$ is discussed via considering three cases: \textbf{Case~1} ($q+r=p-q$), \textbf{Case~2} ($q+r>p-q$), and \textbf{Case~3} ($q+r<p-q$). \textbf{Case~1} is analyzed by the following:

\noindent
$\bullet$ \textbf{Case 1.1} When $S<\kappa \le 2S$, we have ${\text{F}}_Z\left(z\right)$ shown by~\eqref{eq50}.
%(e.g., curves \circled{1} and \circled{2} in Case 1, Fig.~7)
% \begin{align}
% \label{eq50}
%{\text{F}}_Z\left(z\right)&=\text{Pr}\left(Z\le z\right)\notag\\
%&=\begin{cases}
%0,\quad z<0 \\
%\sum\nolimits^{i=\left\lfloor \frac{z}{p-q}\right\rfloor }_{i=0}C^i_{\kappa }a^i{\left(1-a\right)}^{\kappa -i}+\\
%\qquad\sum\nolimits^{i=\kappa}_{i=\left\lceil 2S-\frac{z}{p-q}\right\rceil}{C^i_{\kappa }a^i{\left(1-a\right)}^{\kappa -i}},\\
%\qquad 0\le z\le S\left(p-q\right) \\
%1,\quad z>S\left(p-q\right)
%    \end{cases}
% \end{align}
{\setlength\abovedisplayskip{5pt}
\setlength\belowdisplayskip{5pt}
\begin{align}
 \label{eq50}
{\text{F}}_Z\left(z\right)=
\begin{cases}
0,& z<0 \vspace{-1.5ex}\\
\sum\nolimits^{i=\left\lfloor \frac{z}{p-q}\right\rfloor }_{i=0}C^i_{\kappa }a^i{\left(1-a\right)}^{\kappa -i}+\sum\nolimits^{i=\kappa}_{i=\left\lceil 2S-\frac{z}{p-q}\right\rceil}{C^i_{\kappa }a^i{\left(1-a\right)}^{\kappa -i}}, &0\le z\le S\left(p-q\right)   \vspace{-1.5ex}\\
1,& z>S\left(p-q\right)
    \end{cases}
 \end{align}}$\bullet$ \textbf{Case 1.2} When $\kappa>2S$, we have~the following \eqref{eq51}.
%(e.g., curves \circled{1},\circled{2}, and \circled{3} in Case 1, Fig. 7)
%\begin{align}
%\label{eq51}
%  {\text{F}}_Z\left(z\right)&=\text{Pr}\left(Z\le z\right)\notag\\
% & =\begin{cases}
%0,\quad z<(2S-\kappa)(p-q)\\
%\sum\nolimits^{i=\kappa}_{i=\left\lceil 2S-\frac{z}{p-q}\right\rceil}{C^i_{\kappa }a^i{\left(1-a\right)}^{\kappa -i}},\\
%\qquad (2S-\kappa )(p-q)\le z<0 \\
%\sum\nolimits^{i=\left\lfloor \frac{z}{p-q}\right\rfloor }_{i=0}C^i_{\kappa }a^i{\left(1-\alpha \right)}^{\kappa -i}+\\
%\qquad~\sum\nolimits^{i=\kappa }_{i=\left\lceil 2S-\frac{z}{p-q}\ \right\rceil }{C^i_{\kappa }a^i{\left(1-a\right)}^{\kappa -i}},\\
%\qquad~0\le z\le S\left(p-q\right)\\
%1,\quad z>S\left(p-q\right)
%\end{cases}
%\end{align}
{\setlength\abovedisplayskip{5pt}
\setlength\belowdisplayskip{5pt}
\begin{align}
\label{eq51}
  {\text{F}}_Z\left(z\right)
 & =\begin{cases}
0,~z<(2S-\kappa)(p-q)\\
\sum\nolimits^{i=\kappa}_{i=\left\lceil 2S-\frac{z}{p-q}\right\rceil}{C^i_{\kappa }a^i{\left(1-a\right)}^{\kappa -i}},~(2S-\kappa )(p-q)\le z<0 \\
\sum\nolimits^{i=\left\lfloor \frac{z}{p-q}\right\rfloor }_{i=0}C^i_{\kappa }a^i{\left(1-\alpha \right)}^{\kappa -i}+\sum\nolimits^{i=\kappa }_{i=\left\lceil 2S-\frac{z}{p-q}\ \right\rceil }{C^i_{\kappa }a^i{\left(1-a\right)}^{\kappa -i}},~0\le z\le S\left(p-q\right)\\
1,~z>S\left(p-q\right)
\end{cases}
\end{align}}Consequently, CDF of $Z$ in \textbf{Case 1} is given as~\eqref{eq52} by summarizing (50) and (51):
%\begin{align}
%\label{eq52}
%{\text{F}}_Z\left(z\right)&=\text{Pr}\left(Z\le z\right)\notag\\
%&=\begin{cases}
%0,~ z<{\left(0,(2S-\kappa )(p-q)\right)}^- \\
%\sum\nolimits^{i=\left\lfloor \frac{z}{p-q}\right\rfloor }_{i=0}C^i_{\kappa }a^i{\left(1-a\right)}^{\kappa -i}+\\
%\qquad\sum\nolimits^{i=\kappa }_{i=\left\lceil 2S-\frac{z}{p-q} \right\rceil }{C^i_{\kappa }a^i{\left(1-a\right)}^{\kappa -i}},\\
%\qquad{\left(0,(2S\!-\!\kappa )(p\!-\!q)\right)}^{-} \!\le\! z\!\le\! S\left(p\!-\!q\right)\\
%1,\ z>S\left(p-q\right)
%\end{cases}
%\end{align}
{\setlength\abovedisplayskip{6pt}
\setlength\belowdisplayskip{5pt}
\begin{align}
\label{eq52}
{\text{F}}_Z\left(z\right)
=&\begin{cases}
0,~z<{\left(0,(2S-\kappa )(p-q)\right)}^- \\
\sum\nolimits^{i=\left\lfloor \frac{z}{p-q}\right\rfloor }_{i=0}C^i_{\kappa }a^i{\left(1-a\right)}^{\kappa -i}+\sum\nolimits^{i=\kappa }_{i=\left\lceil 2S-\frac{z}{p-q} \right\rceil }{C^i_{\kappa }a^i{\left(1-a\right)}^{\kappa -i}},\\
{\left(0,(2S\!-\!\kappa )(p\!-\!q)\right)}^{-} \!\le\! z\!\le\! S\left(p\!-\!q\right)\\
1,~z>S\left(p-q\right)
\end{cases}
\end{align}}Due to space limitation, we omit derivations of \textbf{Case 2} ($q+r>p-q$) and \textbf{Case 3 }($q+r<p-q$), which are similar with \textbf{Case 1}. In conclusion, we have ${\text{F}}_Z\left(z\right)$ when $\kappa \ge S$ as \eqref{eq53}.
%(the relevant diagram can be found in Fig. 7)
%\begin{align}
%\label{eq53}
%{\text{F}}_Z\left(z\right)&=\text{Pr}\left(Z\le z\right)\notag\\
%&=\begin{cases}
%0,\ z<{\left(0,S\left(p\!-\!q\right)\!-\!\left(\kappa\!-\!S\right)(q\!+\!r)\right)}^- \\
%\sum\nolimits^{i=\left\lfloor \frac{z}{p-q}\right\rfloor }_{i=0}{C^i_{\kappa }a^i{\left(1-a\right)}^{\kappa -z'}}+\\
%\quad\sum\nolimits^{i=\kappa }_{i=\left\lceil \frac{S\left(p-q\right)-z}{q+r}+S\right\rceil}{C^i_{\kappa }a^i{\left(1-a\right)}^{\kappa-i}},\\
%\quad{\left(0,S\left(p-q\right)-\left(\kappa -S\right)(q+r)\right)}^-\\
%\quad\le z\le S\left(p-q\right) \\
%1,\ z>S\left(p-q\right)
%\end{cases}
{\setlength\abovedisplayskip{5pt}
\setlength\belowdisplayskip{5pt}
\begin{align}
\label{eq53}
{\text{F}}_Z\left(z\right)
&=\begin{cases}
0,~z<{\left(0,S\left(p\!-\!q\right)\!-\!\left(\kappa\!-\!S\right)(q\!+\!r)\right)}^- \\
\sum\nolimits^{i=\left\lfloor \frac{z}{p-q}\right\rfloor }_{i=0}{C^i_{\kappa }a^i{\left(1-a\right)}^{\kappa -i}}+\sum\nolimits^{i=\kappa }_{i=\left\lceil \frac{S\left(p-q\right)-z}{q+r}+S\right\rceil}{C^i_{\kappa }a^i{\left(1-a\right)}^{\kappa-i}},\\
{\left(0,S\left(p-q\right)-\left(\kappa -S\right)(q+r)\right)}^-\le z\le S\left(p-q\right) \\
1,~z>S\left(p-q\right)
\end{cases}
\end{align}}Notably, let $\sum\nolimits^{i=\left\lfloor \frac{z}{p-q}\right\rfloor }_{i=0}{C^i_{\kappa }a^i{\left(1-a\right)}^{\kappa -i}}=0$ when $\frac{z}{p-q}<0$, and $\sum\nolimits^{i=\kappa }_{i=\left\lceil \frac{S\left(p-q\right)-z}{q+r}+S\ \right\rceil }{C^i_{\kappa }a^i{\left(1-a\right)}^{\kappa -i}}=0$ when $\lceil \frac{S\left(p-q\right)-z}{q+r}+S\rceil >\kappa $.
%Correspondingly, risk of the seller upon considering $\kappa \ge S$ is calculated by~\eqref{eq54}, where ${\mathbb{C}}_7=\frac{{\xi}_2\overline{{\mathcal{U}}^{SelF}}\left(p,q,r,\kappa >S,\bm{\mathcal{A}}\right)}{d^{comp}}-q\kappa $ for notational simplicity.
Correspondingly, risk of the seller upon considering $\kappa >S$ can thus be calculated by~\eqref{eq22}, according to (53).
 \section{Derivation of the Optimal Offloading Rate}
\noindent
Under any given price $g_{n}$, we discuss the optimization problem (26b) of maximizing a non-member's (${\alpha }_{n}=1,n\in \left\{\kappa +1,\dots ,\left|\bm{\mathcal{B}}\right|\right\}$) utility in problem $\bm{\mathcal{F}_2}$ by the following cases.

\noindent
$\bullet$ \textbf{Case 1:} when $\frac{{\lambda}_{n}d^{size}}{W{\log}_{2}(1+e^{tran}{\gamma}_{n})}+\frac{{\lambda}_{n}d^{comp}}{f^{s}}\le \frac{(1-{\lambda}_{n})d^{comp}}{f^{b}}$, we have $0\le{\lambda}_{n}\le {\mathbb{C}}_8$ where ${\mathbb{C}}_8=\frac{d^{comp}f^{s}}{\frac{d^{size}f^{s}f^{b}}{W{\log}_{2}(1+e^{tran}{\gamma}_{n})}+d^{comp}f^{b}+d^{comp}f^{s}}$  for notational simplicity. Thus, (26b) is rewritten as $\bm{\mathcal{F}_4}$.
%\begin{align}
%\label{eq55}
%&\bm{{\mathcal{F}_4}}:\argmin\limits_{{\lambda}_{n}\in \left[0,{\mathbb{C}}_8\right]} \bigg(\frac{{\omega}_2e^{tran}d^{size}}{W{\log}_{2}\left(1+e^{tran}{\gamma }_{n}\right)}+\notag\\
%&\qquad g_{n}d^{comp}-\frac{{\omega}_1d^{comp}}{f^{b}}-\frac{{\omega}_2e^{loc}d^{comp}}{f^{b}}\bigg){\lambda}_{n}
%\end{align}
{\setlength\abovedisplayskip{5pt}
\setlength\belowdisplayskip{5pt}
\begin{align}
\label{eq54}
&\bm{\mathcal{F}_4}:\argmin\limits_{{\lambda}_{n}\in \left[0,{\mathbb{C}}_8\right]} \bigg(\frac{{\omega}_2e^{tran}d^{size}}{W{\log}_{2}\left(1+e^{tran}{\gamma }_{n}\right)}+g_{n}d^{comp}-\frac{{\omega}_1d^{comp}}{f^{b}}-\frac{{\omega}_2e^{loc}d^{comp}}{f^{b}}\bigg){\lambda}_{n}
\end{align}}In this case, when $\frac{\partial {\mathcal{U}}^{NonM}_{n}}{\partial {\lambda}_{n}}\le 0$, we have ${\lambda}_{n}=0$; else, we have ${\lambda}_{n}={\mathbb{C}}_8$.

\noindent
$\bullet$ \textbf{Case 2: }when $\frac{{\lambda}_{n}d^{size}}{W{\log}_{2}\left(1+e^{tran}{\gamma }_{n}\right)}+\frac{{\lambda}_{n}d^{comp}}{f^{s}}>\frac{\left(1-{\lambda}_{n}\right)d^{comp}}{f^{b}}$, we have ${\mathbb{C}}_8\le {\lambda}_{n}\le 1$. We reconsider (26b) as ${\bm{\mathcal{F}_5}}$ shown by~(55).
%\begin{align}
%\label{eq56}
%&\bm{{\mathcal{F}}_5}:\argmin\limits_{{\lambda}_{\rm n}\in ({\mathbb{C}}_8,1]} \bigg(\frac{{\omega}_1d^{size}+{\omega}_2e^{tran}d^{size}}{W{\log}_2\left(1+e^{tran}{\gamma }_{n}\right)}+\\
%&\qquad g_{n}d^{comp}+\frac{{\omega}_1d^{comp}}{f^{s}}-\frac{{\omega}_2e^{loc}d^{comp}}{f^{b}}\bigg){\lambda}_{n}
%\end{align}
%{\setlength\abovedisplayskip{5pt}
%\setlength\belowdisplayskip{5pt}
\begin{align}
\label{eq55}
&\bm{\mathcal{F}_5}:\argmin\limits_{{\lambda}_{n}\in ({\mathbb{C}}_8,1]} \bigg(\frac{{\omega}_1d^{size}+{\omega}_2e^{tran}d^{size}}{W{\log}_2\left(1+e^{tran}{\gamma }_{n}\right)}+g_{n}d^{comp}+\frac{{\omega}_1d^{comp}}{f^{s}}-\frac{{\omega}_2e^{loc}d^{comp}}{f^{b}}\bigg){\lambda}_{n}
\end{align}In this case, when  $\frac{\partial {\mathcal{U}}^{NonM}_{n}}{\partial {\lambda}_{n}}\le 0$, we have ${\lambda}_{n}={\mathbb{C}}_8$; else, we have ${\lambda}_{n}=1$. Similarly, problem (27b) can also be solved according to (54) and (55).
%\end{spacing}

%
%\begin{IEEEbiography}{Author 1}
%XXX.
%\end{IEEEbiography}
%
%\begin{IEEEbiographynophoto}{Author 2}
%XXX.
%\begin{IEEEbiographynophoto}{Author 3}
%XXX.


\begin{thebibliography}{99}
\addtolength{\itemsep}{-0.25em} 
  \bibitem{1} C. Yi, J. Cai, and Z. Su, ``A Multi-User Mobile Computation Offloading and Transmission Scheduling Mechanism for Delay-Sensitive Applications,'' \textit{IEEE Trans. Mobile Comput}., vol. 19, no. 1, pp. 29--43, 2020.

  \bibitem{2} J. Yan, S. Bi, Y. J. Zhang, and M. Tao, ``Optimal Task Offloading and Resource Allocation in Mobile-Edge Computing with Inter-User Task Dependency,'' \textit{IEEE Trans. Wireless Commun}., vol. 19, no. 1, pp. 235--250, 2020.

  \bibitem{3} T. X. Tran, and D. Pompili, ``Joint Task Offloading and Resource Allocation for Multi-Server Mobile-Edge Computing Networks,'' \textit{IEEE Trans. Veh. Technol}., vol. 68, no. 1, pp. 856--868, 2019.

  \bibitem{4} E. El Haber, T. M. Nguyen, and C. Assi, ``Joint Optimization of Computational Cost and Devices Energy for Task Offloading in Multi-Tier Edge-Clouds,'' ~\textit{IEEE Trans. Commun}., vol. 67, no. 5, pp. 3407--3421, 2019.

  \bibitem{5} M. Liwang, Z. Gao, and X. Wang, ``Let’s Trade in The Future! A Futures-Enabled Fast Resource Trading Mechanism in Edge Computing-Assisted UAV Networks,'' \textit{IEEE J. Sel. Areas Commun}., pp. 1--1, 2021.

  \bibitem{6} Z. Zhou, X. Chen, E. Li, L. Zeng, K. Luo, and J. Zhang, ``Edge Intelligence: Paving the Last Mile of Artificial Intelligence with Edge Computing,'' \textit{Proc. IEEE}, vol. 107, no. 8, pp. 1738--1762, 2019.

  \bibitem{7} L. Tom\'{a}s, and J. Tordsson, ``An Autonomic Approach to Risk-Aware Data Center Overbooking,'' \textit{IEEE Trans. Cloud Comput}., vol. 2, no. 3, pp. 292-305, 2014.

  \bibitem{8} K. Chard, and K. Bubendorfer, ``High Performance Resource Allocation Strategies for Computational Economies,'' \textit{IEEE Trans. Parallel Distrib. Syst.}, vol. 24, no. 1, pp. 72--84, 2013.

  \bibitem{9} J. Ma, Y. K. Tse, X. Wang, and M. Zhang, ``Examining Customer Perception and Behaviour Through Social Media Research--An Empirical Study of the United Airlines Overbooking Crisis,'' \textit{Transportation Research Part E: Logistics and Transportation Review}, vol. 127, pp.192--205, 2019.

  \bibitem{10} N. Haynes, and D. Egan, ``The Perceptions of Frontline Employees Towards Hotel Overbooking Practices: Exploring Ethical Challenges,'' \textit{J. Revenue Pricing Manage.}, vol. 137, pp. 1--10, 2020.

  \bibitem{11} J. Liu, X. Jiang, and S. Horiguchi, ``Opportunistic Link Overbooking for Resource Efficiency under Per-Flow Service Guarantee,'' \textit{IEEE Trans. Commun.}, vol. 58, no. 6, pp. 1769-1781, 2010.

  \bibitem{12} A. Adebayo, D. B. Rawat, and M. Song, ``Prediction Based Adaptive RF Spectrum Reservation in Wireless Virtualization,'' \textit{IEEE Int. Conf. Commun. (ICC)}, Dublin, Ireland, 2020, pp. 1-6.

  \bibitem{13} M. A. Messous, S. M. Senouci, H. Sedjelmaci, and S. Cherkaoui, ``A Game Theory based Efficient Computation Offloading in an UAV Network,'' \textit{IEEE Trans. Veh. Technol}., vol. 68, no. 5, pp. 4964--4974, 2019.

  \bibitem{14} Y. Wang, P. Lang, D. Tian, J. Zhou, X. Duan, Y. Cao, and D. Zhao, ``A Game-based Computation Offloading Method in Vehicular Multiaccess Edge Computing Networks,'' \textit{IEEE Internet Things J}., vol. 7, no. 6, pp. 4987--4996, 2020.

  \bibitem{15} G. Gao, M. Xiao, J. Wu, H. Huang, S. Wang, and G. Chen, ``Auction-based VM Allocation for Deadline-Sensitive Tasks in Distributed Edge Cloud,'' \textit{IEEE Trans. Serv. Comput}., pp. 1--1, 2019.

  \bibitem{16} M. Liwang, S. Dai, Z. Gao, Y. Tang, and H. Dai, ``A Truthful Reverse-Auction Mechanism for Computation Offloading in Cloud-Enabled Vehicular Network,'' \textit{IEEE Internet Things J.}, vol. 6, no. 3, pp. 4214--4227, 2019.

  \bibitem{17} Z. Gao, M. LiWang, S. Hosseinalipour, H. Dai, and X. Wang, ``A Truthful Auction for Graph Job Allocation in Vehicular Cloud-Assisted Networks,'' \textit{IEEE Trans. Mobile Comput}., pp. 1-1, 2021.
  \bibitem{18} B. Shojaiemehr, A. M. Rahmani, and N. N. Qader, ``A Three-Phase Process for SLA Negotiation of Composite Cloud Services,'' \textit{Computer Standards \& Interfaces}, vol. 64, pp.85-95, 2019.

  \bibitem{19} P. Wang, J. Meng, J. Chen, T. Liu, Y. Zhan, W. Tsai, and Z. Jin, ``Smart Contract-Based Negotiation for Adaptive QoS-Aware Service Composition,'' \textit{IEEE Trans. Parallel Distrib. Syst}., vol. 30, no. 6, pp. 1403-1420, 2019.

  \bibitem{20} S. E. Khatib, and F. D. Galinan, ``Negotiating Bilateral Contracts in Electricity Markets,'' \textit{IEEE Trans. Power Syst.}, vol. 22, no. 2, pp. 553--562, 2007.

  \bibitem{21} A. J. Conejo, R. Garcia-Bertrand, M. Carrion, \'{A}. Caballero, and A. de Andr\'{E}s, ``Optimal Involvement in Futures Markets of a Power Producer,'' \textit{IEEE Trans. Power Syst}., vol. 23, no. 2, pp. 703--711, 2008.

  \bibitem{22} J. M. Morales, S. Pineda, A. J. Conejo, and M. Carrion, ``Scenario Reduction for Futures Market Trading in Electricity Markets,'' \textit{IEEE Trans. Power Syst}., vol. 24, no. 2, pp. 878--888, 2009.

  \bibitem{23} S. Sheng, R. Chen, P. Chen, X. Wang, and L. Wu, ``Futures-based Resource Trading and Fair Pricing in Real-Time IoT Networks,'' \textit{IEEE Wireless Commun. Lett}., vol. 9, no. 1, pp. 125--128, 2020.

  \bibitem{24} H. Li, T. Shu, F. He, and J. B. Song, ``Futures Market for Spectrum Trade in Wireless Communications: Modeling, Pricing and Hedging,'' \textit{IEEE Int. Conf. Global. Commun. (GLOBECOM)}, Atlanta, GA, USA, Dec. 2013, pp. 1--6.

  \bibitem{25} M. Liwang, R. Chen, and X. Wang, ``Resource Trading in Edge Computing-enabled IoV: An Efficient Futures-based Approach,'' \textit{IEEE Trans. Serv. Comput.}, pp. 1--1, 2021.

  \bibitem{26} L. Gao, B. Shou, Y. J. Chen, and J. Huang, ``Combining Spot and Futures Markets: A Hybrid Market Approach to Dynamic Spectrum Access,'' \textit{Operations Res}., vol. 64, no. 4, pp. 794--821, 2016.

  \bibitem{27} K. Vanmechelen, W. Depoorter, and J. Broeckhove, ``Combining Futures and Spot Markets: A Hybrid Market Approach to Economic Grid Resource Management,'' \textit{J. Grid Comput}. Vol. 9, no. 1, pp. 81--94, 2011.

  \bibitem{28} N. Wu, X. Zhou, and M. Sun, ``Incentive Mechanisms and Impacts of Negotiation Power and Information Availability in Multi-Relay Cooperative Wireless Networks,'' \textit{IEEE Trans. Wireless Commun}., vol. 18, no. 7, pp. 3752--3765, 2019.

  \bibitem{29} X. Gao, K. Wang, and Y. Yu, ``To Rent or to Share?,'' \textit{IEEE Int. Conf. Commun., Control, Comput. Technol. Smart Grids (SmartGridComm), } Aalborg, Denmark, Oct. 2018, pp. 1--7.

  \bibitem{30} L. Zanzi, V. Sciancalepore, A. Garcia-Saavedra, and X. Costa-Perez, ``OVNES: Demonstrating 5G network slicing overbooking on real deployments,'' \textit{IEEE Conf. Comput. Commun. Workshops (INFOCOM WKSHPS)}, Honolulu, HI, USA, Apr. 2018, pp. 1--2.

  \bibitem{31} L. Zanzi, J. X. Salvat, V. Sciancalepore, A. G. Saavedra, and X. Costa-Perez, ``Overbooking network slices end-to-end: Implementation and demonstration,'' \textit{ACM SIGCOMM Conf}., New York, NY, USA, Aug. 2018, pp. 144--146.

  \bibitem{32} C. Sexton, N. Marchetti, and L. A. DaSilva, ``On Provisioning Slices and Overbooking Resources in Service Tailored Networks of the Future,'' \textit{IEEE/ACM Trans. Netw}, vol. 28, no. 5, pp. 2106--2119, 2020.

  \bibitem{33} J. Son, A. V. Dastjerdi, R. N. Calheiros, and R. Buyya, ``SLA-Aware and Energy-Efficient Dynamic Overbooking in SDN-Based Cloud Data Centers,'' \textit{IEEE Trans. Sustain. Comput.}, vol. 2, no. 2, pp. 76-89, 2017.

  \bibitem{34} S. Alanazi, and B. Hamdaoui, ``Energy-Aware Resource Management Framework for Overbooked Cloud Data Centers with SLA Assurance,'' \textit{IEEE Int. Conf. Global. Commun. (GLOBECOM)}, Abu Dhabi, United Arab Emirates, Dec. 2018, pp. 1--6.

  \bibitem{35} P. Rahimzadeh, Y. Im, G. Jung, C. Joe-Wong, and S. Ha, ``ECHO: Efficiently Overbooking Applications to Create a Highly Available Cloud,'' \textit{IEEE Int. Conf. Distrib. Comput. Syst. (ICDCS)}, Dallas, TX, USA, July. 2019, pp. 1--11.

  \bibitem{36} M. Yao, D. Chen, and J. Shang, ``Optimal Overbooking Policy for Cloud Service Providers: Profit and Service Quality, `` \textit{IEEE Access}, vol. 7, pp. 96132--96147, 2019.

  \bibitem{37} F. Zhang, Z. Tang, M. Chen, X. Zhou, and W. Jia, ``A Dynamic Resource Overbooking Mechanism in Fog Computing,'' \textit{IEEE Int. Conf. Mobile Ad Hoc and Sensor Syst. (MASS)}, Chengdu, China, Oct. 2018, pp. 89--97.

  \bibitem{38} Y. He, J. Ren, G. Yu, and Y. Cai, ``D2D Communications Meet Mobile Edge Computing for Enhanced Computation Capacity in Cellular Networks,'' \textit{IEEE Trans. Wireless Commun}., vol. 18, no. 3, pp. 1750--1763, 2019.

  \bibitem{39} F. Liu, E. Bala, E. Erkip, M. C. Beluri, and R. Yang, ``Small-Cell Traffic Balancing Over Licensed and Unlicensed Bands,'' \textit{IEEE Trans. Veh. Technol}, vol. 64, no. 12, pp. 5850--5865, 2015.

  \bibitem{40} F. Zhou, and R. Q. Hu, ``Computation Efficiency Maximization in Wireless-Powered Mobile Edge Computing Networks,'' \textit{IEEE Trans. Wireless Commun}., vol. 19, no. 5, pp. 3170--3184, 2020.

  \bibitem{41} B. Zheng, C. You, and R. Zhang, ``Intelligent Reflecting Surface Assisted Multi-User OFDMA: Channel Estimation and Training Design,'' \textit{IEEE Trans. Wireless Commun}., pp. 1--1, 2020.

  \bibitem{42} R. T. Marler, and J. S. Arora, ``The Weighted Sum Method for Multi-Objective Optimization: New Insights,'' \textit{Structural and Multidisciplinary Optimization}, vol. 41, no. 6, pp. 853--862, 2010.

  \bibitem{43} K. Deb, ``Multi-Objective Optimization Using Evolutionary Algorithms,'' \textit{John Wiley \& Sons}, 2001.

  \bibitem{44} C. Zhang, A. K. Qin, W. Shen, L. Gao, K. C. Tan, and X. Li, ``$\epsilon$-Constrained Differential Evolution Using an Adaptive $\epsilon$ -Level Control Method,'' \textit{IEEE Trans. Syst., Man, Cybern., Syst}., pp. 1--17, 2020.

  \bibitem{45} M. Liwang, Z. Gao, and X. Wang, ``Energy-aware Graph Job Allocation in Software Defined Air-Ground Integrated Vehicular Networks,''~\textit{arXiv preprint arXiv:2008.01144, 2020}.

 \bibitem{46} M. Liu, and Y. Liu, ``Price-Based Distributed Offloading for Mobile-Edge Computing with Computation Capacity Constraints,'' \textit{IEEE Wireless Commun. Lett.}, vol. 7, no. 3, pp. 420--423, 2018.

 \bibitem{47} F. Castillo-Zunino, and P. Keskinocak, ``Bi-Criteria Multiple Knapsack Problem with Grouped Items,'' \textit{arXiv preprint arXiv:2006.00322, 2020}.

%  \bibitem{48} Z. Yang, C. Pan, K. Wang, and M. Shikh-Bahaei, ``Energy Efficient Resource Allocation in UAV-Enabled Mobile Edge Computing Networks, `` \textit{IEEE Trans. Wireless Commun}., vol. 18, no. 9, pp. 4576-4589, 2019.

  \bibitem{48}5G Americas, “New Services \& Applications with 5G Ultra-Reliable Low Latency Communications,” White Paper, Nov. 2018.

\end{thebibliography}
\end{document}